\documentstyle[11pt]{article}
\newcommand{\writingdate}{March 29th, 2000}
\newcommand{\room}{\rule[-0.3cm]{0cm}{0.8cm}}

\newcommand{\vsp}{\vspace*{3mm}}

\newcommand{\be}{\begin{equation}}
\newcommand{\ee}{\end{equation}}
\newcommand{\bd}{\begin{displaymath}}
\newcommand{\ed}{\end{displaymath}}
\newcommand{\bea}{\begin{eqnarray}}
\newcommand{\eea}{\end{eqnarray}}
\newcommand{\bean}{\begin{eqnarray*}}
\newcommand{\eean}{\end{eqnarray*}}

\newcommand{\one}{1\!\!{\rm I}}
\newcommand{\sgn}{~{\rm sgn}}

\newcommand{\pprime}{{\prime\prime}}

\newcommand{\bra}{\langle}
\newcommand{\ket}{\rangle}

\newcommand{\order}{{\cal O}}
\newcommand{\minus}{\!-\!}
\newcommand{\plus}{\!+\!}

\newcommand{\bnul}{\mbox{\boldmath $0$}}

\newcommand{\bnabla}{\mbox{\protect\boldmath $\nabla$}}
\newcommand{\ba}{\mbox{\protect\boldmath $a$}}

\newcommand{\bh}{\mbox{\protect\boldmath $h$}}
\newcommand{\bk}{\mbox{\protect\boldmath $k$}}
\newcommand{\bm}{\mbox{\protect\boldmath $m$}}
\newcommand{\bq}{\mbox{\protect\boldmath $q$}}

\newcommand{\bu}{\mbox{\protect\boldmath $u$}}
\newcommand{\bv}{\mbox{\protect\boldmath $v$}}

\newcommand{\bx}{\mbox{\protect\boldmath $x$}}
\newcommand{\by}{\mbox{\protect\boldmath $y$}}

\newcommand{\bA}{\mbox{\protect\boldmath $A$}}

\newcommand{\bC}{\mbox{\protect\boldmath $C$}}
\newcommand{\bD}{\mbox{\protect\boldmath $D$}}

\newcommand{\bF}{\mbox{\protect\boldmath $F$}}
\newcommand{\bG}{\mbox{\protect\boldmath $G$}}
\newcommand{\bH}{\mbox{\protect\boldmath $H$}}

\newcommand{\bK}{\mbox{\protect\boldmath $K$}}

\newcommand{\bM}{\mbox{\protect\boldmath $M$}}

\newcommand{\bQ}{\mbox{\protect\boldmath $Q$}}
\newcommand{\bR}{\mbox{\protect\boldmath $R$}}

\newcommand{\bsigma}{\mbox{\protect\boldmath $\sigma$}}

\newcommand{\bpsi}{\mbox{\protect\boldmath $\psi$}}

\newcommand{\btheta}{\mbox{\protect\boldmath $\theta$}}
\newcommand{\bxi}{\mbox{\protect\boldmath $\xi$}}

\newcommand{\bEta}{\mbox{\protect\boldmath $\eta$}}
\newcommand{\bPhi}{\mbox{\protect\boldmath $\Phi$}}

\newcommand{\bOmega}{\mbox{\protect\boldmath $\Omega$}}
\newcommand{\ha}{\hat{a}}

\newcommand{\hh}{\hat{h}}
\newcommand{\hk}{\hat{k}}
\newcommand{\hm}{\hat{m}}
\newcommand{\hq}{\hat{q}}

\newcommand{\hx}{\hat{x}}
\newcommand{\hy}{\hat{y}}

\newcommand{\hK}{\hat{K}}
\newcommand{\hQ}{\hat{Q}}

\newcommand{\hba}{\hat{\mbox{\protect\boldmath $a$}}}

\newcommand{\hbh}{\hat{\mbox{\protect\boldmath $h$}}}
\newcommand{\hbk}{\hat{\mbox{\protect\boldmath $k$}}}

\newcommand{\hbq}{\hat{\mbox{\protect\boldmath $q$}}}

\newcommand{\hbx}{\hat{\mbox{\protect\boldmath $x$}}}
\newcommand{\hby}{\hat{\mbox{\protect\boldmath $y$}}}

\newcommand{\hbK}{\hat{\mbox{\protect\boldmath $K$}}}
\newcommand{\hbQ}{\hat{\mbox{\protect\boldmath $Q$}}}

\newcommand{\cL}{{\cal L}}

\input epsf.tex
\begin{document}
\title{\Large\bf Statistical Mechanics of Recurrent Neural Networks\\
II. Dynamics}
\author{\large\bf A.C.C. Coolen\\[3mm]
Department of Mathematics, King's College London\\
Strand, London WC2R 2LS, UK}
\date{\writingdate}
\maketitle

\tableofcontents


\clearpage\section{Introduction}

This paper, on solving the dynamics of recurrent neural networks using non-equilibrium
statistical mechanical techniques, is the sequel of \cite{part1},
which was devoted to solving the statics using equilibrium techniques. I
refer to \cite{part1} for a general introduction to recurrent neural networks
and their properties.

{\em Equilibrium} statistical mechanical techniques can provide much detailed
quantitative information on the behaviour of recurrent neural networks,
but they obviously have serious restrictions.
The first one is that, by definition, they will only provide
information on network properties in the stationary state. For associative
memories, for instance, it is not clear how one can calculate quantities like sizes of domains of
attraction without solving the dynamics. The second, and more
serious, restriction is that for equilibrium statistical mechanics to apply the
dynamics of the network under study must obey
detailed balance, i.e. absence of microscopic probability currents in the stationary state.
As we have seen in \cite{part1}, for recurrent networks in which the
dynamics take the form of a stochastic alignment of neuronal firing rates to  post-synaptic
potentials which, in turn, depend linearly on the firing rates,  this
requirement of detailed balance usually implies  symmetry of the
synaptic matrix. From a physiological point of view this
requirement is clearly
unacceptable, since it is violated in any network that obeys Dale's law as soon as an excitatory neuron is
connected to an inhibitory
one. Worse still, we saw in \cite{part1} that in
any network of graded-response neurons detailed balance will always be violated,
even when the synapses are symmetric. The situation will become even worse when we turn to
networks of yet more realistic (spike-based) neurons, such as integrate-and-fire
ones.
In contrast to this, {\em non-equilibrium} statistical mechanical techniques, it will turn out,
do not impose such biologically non-realistic restrictions on neuron types and synaptic symmetry,
and they are consequently the more
appropriate avenue for future theoretical research aimed at
solving biologically more realistic models.

The common strategy of all non-equilibrium statistical mechanical
studies is to derive and solve dynamical
laws for a suitable small set of relevant macroscopic
quantities from
the dynamical laws of the underlying microscopic neuronal system.
In order to make progress, as in equilibrium studies,
one is initially forced to pay the price of having relatively simple
model neurons, and of not having a very complicated
spatial wiring structure in the network under study; the networks
described and analysed in this paper will consequently be either fully connected, or randomly
diluted. When attempting to obtain exact dynamical solutions within this class, one then soon
finds a clear separation
of network models into two distinct complexity classes, reflecting in the dynamics a
separation which we also found in the statics. In statics one could get away with
relatively simple mathematical techniques as long as the number of
attractors of the dynamics was small compared to the number $N$ of neurons. As soon as the
number of attractors became of the order of $N$, on the other hand, one
entered the complex regime, requiring the more complicated formalism of replica theory.
In dynamics we will again find that we can get away with
relatively simple mathematical techniques as long as the number of
attractors remains small, and find closed deterministic
differential equations for macroscopic quantities with just a single
time argument.
As soon as we enter the complex regime, however, we will no longer find closed equations
for one-time macroscopic objects: we will now have to work with
correlation and response functions, which have two time arguments,
and turn to the less trivial generating functional techniques\footnote{A brief note about terminology:
strictly speaking, in this paper we will apply these techniques only to models in which time
is measured in discrete units, so that we should speak about generating functions rather than
generating functionals. However, since these techniques can and
have also been applied intensively to models with continuous time, they are in
literature often referred to as generating functional techniques,
for both discrete and continuous time.}.

In contrast to the situation in statics \cite{part1}, I cannot in this paper give many
references to textbooks on the dynamics, since these are more or
less non-existent. There would appear to be two reasons for this.
Firstly, in most physics departments non-equilibrium statistical mechanics (as a
subject) is generally taught and applied far less intensively than equilibrium statistical mechanics, and
thus the non-equilibrium studies of recurrent neural networks have been considerably less in
number and later in appearance in literature than their equilibrium
counterparts. Secondly, many of the popular textbooks on the
statistical mechanics of neural networks were written around 1989, roughly at the
point in time where non-equilibrium statistical mechanical studies
just started being taken up. When reading such textbooks one could be
forgiven for thinking that solving the dynamics of recurrent neural networks is generally ruled
out, whereas, in fact, nothing could be further from the truth.
Thus the references in this paper will, out of necessity, be mainly to research
papers. I regret that, given constraints on page numbers and
given my aim to explain ideas and techniques in a lecture notes style (rather than display encyclopedic skills),
 I will inevitably have left out relevant references.
Another consequence of the scarce and scattered nature of the literature on
the non-equilibrium statistical mechanics of recurrent
neural networks is that a situation has developed where many mathematical procedures, properties and solutions
are more or less known by the research community, but without there
being a clear reference in literature where these were first formally derived (if at
all). Examples of this are the fluctuation-dissipation theorems for
parallel dynamics and the non-equilibrium analysis of networks
with graded response neurons; often the separating boundary between
accepted general knowledge and {\em published} accepted general
knowledge is somewhat fuzzy.

The structure of this paper mirrors more or less the structure of
\cite{part1}. Again I will start with relatively simple networks,
with a small number of attractors (such as systems with uniform synapses, or with
a small number of patterns stored with Hebbian-type rules), which can be solved with
relatively simple mathematical techniques. These will now also include networks
that do not evolve to a stationary state, and networks
of graded response neurons, which could not be studied within
equilibrium statistical mechanics at all. Next follows a detour on
correlation- and response functions and their relations (i.e. fluctuation-dissipation theorems), which
serves as a prerequisite for the last section on generating
functional methods, which are indeed formulated in the language of
correlation- and response functions. In this last, more mathematically involved,
section I study symmetric and non-symmetric attractor neural networks close to
saturation, i.e. in the complex regime. I will show how to solve
the dynamics of fully connected as well as extremely diluted networks, emphasising
the (again) crucial issue of presence (or absence) of synaptic
symmetry, and compare the predictions of the (exact) generating functional formalism to
both numerical simulations and simple approximate theories.

\section{Attractor Neural Networks with Binary Neurons}

The simplest non-trivial  recurrent neural networks consist of
$N$ binary neurons $\sigma_i\in\{-1,1\}$ (see \cite{part1})
which respond stochastically
to post-synaptic potentials (or local fields) $h_i(\bsigma)$,
with $\bsigma=(\sigma_1,\ldots,\sigma_N)$.
The fields
depend linearly on the instantaneous neuron states,
$h_i(\bsigma)=\sum_{j} J_{ij}\sigma_j+\theta_i$, with the $J_{ij}$
representing synaptic efficacies, and the $\theta_i$ representing
external stimuli and/or neural thresholds.

\subsection{Closed Macroscopic Laws for Sequential Dynamics}

First I show how for sequential dynamics (where neurons are updated one after the other)
 one can calculate, from the
microscopic stochastic laws, differential equations for the probability distribution of
suitably defined macroscopic observables. For mathematical
convenience our starting point will be the continuous-time master
equation for the microscopic probability distribution $p_t(\bsigma)$
\be
\frac{d}{dt}p_t(\bsigma)=\sum_{i}\left\{w_i(F_i\bsigma)p_t(F_i\bsigma)-
w_i(\bsigma)p_t(\bsigma)\right\}
~~~~~~~~~w_i(\bsigma)=\frac{1}{2}[1\minus \sigma_i\tanh[\beta
h_i(\bsigma)]]
\label{eq:sequentialmaster}
\ee
with $F_i\Phi(\bsigma)=\Phi(\sigma_1,\ldots,\sigma_{i-1},-\sigma_i,\sigma_{i+1},\ldots,\sigma_N)$
(see \cite{part1}). I will
discuss the conditions for the evolution of these macroscopic state
variables to become {\em deterministic}
in the limit of infinitely
large networks and, in addition, be governed by a {\em closed} set of equations.
I then turn to specific models, with
and without detailed balance, and show how the macroscopic equations
can be used to illuminate and understand the dynamics of attractor
neural networks away from saturation.
\vsp

\noindent{\em A Toy Model}.
Let me illustrate the basic ideas with the help of a simple
(infinite range) toy model:
$J_{ij}=(J/N)\eta_{i}\xi_{j}$ and $\theta_{i}=0$
(the variables $\eta_{i}$ and $\xi_{i}$ are arbitrary, but may not
depend on $N$). For $\eta_{i}=\xi_{i}=1$ we get a
network with uniform synapses.
For $\eta_{i}=\xi_{i}\in\{\minus1,1\}$ and
$J>0$ we
recover the Hopfield \cite{Hopfield} model with one stored
pattern. Note: the synaptic matrix is non-symmetric
as soon as a pair $(ij)$ exists such that $\eta_{i}\xi_{j}\neq\eta_{j}\xi_{i}$, so in
general equilibrium statistical mechanics will not apply.
The local fields become
$h_{i}(\bsigma)=J\eta_{i}m(\bsigma)$ with $m(\bsigma)=\frac{1}{N}\sum_{k}\xi_{k}\sigma_{k}$. Since they depend
on the microscopic state $\bsigma$ only through
the value of $m$, the latter
quantity appears to constitute a natural macroscopic level of
description. The probability density of finding the macroscopic state $m(\bsigma)=m$ is given by
${\cal P}_{t}[m]=\sum_{\bsigma}p_{t}(\bsigma)\delta[m\minus
m(\bsigma)]$.
Its time derivative follows upon
inserting (\ref{eq:sequentialmaster}):
\bd
\frac{d}{dt}{\cal P}_{t}[m]=
\sum_{\bsigma}\sum_{k=1}^{N}p_{t}(\bsigma)w_{k}(\bsigma)\left\{\delta[m\minus
m(\bsigma)\plus\frac{2}{N}\xi_{k}\sigma_{k}]\minus\delta\left[m\minus
m(\bsigma)\right]\right\}
~~~~~~~~~~~~~~~~~~~~~~~~~~~~~~~~
\vspace*{-2mm}
\ed
\bd
~~~~~~~~~~~~~~~~~~~~~~~
~~~~~~~~~~~~~~~~~~~~=\frac{d}{dm}\left\{\sum_{\bsigma}p_{t}(\bsigma)\delta\left[m\minus
m(\bsigma)\right]\frac{2}{N}\sum_{k=1}^{N}\xi_{k}\sigma_{k}w_{k}(\bsigma)\right\}+\order(\frac{1}{N})
\ed
Inserting our expressions  for the transition rates $w_i(\bsigma)$
and the local
fields $h_i(\bsigma)$ gives:
\bd
\frac{d}{dt}{\cal P}_{t}[m]=
\frac{d}{dm}\left\{{\cal P}_{t}[m]\left[m-\frac{1}{N}\sum_{k=1}^{N}\xi_{k}\tanh[\eta_{k}\beta Jm]\right]\right\}
+\order(N^{\minus1})
\ed
In the limit $N\rightarrow\infty$ only the first term
survives. The general solution of the resulting Liouville equation is
${\cal P}_{t}[m]=\int\! dm_0~{\cal
P}_{0}[m_{0}]\delta\left[m\minus m(t|m_0)\right]$, where $m(t|m_0)$
is the solution of
\be
\frac{d}{dt}m=\lim_{N\rightarrow\infty}\frac{1}{N}\sum_{k=1}^{N}\xi_{k}\tanh[\eta_{k}\beta
Jm]-m~~~~~~~~~~m(0)= m_{0}
\label{eq:magnetisationflow}
\ee
This describes deterministic evolution; the only uncertainty in
the value of $m$ is due to uncertainty in initial conditions. If
at $t=0$ the quantity $m$ is known exactly, this will remain the case
for finite time-scales; $m$ turns out to evolve in time according
to (\ref{eq:magnetisationflow}).
\vsp

\noindent{\em Arbitrary Synapses}.
Let us now allow for less
trivial choices of the synaptic matrix $\{J_{ij}\}$ and try to calculate
the evolution in time of a given set of macroscopic observables
$\bOmega(\bsigma)=(\Omega_{1}(\bsigma),\ldots,\Omega_{n}(\bsigma))$
in the limit $N\rightarrow\infty$. There
are no restrictions yet on the form or the number $n$ of these state
variables; these will, however, arise naturally if we
require the observables $\bOmega$ to obey a closed set
of deterministic laws, as we will see.
The probability density of finding the system in macroscopic state
$\bOmega$ is given by:
\be
{\cal P}_{t}\left[\bOmega\right]=\sum_{\bsigma}p_{t}(\bsigma)\delta\left[\bOmega\minus\bOmega(\bsigma)\right]
\label{eq:macrodensity}
\ee
Its time derivative is obtained
by inserting (\ref{eq:sequentialmaster}). If in those parts of the resulting
expression which contain the operators $F_i$ we
 perform the transformations $\bsigma\rightarrow
F_i\bsigma$, we arrive at
\bd
\frac{d}{dt}{\cal
P}_{t}\left[\bOmega\right]=\sum_i\sum_{\bsigma}p_{t}(\bsigma)w_i(\bsigma)
\left\{\delta\left[\bOmega\minus\bOmega(F_i\bsigma)\right]-\delta\left[\bOmega\minus\bOmega(\bsigma)\right]\right\}
\ed
Upon writing
$\Omega_{\mu}(F_i\bsigma)=\Omega_\mu(\bsigma)\plus\Delta_{i\mu}(\bsigma)$
and making a Taylor expansion in powers of $\{\Delta_{i\mu}(\bsigma)\}$,
we finally obtain the
so-called Kramers-Moyal expansion:
\be
\frac{d}{dt}{\cal P}_{t}\left[\bOmega\right]=
\sum_{\ell\geq
1}\frac{(\minus1)^{\ell}}{\ell!}\sum_{\mu_{1}=1}^{n}\cdots\sum_{\mu_{\ell}=1}^{n}\frac{\partial^{\ell}}
{\partial\Omega_{\mu_{1}}\cdots\partial\Omega_{\mu_{\ell}}}
\left\{{\cal P}_{t}\left[\bOmega\right]
F_{\mu_{1}\cdots \mu_{\ell}}^{(\ell)}\left[\bOmega;t\right]
\right\}
\label{eq:kramersmoyal}
\ee
It involves conditional averages $\bra
f(\bsigma)\ket_{\bOmega;t}$ and the `discrete derivatives' $\Delta_{j\mu}(\bsigma)=
\Omega_{\mu}(F_{j}\bsigma)\minus\Omega_{\mu}(\bsigma)$
\footnote{
Expansion (\ref{eq:kramersmoyal}) is to be interpreted in a distributional sense, i.e. only
to be used in expressions of the form $\int\! d\bOmega {\cal
P}_{t}(\bOmega)G(\bOmega)$ with smooth functions
$G(\bOmega)$, so that
all derivatives are well-defined and finite. Furthermore,
(\ref{eq:kramersmoyal}) will only be useful if the $\Delta_{j\mu}$, which
measure the sensitivity of the macroscopic quantities to single neuron state changes,
are sufficiently
small. This is to be expected: for
finite $N$ {\em any} observable  can only
assume a finite number of possible values; only for
$N\rightarrow\infty$ may we expect smooth probability distributions
for our macroscopic quantities.}:
\be
F_{\mu_{1}\cdots \mu_{l}}^{(l)}\left[\bOmega;t\right]=
\bra\sum_{j=1}^{N}w_{j}(\bsigma)\Delta_{j\mu_{1}}(\bsigma)\cdots\Delta_{j\mu_{\ell}}(\bsigma)\ket_{\bOmega;t}
~~~~~~~~~~~
\bra f(\bsigma)\ket_{\bOmega;t}=\frac{
\sum_{\bsigma}p_{t}(\bsigma)\delta\left[\bOmega\minus\bOmega(\bsigma)\right]
f(\bsigma)}{
\sum_{\bsigma}p_{t}(\bsigma)\delta\left[\bOmega\minus\bOmega(\bsigma)\right]}
\label{eq:KMterms}
\ee
Retaining only the $\ell=1$
term in (\ref{eq:kramersmoyal}) would lead us to a Liouville equation, which
describes deterministic flow in $\bOmega$ space. Including also the $\ell=2$
term leads us to a Fokker-Planck equation which, in addition to
flow, describes diffusion of the macroscopic
probability density. Thus a sufficient condition for the observables $\bOmega(\bsigma)$
to evolve in time deterministically in the
limit $N\rightarrow\infty$ is:
\be
\lim_{N\rightarrow\infty}\sum_{\ell\geq
2}\frac{1}{\ell!}\sum_{\mu_{1}=1}^{n}\cdots\sum_{\mu_{\ell}=1}^{n}
\sum_{j=1}^{N}\bra|\Delta_{j\mu_{1}}(\bsigma)\cdots\Delta_{j\mu_{\ell}}(\bsigma)|\ket_{\bOmega;t}=0
\label{eq:conditionliouville}
\ee
In the simple case where all observables $\Omega_{\mu}$
scale similarly in the sense that all `derivatives' $\Delta_{j\mu}=\Omega_\mu(F_i\bsigma)\minus\Omega_\mu(\bsigma)$
are of the same
order in  $N$ (i.e. there is a monotonic function
$\tilde{\Delta}_N$ such that $\Delta_{j\mu}=\order(\tilde{\Delta}_{N})$ for all
$j\mu$), for instance, criterion (\ref{eq:conditionliouville}) becomes:
\be
\lim_{N\rightarrow\infty} n\tilde{\Delta}_{N}\sqrt{N}=0
\label{eq:criterion}
\ee
If for a given set of observables condition
(\ref{eq:conditionliouville}) is satisfied, we can for large $N$
describe the evolution of the macroscopic probability
density by a Liouville equation:
\bd
\frac{d}{dt}{\cal P}_{t}\left[\bOmega\right]=
-\sum_{\mu=1}^{n}\frac{\partial}{\partial\Omega_{\mu}}
\left\{{\cal P}_{t}\left[\bOmega\right]
F_{\mu}^{(1)}\left[\bOmega;t\right]
\right\}
\ed
\noindent
whose solution describes deterministic flow:
${\cal P}_{t}[\bOmega]=\int d\bOmega_{0}{\cal
P}_{0}[\bOmega_{0}]\delta[\bOmega\minus\bOmega(t|\bOmega_0)]$
with $\bOmega(t|\bOmega_0)$ given, in turn, as the solution of
\be
\frac{d}{dt}\bOmega(t)=\bF^{(1)}\left[\bOmega(t);t\right]~~~~~~~~\bOmega(0)=\bOmega_{0}
\label{eq:flowitself}
\ee
In taking the limit $N\rightarrow\infty$, however, we have to
keep in mind that the resulting deterministic theory is obtained by
taking this limit for {\em finite} $t$. According to
(\ref{eq:kramersmoyal}) the $\ell>1$ terms do come into play for
sufficiently large times $t$; for $N\rightarrow\infty$, however, these
times diverge by virtue of (\ref{eq:conditionliouville}).
\vsp

\noindent{\em The Issue of Closure.}
Equation (\ref{eq:flowitself}) will in general not be
autonomous; tracing back the origin of the explicit time dependence in
the right-hand side of (\ref{eq:flowitself}) one finds that to
calculate $\bF^{(1)}$ one needs to know the microscopic probability
density $p_{t}(\bsigma)$. This, in turn, requires
solving equation (\ref{eq:sequentialmaster}) (which is exactly
what one tries to avoid). We will now discuss a mechanism via which to eliminate
the offending explicit time dependence, and to turn the observables $\bOmega(\bsigma)$
into an
autonomous level of description, governed by {\em closed} dynamic
laws.
The idea is to choose the observables $\bOmega(\bsigma)$ in such a way that there is no explicit time
dependence in the flow field $\bF^{(1)}\left[\bOmega;t\right]$ (if possible). According to
 (\ref{eq:KMterms}) this implies making sure that there
exist functions $\Phi_\mu\left[\bOmega\right]$ such that
\be
\lim_{N\rightarrow\infty}\sum_{j=1}^{N}w_{j}(\bsigma)\Delta_{j\mu}(\bsigma)=\Phi_\mu\left[\bOmega(\bsigma)\right]
\label{eq:conditionautonomous}
\ee
in which case the time dependence of $\bF^{(1)}$ indeed drops out and the
macroscopic state vector simply evolves in time according to:
\bd
\frac{d}{dt}\bOmega=\bPhi\left[\bOmega\right],~~~~~~~~~~~~~\bPhi[\bOmega]=(\Phi_1[\bOmega],\ldots,\Phi_n[\bOmega])
\ed
Clearly, for this closure method to apply,
a suitable separable structure of the synaptic matrix is required.
If, for instance, the macroscopic observables $\Omega_{\mu}$ depend
linearly on the microscopic state variables $\bsigma$ (i.e.
$\Omega_\mu(\bsigma)=\frac{1}{N}\sum_{j=1}^{N}\omega_{\mu j}\sigma_{j}$), we
obtain with the
transition rates defined in (\ref{eq:sequentialmaster}):
\be
\frac{d}{dt}\Omega_\mu=
\lim_{N\rightarrow\infty}\frac{1}{N}\sum_{j=1}^{N}\omega_{\mu j}\tanh(\beta
h_{j}(\bsigma))-\Omega_\mu
\label{eq:linear_observables}
\ee
\noindent
in which case the only further condition
for (\ref{eq:conditionautonomous}) to hold is that all local fields
$h_{k}(\bsigma)$ must
(in leading order in $N$) depend on the microscopic state $\bsigma$ only through the
values of the  observables $\bOmega$; since the
local fields depend linearly on $\bsigma$ this, in turn, implies that
the synaptic matrix must be separable: if $J_{ij}=\sum_{\mu}K_{i\mu}\omega_{\mu j}$
then indeed $h_i(\bsigma)=\sum_\mu
K_{i\mu}\Omega_\mu(\bsigma)\plus\theta_i$.
Next I will show how this approach can be applied to networks
for which the matrix of synapses has a
separable form (which includes most symmetric and non-symmetric
Hebbian type attractor models). I will restrict myself to models
with $\theta_{i}=0$; introducing non-zero thresholds is
straightforward and does not pose new problems.

\subsection{Application to Separable Attractor Networks}

\noindent{\em Separable models: Description at the Level of Sublattice
Activities}.
We consider the following class of models, in which the interaction
matrices have the form
\be
J_{ij}=\frac{1}{N}Q(\bxi_{i};\bxi_{j})~~~~~~~~~~
\bxi_{i}=(\xi_{i}^{1},\ldots,\xi_{i}^{p})
\label{eq:vanhemmenmatrix}
\ee
The components $\xi_{i}^{\mu}$, representing the information ('patterns') to be stored or
processed, are assumed to be drawn from a
finite discrete set $\Lambda$, containing $n_{\Lambda}$ elements
(they are not allowed to depend on
$N$). The Hopfield model \cite{Hopfield} corresponds to choosing
$Q(\bx;\by)=\bx\cdot\by$ and $\Lambda\equiv\{-1,1\}$. One now
introduces a partition of the system $\{1,\ldots,N\}$ into
$n_{\Lambda}^{p}$ so-called sublattices $I_{\bEta}$:
\be
I_{\bEta}=\{i|~\bxi_{i}=\bEta\}~~~~~~~~~~~~\{1,\ldots,N\}=\bigcup_{\bEta}I_{\bEta}
~~~~~~~~~~~~\bEta\in\Lambda^{p}
\label{eq:sublattices}
\ee
The number of neurons in sublattice $I_{\bEta}$ is denoted by
$|I_{\bEta}|$ (this number will have to be large).
If we choose as our macroscopic observables the average activities (`magnetisations') within
these sublattices, we are able to express the local fields $h_{k}$
solely in terms of macroscopic quantities:
\be
m_{\bEta}(\bsigma)=\frac{1}{|I_{\bEta}|}\sum_{i\in
I_{\bEta}}\sigma_{i},~~~~~~~~~~~~~
h_{k}(\bsigma)=\sum_{\bEta}p_{\bEta}Q\left(\bxi_{k};\bEta\right)m_{\bEta}
\label{eq:sublatticemagnetisations}
\ee
with the relative sublattice sizes
$p_{\bEta}= |I_{\bEta}|/N$.
If all $p_{\bEta}$ are of the
same order in $N$ (which, for example, is the case if the vectors
$\bxi_{i}$ have been drawn at random from the set $\Lambda^{p}$) we
may write $\Delta_{j\bEta}=\order(n_{\Lambda}^{p}N^{\minus1})$ and use
(\ref{eq:criterion}). The evolution in time of
the sublattice activities is then found to be deterministic in the $N\to\infty$ limit if
$\lim_{N\rightarrow\infty}p/\log N=0$.
Furthermore, condition (\ref{eq:conditionautonomous}) holds, since
\bd
\sum_{j=1}^{N}w_{j}(\bsigma)\Delta_{j\bEta}(\bsigma)=
\tanh[
\beta\sum_{\bEta^{\prime}}p_{\bEta^{\prime}}Q\left(\bEta;\bEta^{\prime}\right)m_{\bEta^{\prime}}]
-m_{\bEta}
\ed
\noindent
We may conclude that the situation is that described by (\ref{eq:linear_observables}), and that
 the evolution in time of the sublattice
activities is governed by the following autonomous set of differential
equations \cite{Riedeletal}:
\be
\frac{d}{dt}m_{\bEta}=
\tanh[
\beta\sum_{\bEta^{\prime}}p_{\bEta^{\prime}}Q\left(\bEta;\bEta^{\prime}\right)m_{\bEta^{\prime}}]
-m_{\bEta}
\label{eq:sublatticeflow}
\ee
We see that, in contrast to the equilibrium techniques as described
in \cite{part1}, here there is no need at all to require symmetry of the interaction
matrix or absence of self-interactions. In the symmetric case $Q(\bx;\by)=Q(\by;\bx)$
the system will approach equilibrium; if the kernel $Q$ is
positive definite this can be shown, for instance, by inspection of the
Lyapunov function\footnote{A function of the state variables
which is bounded from below and whose value decreases monotonically during the
dynamics, see e.g. \cite{Khalil}. Its existence guarantees evolution towards
a stationary state (under some weak conditions).} ${\cal L}\{m_{\bEta}\}$:
\bd
{\cal L}\{m_{\bEta}\}=
\frac{1}{2}\sum_{\bEta\bEta^{\prime}}p_{\bEta}m_{\bEta}Q(\bEta;\bEta^{\prime})m_{\bEta^{\prime}}p_{\bEta^{\prime}}
-
\frac{1}{\beta}\sum_{\bEta}p_{\bEta}\log\cosh[\beta\sum_{\bEta^{\prime}}Q(\bEta;\bEta^{\prime})m_{\bEta^{\prime}}
p_{\bEta^{\prime}}]
\ed
\noindent
which is bounded from below and obeys:
\be
\frac{d}{dt}{\cal
L}=-\sum_{\bEta\bEta^{\prime}}\left[p_{\bEta}\frac{d}{dt}m_{\bEta}\right]Q(\bEta;\bEta^{\prime})\left[p_{\bEta^{\prime}}\frac{d}{dt}m_{\bEta^{\prime}}\right]\leq0
\label{eq:liapunov}
\ee
Note that from the sublattice activities, in turn, follow
the `overlaps' $m_\mu(\bsigma)$ (see \cite{part1}):
\be
m_{\mu}(\bsigma)=\frac{1}{N}\sum_{i=1}^{N}\xi_{i}^{\mu}\sigma_{i}=
\sum_{\bEta}p_{\bEta}\eta_{\mu}m_{\bEta}
\label{eq:linklevels}
\ee
Simple examples of relevant models of the type (\ref{eq:vanhemmenmatrix}),
the dynamics of which are for large $N$ described by equation
(\ref{eq:sublatticeflow}), are
for instance the ones where one applies a non-linear operation $\Phi$ to the
standard Hopfield-type \cite{Hopfield} (or Hebbian-type) interactions .
This non-linearity could result from e.g. a clipping procedure
or from retaining only the {\em sign} of the
Hebbian values:
\bd
J_{ij}=\frac{1}{N}\Phi(\sum_{\mu\leq p}\xi_i^\mu
\xi_j^\mu):
~~~~~~{\rm e.g.}~~~~~~
\Phi(x)= \left\{\begin{array}{ccc}
-K & {\rm for} & x\leq K\\
x & {\rm for} & -K < x < K\\
K & {\rm for} & x\geq K
\end{array}\right.
~~~~~~{\rm or}~~~~~~
\Phi(x)=\sgn(x)
\ed
The effect of introducing such non-linearities is found to be of a
quantitative nature, giving rise to little more than a re-scaling of
critical noise levels and storage capacities. I will not go
into full details, these can be found in e.g. \cite{DHS1}, but illustrate this statement
by working out the $p=2$ equations for randomly drawn pattern
bits $\xi_i^\mu\in\{\minus 1,1\}$, where there are only four
sub-lattices, and where $p_{\bEta}=\frac{1}{4}$ for all $\bEta$.
Using $\Phi(0)=0$ and
$\Phi(-x)=-\Phi(x)$ (as with the above examples) we obtain from (\ref{eq:sublatticeflow}):
\be
\frac{d}{dt}m_{\bEta}=
\tanh[
\frac{1}{4}\beta\Phi(2)(m_{\bEta}-m_{\minus\bEta})]
-m_{\bEta}
\label{eq:sublatticeexample}
\ee
Here the choice made for $\Phi(x)$ shows up only as a re-scaling of the temperature.
From (\ref{eq:sublatticeexample}) we further obtain $\frac{d}{dt}(m_{\bEta}\plus
m_{-\bEta})=-
(m_{\bEta}\plus m_{-\bEta})$. The system decays exponentially towards
a state where, according to (\ref{eq:linklevels}),  $m_{\bEta}=-m_{-\bEta}$ for
all $\bEta$. If
at $t=0$ this is already the case, we find (at least for $p=2$) decoupled equations
for the sub-lattice activities.
\vsp

\noindent{\em Separable Models: Description at the Level of
Overlaps}.
Equations (\ref{eq:sublatticeflow},\ref{eq:linklevels}) suggest that at the level of
overlaps there will be, in turn, closed laws if the kernel $Q$ is bi-linear:\footnote{Strictly speaking, it is already sufficient to have a
kernel which is linear in $\by$ only, i.e.
$Q(\bx;\by)=\sum_\nu f_\nu(\bx)y_{\nu}$},
$Q(\bx;\by)=\sum_{\mu\nu}x_{\mu}A_{\mu\nu}y_{\nu}$, or:
\be
J_{ij}=\frac{1}{N}\sum_{\mu\nu=1}^{p}\xi_{i}^{\mu}A_{\mu\nu}\xi_{j}^{\nu}~~~~~~~~~~
\bxi_{i}=(\xi_{i}^{1},\ldots,\xi_{i}^{p})
\label{eq:coolenmatrix}
\ee
We will see that now the $\xi_{i}^{\mu}$ need not be drawn from a
finite discrete set (as long as they do not depend on
$N$). The Hopfield model corresponds to
$A_{\mu\nu}=\delta_{\mu\nu}$ and $\xi_{i}^{\mu}\in\{-1,1\}$.
The fields $h_{k}$ can now be written in terms of the
overlaps $m_{\mu}$:
\be
h_{k}(\bsigma)=\bxi_{k}\cdot A\bm(\bsigma)
~~~~~~~~~
\bm=(m_{1},\ldots,m_{p})
~~~~~~~~~
m_{\mu}(\bsigma)=\frac{1}{N}\sum_{i=1}^{N}\xi_{i}^{\mu}\sigma_{i}
\label{eq:coolenlevel}
\ee
For this choice of macroscopic variables we find
$\Delta_{j\mu}=\order(N^{\minus1})$, so the evolution of
the vector $\bm$ becomes deterministic for $N\to\infty$ if, according to
(\ref{eq:criterion}),
$\lim_{N\rightarrow\infty}p/\sqrt{N}=0$.
Again (\ref{eq:conditionautonomous}) holds, since
\bd
\sum_{j=1}^{N}w_{j}(\bsigma)\Delta_{j\mu}(\bsigma)=
\frac{1}{N}\sum_{k=1}^{N}\bxi_{k}\tanh\left[\beta\bxi_{k}\cdot
A\bm\right]-\bm
\ed
\noindent
Thus the evolution in time of the overlap vector
$\bm$ is governed by a closed set of differential equations:
\be
\frac{d}{dt}\bm=
\bra\bxi\tanh\left[\beta\bxi\cdot
A\bm\right]\ket_{\bxi}-\bm~~~~~~~~~~~~
\bra\Phi(\bxi)\ket_{\bxi}=\int
d\bxi~\rho(\bxi)\Phi(\bxi)
\label{eq:overlapflow}
\ee
with $\rho(\bxi)=\lim_{N\to\infty}N^{-1}\sum_i\delta[\bxi\minus\bxi_i]$.
Symmetry of the synapses is not required.
For certain non-symmetric matrices $A$ one finds stable
limit-cycle solutions of (\ref{eq:overlapflow}).
In the symmetric case
$A_{\mu\nu}=A_{\nu\mu}$ the system will approach equilibrium; the
Lyapunov function (\ref{eq:liapunov}) for positive definite matrices $A$ now becomes:
\bd
{\cal L}\{\bm\}=
\frac{1}{2}\bm\cdot A\bm -
\frac{1}{\beta}\bra\log\cosh\left[\beta\bxi\cdot
A\bm\right]\ket_{\bxi}
\ed
\begin{figure}[t]
\begin{center}\vspace*{-2mm}
\epsfxsize=95mm\epsfbox{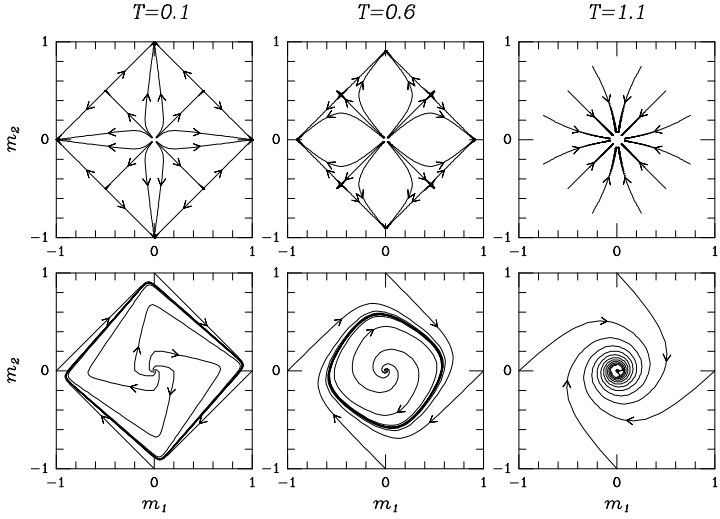}
\end{center}
\vspace*{-5mm}
\caption{Flow diagrams obtained by
numerically solving equations (\protect{\ref{eq:overlapflow}}) for $p=2$. Upper row:
$A_{\mu\nu}=\delta_{\mu\nu}$ (the Hopfield model); lower row:
$A=(\!{\protect\scriptsize
\protect\begin{array}{c}~1~~1 \protect\\
-1~~1\protect\end{array}}\!)$
(here the critical noise level is $T_c=1$).}
\label{fig:ruijgrokflows}
\end{figure}
\begin{figure}[t]
\begin{center}\vspace*{-2mm}
\epsfxsize=95mm\epsfbox{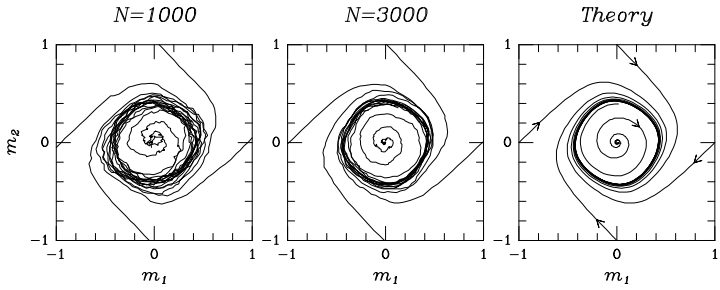}
\end{center}
\vspace*{-5mm}
\caption{Comparison between simulation results
for finite systems ($N=1000$ and $N=3000$) and the $N=\infty$ analytical
prediction (\protect{\ref{eq:overlapflow}}), for $p=2$, $T=0.8$ and
$A=(\!{\protect\scriptsize
\protect\begin{array}{c}~1~~1 \protect\\
-1~~1\protect\end{array}}\!)$.}
\label{fig:ruijgroksimu}
\end{figure}
Figure \ref{fig:ruijgrokflows} shows in the $m_{1},m_{2}$-plane the result of solving the macroscopic laws
(\ref{eq:overlapflow}) numerically for $p=2$, randomly drawn pattern
bits $\xi_i^\mu\in\{-1,1\}$, and two choices of the matrix $A$.
The first choice (upper row) corresponds to the Hopfield model; as the
noise level $T=\beta^{-1}$ increases the amplitudes of the four attractors (corresponding to the two
patterns $\bxi^{\mu}$ and their mirror images $\minus\bxi^{\mu}$)
continuously decrease, until at the critical noise level $T_{c}=1$
(see also \cite{part1})
they merge into the trivial attractor $\bm=(0,0)$. The second choice
corresponds to a non-symmetric model (i.e. without detailed balance);
at the macroscopic level of description (at finite time-scales) the system clearly does not
approach equilibrium; macroscopic order now manifests itself in the
form of a limit-cycle (provided the noise level $T$ is below the
critical value $T_c=1$ where this limit-cycle is destroyed).
To what extent the laws (\ref{eq:overlapflow}) are in agreement with
the result of performing the actual simulations in finite systems is
illustrated in figure \ref{fig:ruijgroksimu}. Other examples can
be found in \cite{BuhmannSchulten,CoolenRuijgrok}.
\begin{figure}[t]
\centering
\vspace*{87mm}
\hbox to \hsize{\hspace*{40mm}\includegraphics{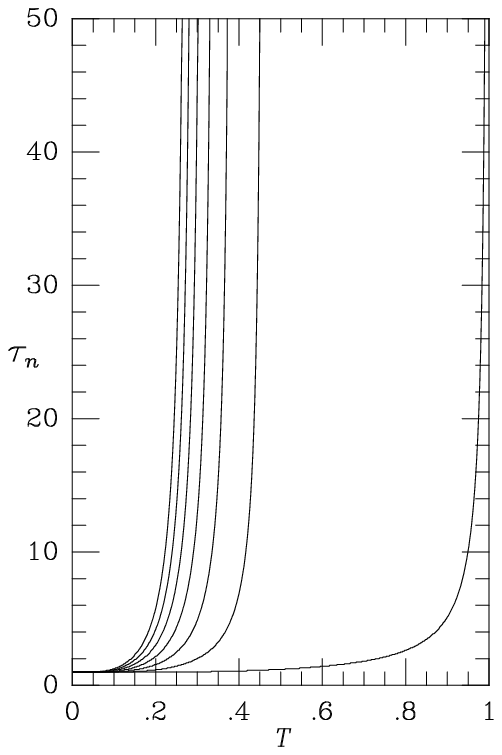}\hspace*{-40mm}}
\vspace*{-12mm}
\caption{Asymptotic relaxation times $\tau_n$ of the mixture states of the Hopfield model as
a function of the noise level $T=\beta^{-1}$. From bottom to top: $n=1,3,5,7,9,11,13$.}
\label{fig:mixrelaxtimes}
\end{figure}

As a second simple application of the flow equations (\ref{eq:overlapflow}) we turn to the
relaxation times corresponding to the attractors of the
Hopfield model (where $A_{\mu\nu}=\delta_{\mu\nu}$). Expanding
(\ref{eq:overlapflow}) near a stable fixed-point $\bm^{*}$, i.e.
$\bm(t)=\bm^*+\bx(t)$ with $|\bx(t)|\ll 1$, gives the
linearised equation
\be
\frac{d}{dt}x_\mu=[\beta\sum_\nu\bra
\xi_\mu\xi_\nu\tanh[\beta\bxi\cdot\bm^*]\ket_{\bxi}-\delta_{\mu\nu}]x_\nu
+\order(\bx^2)
\label{eq:linearisedformixture}
\ee
The Jacobian of
(\ref{eq:overlapflow}), which determines the linearised equation
(\ref{eq:linearisedformixture}), turns out to be {\em minus} the curvature
matrix of the free energy surface at
 the fixed-point (c.f. the derivations in \cite{part1}).
 The asymptotic relaxation towards any stable
attractor is generally exponential, with a characteristic time
$\tau$ given by the inverse of the smallest eigenvalue of the
curvature matrix.
If, in particular, for the fixed point $\bm^*$ we substitute an $n$-mixture state,
i.e. $m_\mu=m_n~ (\mu\leq n)$ and $m_\mu=0~(\mu>n)$,
and transform
(\ref{eq:linearisedformixture}) to the basis where the corresponding
curvature matrix $\bD^{(n)}$  (with eigenvalues
$D_\lambda^n$) is
diagonal, $\bx\rightarrow\tilde{\bx}$, we obtain
\bd
\tilde{x}_\lambda(t)=\tilde{x}_\lambda(0)e^{-t D_\lambda^n}+\ldots
\ed
so $\tau^{-1}=\min_\lambda D_\lambda^n$, which we have already calculated
(see \cite{part1}) in determining the character of the
saddle-points of the free-energy surface.
The result is shown in figure  \ref{fig:mixrelaxtimes}. The relaxation
time for the $n$-mixture attractors decreases
monotonically with the degree of mixing $n$, for any noise level.
At the transition where a macroscopic state $\bm^*$ ceases to
correspond to a local
minimum of the free energy surface, it also de-stabilises in terms of
the linearised dynamic equation (\ref{eq:linearisedformixture}) (as it
should). The Jacobian develops a zero eigenvalue, the relaxation time
diverges, and the long-time
behaviour is no longer obtained from the linearised equation. This
gives rise to critical slowing down (power law relaxation as
opposed to exponential relaxation). For instance, at the transition
temperature $T_c=1$ for the $n=1$ (pure) state, we find by expanding (\ref{eq:overlapflow}):
\bd
\frac{d}{dt}m_\mu=m_\mu[\frac{2}{3}m_\mu^2-\bm^2]+\order(\bm^5)
\ed
which gives rise to a relaxation towards the trivial fixed-point of the form
$\bm\sim t^{-\frac{1}{2}}$.

If one is willing to restrict oneself to the
limited class of models (\ref{eq:coolenmatrix}) (as opposed to the
more general class (\ref{eq:vanhemmenmatrix})) and to the more global
level of description in terms of $p$ overlap parameters $m_{\mu}$ instead of
$n_{\Lambda}^{p}$ sublattice activities $m_{\bEta}$, then there are two rewards.
Firstly there will be no restrictions on the stored pattern
components
$\xi_{i}^{\mu}$ (for instance, they are allowed to be
real-valued); secondly the number $p$ of patterns stored can be much
larger for the deterministic autonomous dynamical laws to hold ($p\ll\sqrt{N}$ instead of
$p\ll\log N$, which from a biological point of view is not impressive.

\subsection{Closed Macroscopic Laws for Parallel Dynamics}

We now turn to the parallel dynamics counterpart of (\ref{eq:sequentialmaster}), i.e. the
Markov chain
\be
p_{\ell+1}(\bsigma)=\sum_{\bsigma^{\prime}}
W\left[\bsigma;\bsigma^{\prime}\right]
p_\ell(\bsigma^{\prime})
~~~~~~~~~
W\left[\bsigma;\bsigma^{\prime}\right]=
\prod_{i=1}^{N}\frac{1}{2}\left[1+\sigma_i \tanh[\beta
h_i(\bsigma^\prime)]\right]
\label{eq:parallelmarkov}
\ee
(with $\sigma_i\in\{-1,1\}$, and with local fields $h_i(\bsigma)$ defined in the usual way).
The evolution of macroscopic probability densities
will here be described by discrete mappings, in stead of
differential equations.
\vsp

\noindent{\em The Toy Model}.
Let us first see what happens to our previous toy model:
$J_{ij}=(J/N)\eta_i\xi_j$ and $\theta_i=0$. As before we try to describe the dynamics at the
(macroscopic) level of the quantity
$m(\bsigma)=\frac{1}{N}\sum_{k}\xi_{k}\sigma_{k}$. The evolution of the macroscopic probability
density
${\cal P}_t[m]$ is obtained by
inserting (\ref{eq:parallelmarkov}):
\be
{\cal P}_{t+1}[m]=
\sum_{\bsigma\bsigma^{\prime}}\delta\left[m\minus
m(\bsigma)\right]W\left[\bsigma;\bsigma^{\prime}\right]p_t(\bsigma^{\prime})
= \int\!dm^{\prime}~\tilde{W}_t\left[m,m^{\prime}\right]{\cal P}_t[m^\prime]
\label{eq:toymacromarkov}
\vspace*{-3mm}
\ee
with
\bd
\tilde{W}_t\left[m,m^\prime\right]=\frac
{\sum_{\bsigma\bsigma^{\prime}}
\delta\left[m\minus m(\bsigma)\right]\delta\left[m^\prime\minus m(\bsigma^\prime)\right]
W\left[\bsigma;\bsigma^{\prime}\right]p_t(\bsigma^{\prime})}
{\sum_{\bsigma^{\prime}}
\delta\left[m^\prime\minus m(\bsigma^\prime)\right]
p_t(\bsigma^{\prime})}
\ed
We now insert our expression for the transition
probabilities $W[\bsigma;\bsigma^\prime]$ and for the local fields.
Since the fields depend on the microscopic state $\bsigma$
only through $m(\bsigma)$, the distribution
$p_t(\bsigma)$ drops out of the above expression for
$\tilde{W}_t$ which thereby loses its explicit time-dependence, $\tilde{W}_t\left[m,m^\prime\right]\rightarrow \tilde{W}\left[m,m^\prime\right]$:
\bd
\tilde{W}\left[m,m^\prime\right]=
e^{-\sum_i\log\cosh(\beta Jm^\prime \eta_i)}\bra \delta\left[m\minus
m(\bsigma)\right] e^{\beta J m^\prime
\sum_i\eta_i\sigma_i}\ket_{\bsigma}\room
~~~~~~{\rm with}~~~~~~\bra \ldots\ket_{\bsigma}=2^{-N}\sum_{\bsigma}\ldots
\ed
Inserting the
integral representation for the $\delta$-function allows us to
perform the average:
\bd
\tilde{W}\left[m,m^\prime\right]=\left[\frac{\beta N}{2\pi}\right]\int\!dk~e^{N\Psi(m,m^\prime,k)}
\ed
\bd
\Psi=i\beta km+\bra\log\cosh\beta[J\eta m^\prime\minus
ik\xi]\ket_{\eta,\xi}-\bra\log\cosh\beta[J\eta m^\prime]\ket_\eta\room
\ed
Since $\tilde{W}\left[m,m^\prime\right]$ is (by construction) normalised,
$\int\!dm~\tilde{W}\left[m,m^\prime\right]=1$, we find that for
$N\rightarrow\infty$ the expectation
value with respect to $\tilde{W}\left[m,m^\prime\right]$ of any
sufficiently smooth function $f(m)$ will be determined only by the value
$m^*(m^\prime)$ of $m$ in the relevant
saddle-point of $\Psi$:
\bd
\int\!dm~f(m)\tilde{W}\left[m,m^\prime\right]=\frac{\int\!dmdk~f(m)e^{N\Psi(m,m^\prime,k)}}
{\int\!dmdk~e^{N\Psi(m,m^\prime,k)}}\rightarrow f(m^*(m^\prime))
~~~~~~~~(N\rightarrow\infty)
\ed
Variation of $\Psi$ with respect to $k$ and $m$ gives the two
saddle-point equations:
\bd
m=\bra \xi\tanh\beta[J\eta m^\prime\minus \xi k]\ket_{\eta,\xi},~~~~~~~~
k=0
\ed
We may now conclude that
$\lim_{N\rightarrow\infty}\tilde{W}\left[m,m^\prime\right]=\delta\left[m\minus
m^*(m^\prime)\right]$
with $m^*(m^\prime)=\bra \xi\tanh(\beta J\eta m^\prime)\ket_{\eta,\xi}$,
and that the macroscopic equation (\ref{eq:toymacromarkov}) becomes:
\bd
{\cal P}_{t+1}[m]=\int\!dm^\prime ~ \delta\left[m\minus \bra \xi\tanh(\beta
J \eta m^\prime)\ket_{\eta\xi}\right]{\cal
P}_t[m^\prime]~~~~~~(N\rightarrow\infty)
\ed
This describes deterministic evolution. If
at $t=0$ we know $m$ exactly, this will remain the case
for finite time-scales, and $m$ will evolve according to a discrete version of
the sequential dynamics law (\ref{eq:magnetisationflow}):
\be
m_{t+1}=\bra \xi\tanh[\beta
J \eta m_t]\ket_{\eta,\xi}\room
\label{eq:toymapping}
\ee
\vsp

\noindent{\em Arbitrary Synapses}. We now try to
generalise the above approach to less trivial classes of models.
As for the sequential case we
will find in the limit $N\rightarrow\infty$ closed deterministic evolution equations for a more general
set of intensive macroscopic state
variables $\bOmega(\bsigma)=(\Omega_1(\bsigma),\ldots,\Omega_n(\bsigma)$ if the local fields
$h_i(\bsigma)$ depend on the microscopic state $\bsigma$ only
through the values of $\bOmega(\bsigma)$,  and if the number $n$ of these state variables
necessary to do so is not too large.
The evolution of the ensemble probability density (\ref{eq:macrodensity})
is now obtained by
inserting the Markov equation (\ref{eq:parallelmarkov}):
\be
{\cal P}_{t+1}\left[\bOmega\right]=
\int\!d\bOmega^{\prime}~\tilde{W}_t\left[\bOmega,\bOmega^{\prime}\right]{\cal P}_t\left[\bOmega^\prime\right]
\label{eq:parmacromarkov}
\ee
\bd
\tilde{W}_t\left[\bOmega,\bOmega^\prime\right]
=\frac{\sum_{\bsigma\bsigma^{\prime}}
\delta\left[\bOmega\minus\bOmega(\bsigma)\right]\delta\left[\bOmega^\prime\minus \bOmega(\bsigma^\prime)\right]
W\left[\bsigma;\bsigma^{\prime}\right]p_t(\bsigma^{\prime})}
{\sum_{\bsigma^{\prime}}
\delta\left[\bOmega^\prime\minus \bOmega(\bsigma^\prime)\right]
p_t(\bsigma^{\prime})}
\ed
\be
=\bra \delta\left[\bOmega\minus\bOmega(\bsigma)\right]
\bra
e ^{\sum_i\left[\beta\sigma_i h_i(\bsigma^\prime)-\log\cosh(\beta h_i(\bsigma^\prime))\right]}\ket_{\bOmega^\prime;t}\ket_{\bsigma}\room
\label{eq:parmacrokernel}
\ee
with $\bra\ldots\ket_{\bsigma}=2^{-N}\sum_{\bsigma}\ldots$, and
with the
conditional (or sub-shell) average defined as in
(\ref{eq:KMterms}).
It is clear from (\ref{eq:parmacrokernel}) that in order to find
autonomous macroscopic laws, i.e. for the distribution $p_t(\bsigma)$ to
drop out, the local fields must depend on the
microscopic state $\bsigma$  only through the macroscopic
quantities $\bOmega(\bsigma)$:
$h_i(\bsigma)=h_i[\bOmega(\bsigma)]$.
In this case $\tilde{W}_t$ loses its explicit time-dependence,
$\tilde{W}_t\left[\bOmega,\bOmega^\prime\right]\rightarrow
\tilde{W}\left[\bOmega,\bOmega^\prime\right]$. Inserting
integral representations for the $\delta$-functions leads to:
\bd
\tilde{W}\left[\bOmega,\bOmega^\prime\right]=
\left[\frac{\beta N}{2\pi}\right]^n\int\!d\bK~e^{N\Psi(\bOmega,\bOmega^\prime,\bK)}
\ed
\bd
\Psi=i\beta\bK\cdot\bOmega+\frac{1}{N}\log\bra
e^{\beta\left[\sum_i\sigma_i h_i[\bOmega^\prime]-
iN\bK\cdot\bOmega(\bsigma)\right]}\ket_{\bsigma}
-\frac{1}{N}\sum_i \log \cosh[\beta h_i[\bOmega^\prime]]\room
\ed
Using the normalisation
$\int\!d\bOmega~\tilde{W}\left[\bOmega,\bOmega^\prime\right]=1$, we
can write expectation
values with respect to $\tilde{W}\left[\bOmega,\bOmega^\prime\right]$ of
macroscopic quantities $f[\bOmega]$ as
\be
\int\!d\bOmega~f[\bOmega]\tilde{W}\left[\bOmega,\bOmega^\prime\right]=\frac{\int\!d\bOmega
d\bK~f[\bOmega] e^{N\Psi(\bOmega,\bOmega^\prime,\bK)}}
{\int\!d\bOmega d\bK ~e^{N\Psi(\bOmega,\bOmega^\prime,\bK)}}
\label{eq:parmacroaverages}
\ee
For saddle-point arguments to apply in determining the leading order in $N$
of (\ref{eq:parmacroaverages}), we encounter restrictions on the number $n$ of our macroscopic
quantities (as expected),
since $n$ determines the dimension of the integrations in (\ref{eq:parmacroaverages}).
The restrictions can be found by expanding $\Psi$
around its maximum $\Psi^*$. After defining $\bx=(\bOmega,\bK)$, of
dimension $2n$, and after translating the location of the
maximum to the origin, one has
\bd
\Psi(\bx)=\Psi^*-\frac{1}{2}\sum_{\mu\nu}x_\mu x_\nu H_{\mu\nu}
+\sum_{\mu\nu\rho}x_\mu x_\nu x_\rho L_{\mu\nu\rho}+\order(\bx^4)
\vspace*{-3mm}
\ed
giving
\bd
\frac{\int\!d\bx~g(\bx)e^{N\Psi(\bx)}}{\int\!d\bx~g(\bx)e^{N\Psi(\bx)}}-
g(\bnul)=\frac{\int\!d\bx~[g(\bx)-g(\bnul)]e^{-\frac{1}{2}N\bx\cdot
\bH\bx+N\sum_{\mu\nu\rho}x_\mu x_\nu x_\rho
L_{\mu\nu\rho}+\order(N\bx^4)}}
{\int\!d\bx~e^{-\frac{1}{2}N\bx\cdot
\bH\bx+N\sum_{\mu\nu\rho}x_\mu x_\nu x_\rho
L_{\mu\nu\rho}+\order(N\bx^4)}}
\ed
\bd
=\frac{\int\!d\by~[g(\by/\sqrt{N})-g(\bnul)]e^{-\frac{1}{2}\by\cdot
\bH\by+\sum_{\mu\nu\rho}y_\mu y_\nu y_\rho
L_{\mu\nu\rho}/\sqrt{N}+\order(\by^4/N)}}
{\int\!d\by~e^{-\frac{1}{2}\by\cdot
\bH\by+\sum_{\mu\nu\rho}y_\mu y_\nu y_\rho
L_{\mu\nu\rho}/\sqrt{N}+\order(\by^4/N)}}
\ed
\bd
=\frac{\int\!d\by~\left[N^{-\frac{1}{2}}\by\cdot\bnabla g(\bnul) +\order(\by^2/N)\right]
e^{-\frac{1}{2}\by\cdot
\bH\by}\left[1+\sum_{\mu\nu\rho}y_\mu y_\nu y_\rho
L_{\mu\nu\rho}/\sqrt{N}+\order(\by^6/N)\right]}
{\int\!d\by~
e^{-\frac{1}{2}\by\cdot
\bH\by}\left[1+\sum_{\mu\nu\rho}y_\mu y_\nu y_\rho
L_{\mu\nu\rho}/\sqrt{N}+\order(\by^6/N)\right]}
\ed
\bd
=\order(n^2/N)+\order(n^4/N^2) + ~{\rm
non\!-\!dominant~terms}~~~~~~~(N,n\to\infty)
\ed
with $\bH$ denoting the Hessian (curvature)
matrix of the surface $\Psi$ at the minimum $\Psi^*$. We thus find
\bd
\lim_{N\rightarrow\infty} n/\sqrt{N}=0:~~~~~~
\lim_{N\rightarrow\infty}\int\!d\bOmega~f[\bOmega]\tilde{W}\left[\bOmega,\bOmega^\prime\right]=f\left[\bOmega^*(\bOmega^\prime)\right]
\ed
where $\bOmega^*(\bOmega^\prime)$ denotes the value of $\bOmega$ in
the saddle-point where $\Psi$ is minimised.
Variation of $\Psi$ with respect to $\bOmega$ and $\bK$ gives the
saddle-point equations:
\bd
\bOmega=
\frac{\bra \bOmega(\bsigma) e^{\beta\left[\sum_i\sigma_i
h_i[\bOmega^\prime]-iN\bK\cdot\bOmega(\bsigma)\right]}\ket_{\bsigma}}
{\bra e^{\beta\left[\sum_i\sigma_i h_i[\bOmega^\prime]-iN\bK\cdot\bOmega(\bsigma)\right]}\ket_{\bsigma}},~~~~~~~~~~
\bK=0
\ed
We may now conclude that
$\lim_{N\rightarrow\infty}\tilde{W}\left[\bOmega,\bOmega^\prime\right]=\delta\left[\bOmega\minus
\bOmega^*(\bOmega^\prime)\right]$, with
\bd
\bOmega^*(\bOmega^\prime)=
\frac{\bra \bOmega(\bsigma) e^{\beta\sum_i\sigma_i
h_i[\bOmega^\prime]}\ket_{\bsigma}}
{\bra e^{\beta\sum_i\sigma_i
h_i[\bOmega^\prime]}\ket_{\bsigma}}
\ed
and that for $N\rightarrow\infty$ the macroscopic equation (\ref{eq:parmacromarkov}) becomes
${\cal P}_{t+1}[\bOmega]=\int\!d\bOmega^\prime ~
\delta[\bOmega\minus\bOmega^*(\bOmega^\prime)]
{\cal P}_t[\bOmega^\prime]$.
This relation again describes deterministic evolution. If
at $t=0$ we know $\bOmega$ exactly, this will remain the case
for finite time-scales and $\bOmega$ will evolve according to
\be
\bOmega(t+1)=
\frac{\bra \bOmega(\bsigma) e^{\beta\sum_i\sigma_i
h_i[\bOmega(t)]}\ket_{\bsigma}}
{\bra e^{\beta\sum_i\sigma_i
h_i[\bOmega(t)]}\ket_{\bsigma}}
\label{eq:parfinalresult}
\ee
As with the sequential case, in taking the limit
$N\rightarrow\infty$ we have to
keep in mind that the resulting laws apply to
finite $t$, and that for sufficiently large times terms of
higher order in $N$ do come into play.
As for the sequential case, a more rigorous and tedious analysis shows that the restriction
$n/\sqrt{N}\rightarrow 0$ can in fact be weakened to $n/N\to 0$.
Finally, for macroscopic quantities $\bOmega(\bsigma)$ which are
linear in $\bsigma$, the remaining $\bsigma$-averages become trivial, so
that \cite{Bernier}:
\be
\Omega_\mu(\bsigma)=\frac{1}{N}\sum_i \omega_{\mu i}\sigma_i:~~~~~~~~
\Omega_\mu(t+1)=
\lim_{N\rightarrow\infty}\frac{1}{N}\sum_i\omega_{\mu i}\tanh\left[\beta
h_i[\bOmega(t)]\right]
\label{eq:parlinearresult}
\ee
(to be compared with (\ref{eq:linear_observables}), as derived for
sequential dynamics).

\subsection{Application to Separable Attractor Networks}

\begin{figure}[t]
\begin{center}\vspace*{-3mm}
\epsfxsize=95mm\epsfbox{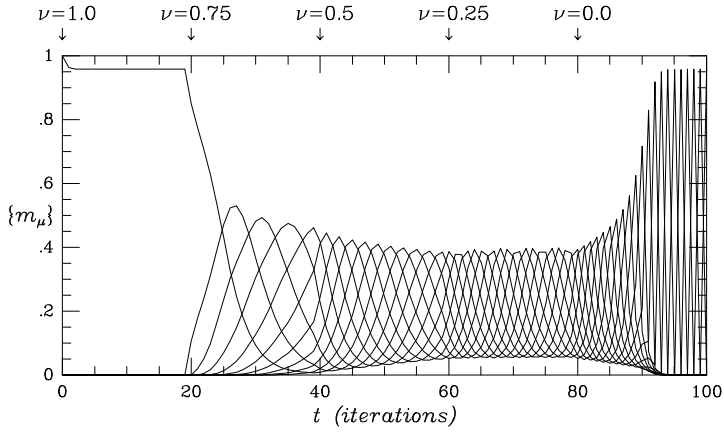}
\end{center}
\vspace*{-5mm}
\caption{Evolution of overlaps
$m_\mu(\bsigma)$, obtained by
numerical iteration of the macroscopic parallel dynamics laws
(\ref{eq:overlapmap}), for the synapses
$J_{ij}=\frac{\nu}{N}\sum_{\mu}\xi_i^\mu\xi_j^\mu+\frac{1\minus \nu}{N}\sum_{\mu}\xi_i^{\mu+1}\xi_j^\mu
$, with $p=10$ and $T=0.5$.}
\label{fig:cycles}
\end{figure}

\noindent{\em Separable models: Sublattice Activities and
Overlaps}. The separable attractor models
(\ref{eq:vanhemmenmatrix}), described at the level of sublattice
activities (\ref{eq:sublatticemagnetisations}), indeed have the
property that all local fields can be written in terms of the macroscopic
observables. What remains
to ensure deterministic evolution is meeting the condition on the number of sublattices.
If all relative sublattice sizes $p_{\bEta}$ are of the
same order in $N$ (as for randomly drawn patterns) this condition again translates into
$\lim_{N\rightarrow\infty}p/\log N=0$ (as for sequential
dynamics).
Since the sublattice activities are linear functions of the $\sigma_i$,
their evolution in time is governed by equation
(\ref{eq:parlinearresult}), which acquires the form:
\be
m_{\bEta}(t+1)=
\tanh[\beta
\sum_{\bEta^\prime}p_{\bEta^\prime}Q\left(\bEta;\bEta^\prime\right)m_{\bEta^\prime}(t)]
\label{eq:sublatticemap}
\ee
As for sequential dynamics, symmetry of the interaction
matrix does not play a role.

At the more global level of overlaps $m_\mu(\bsigma)=N^{-1}\sum_i\xi_i^\mu\sigma_i$
we, in turn, obtain autonomous deterministic laws
if the local fields $h_i(\bsigma)$ can be expressed in terms if $\bm(\bsigma)$ only, as for
the models (\ref{eq:coolenmatrix}) (or, more generally, for all models
in which the
interactions are of the form
$J_{ij}=\sum_{\mu\leq p} f_{i\mu}\xi_j^\mu$), and with the  following restriction on the
number $p$ of embedded patterns: $\lim_{N\rightarrow\infty}p/\sqrt{N}=0$
(as with sequential dynamics).
For the  bi-linear models (\ref{eq:coolenmatrix}), the evolution in time of the overlap vector
$\bm$ (which depends linearly on the $\sigma_i$) is governed by
(\ref{eq:parlinearresult}), which now translates into the
iterative map:
\be
\bm(t+1)=
\bra\bxi\tanh[\beta\bxi\cdot
A\bm(t)]\ket_{\bxi}~~~~~~~~~~~~~~~~
\label{eq:overlapmap}
\ee
with $\rho(\bxi)$ as defined in (\ref{eq:overlapflow}).
Again symmetry of the synapses is not required. For parallel dynamics it is far more difficult
than for sequential dynamics to construct Lyapunov functions, and prove that the macroscopic
laws (\ref{eq:overlapmap}) for symmetric systems evolve towards a stable fixed-point
(as one would expect), but it can still be done.
 For non-symmetric systems the macroscopic
laws (\ref{eq:overlapmap}) can in principle display all the interesting, but
complicated, phenomena of non-conservative non-linear systems.
Nevertheless, it is also not uncommon that the equations (\ref{eq:overlapmap}) for
 non-symmetric systems can be mapped by a time-dependent transformation
onto the equations for related symmetric systems (mostly
variants of the original Hopfield model).

As an example we
show in figure \ref{fig:cycles} as functions of time the values of the
overlaps $\{m_\mu\}$ for $p=10$ and $T=0.5$, resulting
from numerical iteration of the macroscopic laws
(\ref{eq:overlapmap})
for the model
\bd
J_{ij}=\frac{\nu}{N}\sum_{\mu}\xi_i^\mu \xi_j^\mu+\frac{1\minus
\nu}{N}\sum_{\mu}\xi_i^{\mu+1} \xi_j^\mu~~~~~~(\mu:~{\rm mod}~p)
\ed
i.e. $A_{\lambda\rho}=\nu\delta_{\lambda\rho}+(1\minus \nu)\delta_{\lambda,\rho+1}~(\lambda,\!\rho:~{\rm
mod}~p)$,
with randomly drawn pattern bits
$\xi_i^\mu\in\{-1,1\}$
The initial state is chosen to be the pure state
$m_\mu=\delta_{\mu,1}$. At intervals of $\Delta t= 20$ iterations the
parameter $\nu$ is reduced in $\Delta\nu=0.25$ steps from
$\nu=1$ (where one recovers the symmetric Hopfield model) to $\nu=0$
(where one obtains a
non-symmetric model which processes the $p$ embedded patterns in
strict sequential
order as a period-$p$ limit-cycle).
The analysis of the
equations (\ref{eq:overlapmap}) for the pure sequence processing case
$\nu=0$ is greatly simplified by mapping the model onto the ordinary
($\nu=1$) Hopfield model, using the index permutation symmetries of
the present pattern distribution, as follows (all pattern indices
are periodic, mod $p$). Define $m_\mu(t)=M_{\mu-t}(t)$, now
\bd
M_\mu(t\plus1)
=\bra\xi_{\mu+t+1}\tanh[\beta\sum_\rho
\xi_{\rho+1}M_{\rho-t}(t)]\ket_{\bxi}
= \bra\xi_{\mu}
\tanh[\beta \bxi\cdot \bM(t)]\ket_{\bxi}
\ed
We can now immediately infer, in particular, that
to each stable macroscopic fixed-point attractor of the original
Hopfield model
corresponds a stable period-$p$ macroscopic
limit-cycle attractor in the $\nu=1$ sequence processing model (e.g. pure
states  $\leftrightarrow$ pure sequences, mixture states $\leftrightarrow$ mixture sequences),
with identical amplitude as a function
of the noise level. Figure \ref{fig:cycles} shows for
$\nu=0$ (i.e. $t>80$) a relaxation towards such a pure sequence.

Finally we note that the fixed-points of the macroscopic equations
(\ref{eq:sublatticeflow}) and (\ref{eq:overlapflow}) (derived for
sequential dynamics) are identical to those of
(\ref{eq:sublatticemap}) and (\ref{eq:overlapmap}) (derived for
parallel dynamics). The stability properties of these fixed
points, however, need not be the same, and have to be assessed on
a case-by-case basis. For the Hopfield model, i.e. equations
(\ref{eq:overlapflow},\ref{eq:overlapmap}) with
$A_{\mu\nu}=\delta_{\mu\nu}$, they are found to be the same, but
already for $A_{\mu\nu}=-\delta_{\mu\nu}$ the two types of
dynamics would behave differently.


\section{Attractor Neural Networks with Continuous Neurons}

\subsection{Closed Macroscopic Laws}

\noindent{\em General Derivation.}
We have seen in \cite{part1}) that models of recurrent neural networks with continuous
neural variables (e.g. graded response neurons or coupled oscillators) can
often be described by a  Fokker-Planck equation for
the microscopic state probability density $p_t(\bsigma)$:
\be
\frac{d}{dt}p_t(\bsigma)=-\sum_i
\frac{\partial}{\partial\sigma_i}\left[p_t(\bsigma)
f_i(\bsigma)\right]+T\sum_i\frac{\partial^2}{\partial\sigma_i^2}p_t(\bsigma)
\label{eq:fokkerplanck}
\ee
Averages over $p_t(\bsigma)$ are denoted by $\bra
G\ket=\int\!d\bsigma~p_t(\bsigma)G(\bsigma,t)$.
From (\ref{eq:fokkerplanck}) one obtains directly (through integration
by
parts) an equation for the time derivative of averages:
\be
\frac{d}{dt}\bra G\ket=\bra\frac{\partial G}{\partial t}\ket
+\bra\sum_i
\left[f_i(\bsigma)
+T\frac{\partial}{\partial \sigma_i}\right]
\frac{\partial G}{\partial\sigma_i}\ket
\label{eq:derivatives}
\ee
In particular, if we apply (\ref{eq:derivatives}) to
$G(\bsigma,t)=\delta[\bOmega-\bOmega(\bsigma)]$,
for any set of macroscopic observables
$\bOmega(\bsigma)=
(\Omega_1(\bsigma),\ldots,\Omega_n(\bsigma))$ (in the spirit of the previous section),
we obtain a dynamic equation for the macroscopic probability
density
$P_t(\bOmega)=\bra\delta[\bOmega-\bOmega(\bsigma)]\ket$,
which is again of the Fokker-Planck form:
\bd
\frac{d}{dt}P_t(\bOmega)= -\sum_\mu
\frac{\partial}{\partial\Omega_\mu}
\left\{P_t(\bOmega)~\bra
\sum_i \left[f_i(\bsigma)
+T\frac{\partial}{\partial\sigma_i}\right]\frac{\partial}{\partial\sigma_i}
\Omega_\mu(\bsigma)
\ket_{\bOmega;t} \right\}
~~~~~~~~~~~~~~~~~~~~
\ed
\be
~~~~~~~~~~~~~~~~~~~~
+ T\sum_{\mu\nu}
\frac{\partial^2}{\partial\Omega_\mu\partial\Omega_\nu}
\left\{P_t(\bOmega)~
\bra\sum_i\left[\frac{\partial}{\partial\sigma_i}\Omega_\mu(\bsigma)\right]
\left[\frac{\partial}{\partial\sigma_i}\Omega_\nu(\bsigma)\right]
\ket_{\bOmega;t}
\right\}
\label{eq:macrofokkerplanck}
\ee
with the conditional (or sub-shell) averages:
\be
\bra G(\bsigma)\ket_{\bOmega,t}=
\frac{\int\!d\bsigma~p_t(\bsigma)\delta[\bOmega-\bOmega(\bsigma)]
G(\bsigma)}{\int\!d\bsigma~p_t(\bsigma)\delta[\bOmega-\bOmega(\bsigma)]}
\label{eq:generalsubshellaverage}
\ee
From (\ref{eq:macrofokkerplanck}) we infer that a sufficient condition
for the
observables $\bOmega(\bsigma)$ to evolve in time
deterministically (i.e. for having vanishing diffusion matrix elements in (\ref{eq:macrofokkerplanck})) in
the limit $N\to\infty$ is
\be
\lim_{N\to\infty}
\bra\sum_i\left[\sum_\mu\left|\frac{\partial}{\partial\sigma_i}\Omega_\mu(\bsigma)\right|\right]^2
\ket_{\bOmega;t}=0
\label{eq:conditiondeterministic}
\ee
If (\ref{eq:conditiondeterministic}) holds, the macroscopic
Fokker-Planck equation (\ref{eq:macrofokkerplanck}) reduces for
$N\to\infty$ to a Liouville equation, and the observables
$\bOmega(\bsigma)$ will evolve in time according to the
coupled deterministic
equations:
\be
\frac{d}{dt}\Omega_\mu=\lim_{N\to\infty}
\bra\sum_i \left[f_i(\bsigma)
+T\frac{\partial}{\partial\sigma_i}\right]\frac{\partial}{\partial\sigma_i}
\Omega_\mu(\bsigma)
\ket_{\bOmega;t}
\label{eq:deterministicflow}
\ee
The deterministic
macroscopic equation (\ref{eq:deterministicflow}), together with
its
associated condition for validity
(\ref{eq:conditiondeterministic})
will form the basis for the subsequent analysis.
\vsp

\noindent{\em Closure: A Toy Model Again.}
The general derivation given above went smoothly. However,
the equations (\ref{eq:deterministicflow}) are not yet closed. It turns out that
to achieve closure even for simple
 continuous networks we can no longer get away with just a
 finite (small) number of macroscopic observables (as with binary neurons).
 This I will now
 illustrate with a simple toy network of graded response neurons:
\be
\frac{d}{dt}u_i(t)
=\sum_{j} J_{ij}~g[ u_j(t)]-u_i(t) +\eta_i(t)
\label{eq:graded_response}
\ee
with $g[z]=\frac{1}{2}[\tanh(\gamma z)\plus 1]$ and with the standard Gaussian white noise $\eta_i(t)$ (see \cite{part1}). In the language of (\ref{eq:fokkerplanck})
this means $f_i(\bu)\!=\!\sum_{j} J_{ij}g[ u_j]-u_i$.  We
choose uniform synapses
$J_{ij}=J/N$, so $f_i(\bu)\to (J/N)\sum_{j}g[
u_j]-u_i$. If (\ref{eq:conditiondeterministic})
were to hold, we would find the deterministic macroscopic laws
\be
\frac{d}{dt}\Omega_\mu=\lim_{N\to\infty}
\bra\sum_i [
\frac{J}{N}\sum_{j}g[u_j]-u_i
+T\frac{\partial}{\partial u_i}]\frac{\partial}{\partial u_i}
\Omega_\mu(\bu)
\ket_{\bOmega;t}
\label{eq:toygraded}
\ee
In contrast to similar models
with binary neurons, choosing as our macroscopic level of
description $\bOmega(\bu)$ again simply the average $m(\bu)=N^{-1}\sum_i u_i$
now leads to
an equation which fails to close:
\bd
\frac{d}{dt}m=\lim_{N\to\infty}
J~\bra \frac{1}{N}\sum_{j}g[u_j]\ket_{m;t}-m
\ed
The term $N^{-1}\sum_j g[ u_j]$ cannot be written as
a function of $N^{-1}\sum_i u_i$. We might be tempted to try dealing with this problem
by just including the offending term in our macroscopic set, and choose
$\bOmega(\bu)=(N^{-1}\sum_i u_i,N^{-1}\sum_i
g[u_i])$.
This would indeed solve our closure problem for the $m$-equation,
but we would now find a new closure problem in the equation for the
newly introduced observable. The only way out is to
choose an observable {\em function}, namely the distribution of
potentials
\be
\rho(u;\bu)=\frac{1}{N}\sum_i\delta[u-u_i],~~~~~~~~
\rho(u)=\bra \rho(u;\bu)\ket=\bra
\frac{1}{N}\sum_i\delta[u-u_i]\ket
\label{eq:singlesitedist}
\ee
This is to be done with care, in view of our restriction on
the number of observables: we evaluate
(\ref{eq:singlesitedist}) at first only for $n$ specific values
$u_\mu$ and take the limit $n\to\infty$ only after the limit
$N\to\infty$.
Thus we define $\Omega_\mu(\bu)=\frac{1}{N}\sum_i\delta[u_\mu\minus u_i]$,
condition (\ref{eq:conditiondeterministic}) reduces to the
familiar expression
$\lim_{N\to\infty}n/\sqrt{N}=0$, and
we get for $N\to\infty$ and $n\to\infty$ (taken in that order) from (\ref{eq:toygraded})
a diffusion equation for the distribution of membrane potentials (describing a so-called
`time-dependent Ornstein-Uhlenbeck process' \cite{VanKampen,Gardiner}):
\be
\frac{d}{dt}\rho(u)=
-\frac{\partial}{\partial u}
\left\{
\rho(u)\left[
J\!\int\!du^\prime ~\rho(u^\prime)g[u^\prime]-u\right]
\right\}
+T\frac{\partial^2}{\partial u^2} \rho(u)
\label{eq:TDOU}
\ee
\begin{figure}[t]
\begin{center}\vspace*{-10mm}
\epsfxsize=57mm\epsfbox{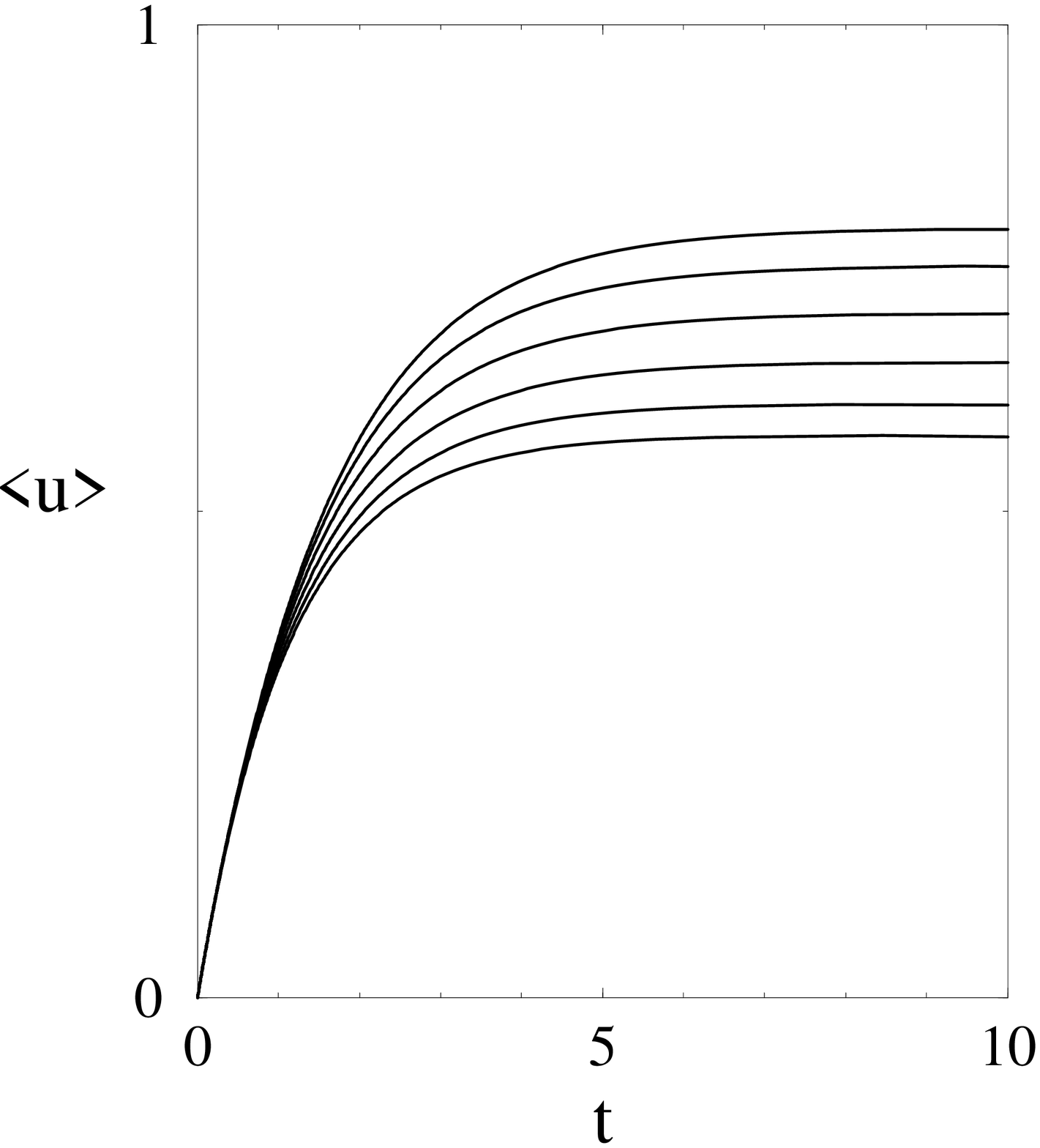}\hspace*{-2mm}
\epsfxsize=57mm\epsfbox{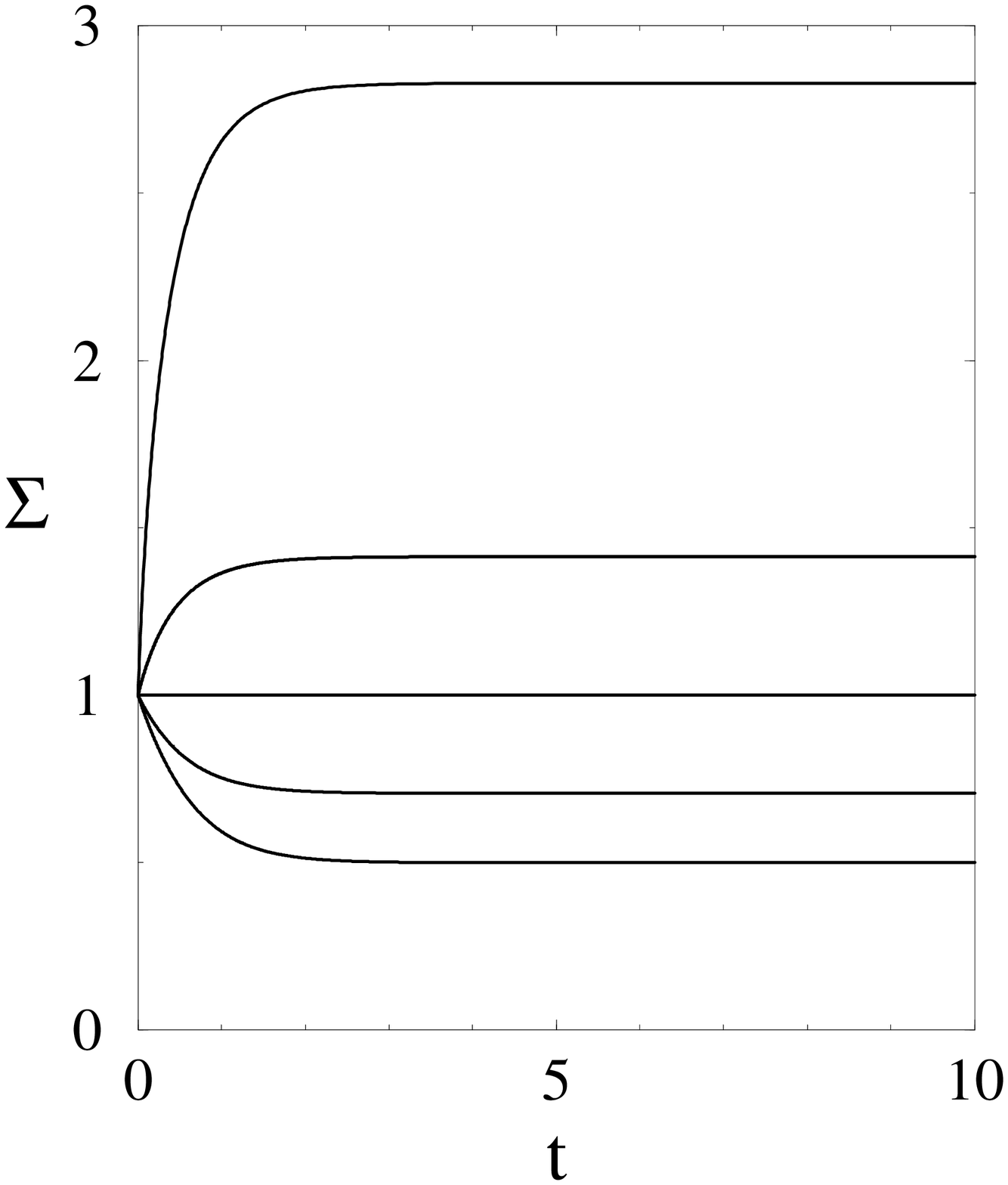}\hspace*{-2mm}
\epsfxsize=57mm\epsfbox{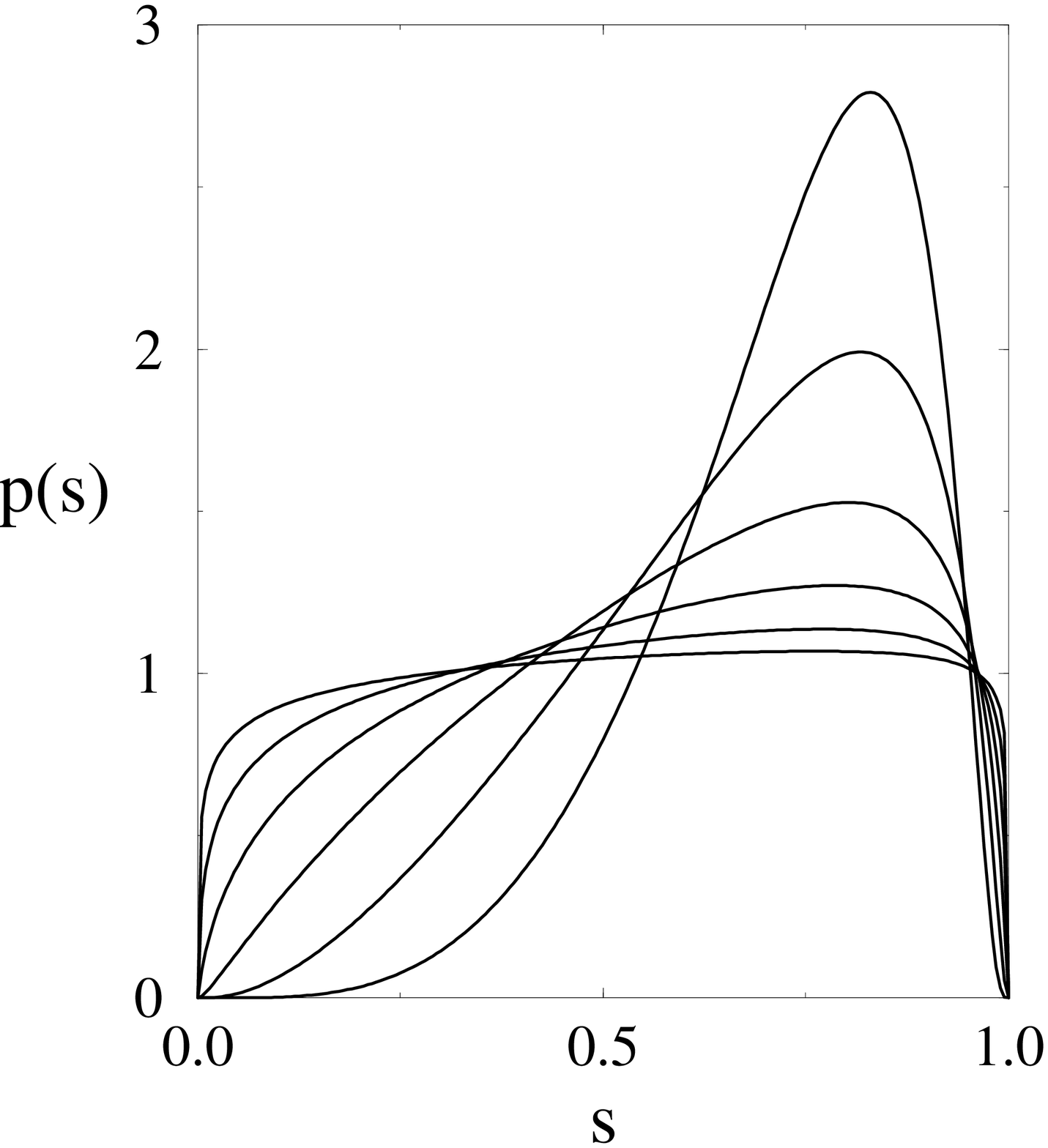}
\end{center}
\vspace*{-9mm}
\caption{Dynamics of a simple network of $N$ graded response neurons
(\ref{eq:graded_response}) with synapses $J_{ij}\!=\!J/N$ and
non-linearity $g[z]\!=\!\frac{1}{2}[1\plus\tanh(\gamma z)]$, for
$N\!\to\!\infty$, $\gamma\!=\!J\!=\!1$, and $T\!\in\!\{0.25,0.5,1,2,4\}$. Left: evolution of average
membrane potential $\bra u\ket=\overline{u}$, with noise levels $T$
increasing from top graph ($T=0.25$) to bottom graph ($T=4$).
Middle: evolution of the width $\Sigma$ of
 the membrane potential distribution, $\Sigma^2=\bra u^2\ket-\bra
 u\ket^2$,
with noise levels decreasing from top graph ($T=4$) to bottom graph ($T=0.25$).
Right: asymptotic ($t=\infty$) distribution of neural firing activities
$p(s)=\bra \delta[s\minus g[u]]\ket$,
with noise levels increasing from the sharply peaked curve ($T=0.25$) to the almost flat curve ($T=4$).}
\label{fig:gradedtoy}
\end{figure}
The natural\footnote{For non-Gaussian initial conditions $\rho_0(u)$  the solution
of (\ref{eq:TDOU}) would in time converge towards the Gaussian solution.}
 solution of (\ref{eq:TDOU}) is the Gaussian distribution
\be
\rho_t(u)=[2\pi\Sigma^2(t)]^{-\frac{1}{2}}
e^{-\frac{1}{2}[u-\overline{u}(t)]^2/\Sigma^2(t)}
\label{eq:TDOUsolution}
\ee
in which $\Sigma=[T\plus(\Sigma^2_0\minus T) e^{-2t}]^{\frac{1}{2}}$, and $\overline{u}$ evolves in time according to
\be
\frac{d}{dt}\overline{u}=J\int\!Dz ~g[\overline{u}+\Sigma z]
-\overline{u}
\label{eq:TDOUaverage}
\ee
(with $Dz=(2\pi)^{-\frac{1}{2}}e^{-\frac{1}{2}z^2}dz$).
We can now also calculate the distribution $p(s)$ of
neuronal firing activities $s_i=g[u_i]$ at any time:
\bd
p(s)=\int\!du~\rho(u)~\delta[s\minus g[u]]
=\frac{\rho(g^{\rm inv}[s])}{\int_0^1\!ds^\prime ~\rho(g^{\rm inv}[s^\prime])}
\ed
For our choice $g[z]=\frac{1}{2}+\frac{1}{2}\tanh[\gamma z]$ we have $g^{\rm
inv}[s]=\frac{1}{2\gamma}\log[s/(1\minus s)]$, so in
combination with
(\ref{eq:TDOUsolution}):
\be
0<s<1:~~~~~~~~~~~~
p(s)=
\frac{e^{-\frac{1}{2}[(2\gamma)^{-1}\log[s/(1-s)]-\overline{u}]^2/\Sigma^2}}
{\int_0^1\!ds^\prime~e^{-\frac{1}{2}[(2\gamma)^{-1}\log[s^\prime/(1-s^\prime)]-\overline{u}]^2/\Sigma^2}}
\label{eq:TDOUactivities}
\ee
The results of solving and integrating numerically (\ref{eq:TDOUaverage})
and (\ref{eq:TDOUactivities}) is shown in figure
\ref{fig:gradedtoy}, for Gaussian initial conditions (\ref{eq:TDOUsolution}) with $\overline{u}_0=0$
and $\Sigma_0=1$, and with parameters $\gamma=J=1$ and
different noise levels $T$. For low noise levels we find
high average membrane potentials, low membrane potential variance, and high firing
rates; for high noise levels the picture changes to lower average
membrane potentials, higher potential variance, and uniformly
distributed (noise-dominated) firing activities.
The extreme cases
$T=0$ and $T=\infty$ are easily extracted from our equations. For $T=0$
one finds $\Sigma(t)=\Sigma_0 e^{-t}$ and $\frac{d}{dt}\overline{u}=J
g[\overline{u}]-\overline{u}$.
This leads to a final state where $\overline{u}=\frac{1}{2}J+\frac{1}{2}J\tanh[\gamma \overline{u}]$
and where $p(s)=\delta[s\minus \overline{u}/J]$. For $T=\infty$
one finds $\Sigma=\infty$ (for any $t>0$) and
$\frac{d}{dt}\overline{u}=\frac{1}{2}J-\overline{u}$.
This leads to an final state where $\overline{u}=\frac{1}{2}J$
and where $p(s)=1$ for all $0<s<1$.

None of the above results (not even those on the
stationary state) could have been obtained within equilibrium
statistical mechanics, since any network of connected graded
response neurons will violate detailed balance \cite{part1}.
Secondly, there appears to be a qualitative difference
between simple networks (e.g. $J_{ij}=J/N$) of binary neurons versus those of continuous
neurons, in terms of the types of macroscopic
observables needed for deriving closed deterministic laws: a single
number $m=N^{-1}\sum_i\sigma_i$ versus a distribution
$\rho(\sigma)=N^{-1}\sum_i\delta[\sigma\minus\sigma_i]$.
Note, however, that in the binary case the latter distribution would in fact have been
been characterised fully by a single number: the average $m$,
since $\rho(\sigma)=\frac{1}{2}[1\plus m]\delta[\sigma\minus 1]+\frac{1}{2}[1\minus m]\delta[\sigma\plus
1]$. In other words: there we were just lucky.

\subsection{Application to Graded Response Attractor Networks}

\noindent{\em Derivation of Closed Macroscopic Laws.}
I will now turn to attractor networks with graded response neurons of the
type (\ref{eq:graded_response}), in which
$p$ binary patterns $\bxi^\mu=(\xi_1^\mu,\ldots,\xi_N^\mu)\in\{-1,1\}^N$
have been stored via separable Hebbian-type synapses (\ref{eq:coolenmatrix}): $J_{ij}=(2/N)\sum_{\mu\nu=1}^p
\xi_i^\mu A_{\mu\nu} \xi_j^\nu$ (the extra factor 2 is inserted for future convenience).
Adding suitable thresholds $\theta_i=-\frac{1}{2}\sum_j J_{ij}$ to
the right-hand sides of (\ref{eq:graded_response}), and choosing the non-linearity
$g[z]=\frac{1}{2}(1\plus \tanh[\gamma z])$ would then give us
\bd
\frac{d}{dt}u_i(t)
=\sum_{\mu\nu} \xi_i^\mu A_{\mu\nu}
\frac{1}{N}\sum_{j} \xi_j^\nu\tanh[\gamma u_j(t)]-u_i(t) +\eta_i(t)
\ed
so the deterministic forces are $f_i(\bu)=
N^{-1}\sum_{\mu\nu}\xi_i^\mu A_{\mu\nu}\sum_j \xi_j^\nu \tanh[\gamma u_j]-u_i$.
Choosing our macroscopic observables $\bOmega(\bu)$ such that (\ref{eq:conditiondeterministic})
holds, would lead to the deterministic macroscopic laws
\be
\frac{d}{dt}\Omega_\mu=\lim_{N\to\infty}
\sum_{\mu\nu}A_{\mu\nu}\bra \left[\frac{1}{N}\sum_j \xi_j^\nu \tanh[\gamma u_j]\right]
\left[\sum_i \xi_i^\mu
\frac{\partial}{\partial u_i}
\Omega_\mu(\bu)\right]
\ket_{\bOmega;t}
+
\lim_{N\to\infty}
\bra\sum_i \left[T\frac{\partial}{\partial u_i}\minus
u_i\right]\frac{\partial}{\partial u_i}
\Omega_\mu(\bu)
\ket_{\bOmega;t}
\label{eq:attdeterministicflow}
\ee
As with the uniform synapses case, the main problem to be dealt
with
is how to choose the $\Omega_\mu(\bu)$ such that
(\ref{eq:attdeterministicflow}) closes. It turns out that the
canonical choice is to turn to the distributions of membrane
potentials within each of the $2^p$ sub-lattices, as introduced in (\ref{eq:sublattices}):
\be
I_{\bEta}=\{i|~\bxi_{i}=\bEta\}:
~~~~~~~~~~\rho_{\bEta}(u;\bu)=\frac{1}{|I_{\bEta}|}\sum_{i\in {I_{\bEta}}}\delta[u-u_i],~~~~~~~~
\rho_{\bEta}(u)=\bra \rho_{\bEta}(u;\bu)\ket
\label{eq:sublatticedist}
\ee
with $\bEta\in\{-1,1\}^{p}$ and $\lim_{N\to\infty}|I_{\bEta}|/N=p_{\bEta}$. Again we evaluate the distributions in
(\ref{eq:sublatticedist}) at first only for $n$ specific values
$u_\mu$ and send $n\to\infty$ after $N\to\infty$.
Now condition (\ref{eq:conditiondeterministic}) reduces to
$\lim_{N\to\infty}2^p/\sqrt{N}=0$. We will keep $p$ finite, for simplicity.
Using identities such as
$\sum_i\ldots=\sum_{\bEta}\sum_{i\in I_{\bEta}}\ldots$ and
\bd
i\in I_{\bEta}:~~~~~~
\frac{\partial}{\partial u_i}\rho_{\bEta}(u;\bu)=
-|I_{\bEta}|^{-1}\frac{\partial}{\partial u}\delta[u\minus u_i],
~~~~~~~~
\frac{\partial^2}{\partial u^2_i}\rho_{\bEta}(u;\bu)=
|I_{\bEta}|^{-1}\frac{\partial^2}{\partial u^2}
\delta[u\minus u_i],
\ed
we then obtain for $N\to\infty$ and $n\to\infty$ (taken in that order) from equation (\ref{eq:attdeterministicflow})
$2^p$ coupled diffusion equations for the distributions $\rho_{\bEta}(u)$
of membrane potentials in each of the $2^p$ sub-lattices
$I_{\bEta}$:
\be
\frac{d}{dt}\rho_{\bEta}(u)=
-\frac{\partial}{\partial u}
\left\{
\rho_{\bEta}(u)\left[
\sum_{\mu\nu=1}^p \eta_\mu A_{\mu\nu}
\sum_{\bEta^\prime}p_{\bEta^\prime}\eta_\nu^\prime
\int\!du^\prime ~\rho_{\bEta^\prime}(u^\prime)\tanh[\gamma u^\prime]-u\right]
\right\}
+T\frac{\partial^2}{\partial u^2} \rho_{\bEta}(u)
\label{eq:attTDOU}
\ee
Equation (\ref{eq:attTDOU}) is the basis for our further analysis.
It can be simplified only if we make additional assumptions on the
system's initial conditions, such as $\delta$-distributed or Gaussian distributed
$\rho_{\bEta}(u)$ at $t=0$ (see below);
otherwise it will have to be solved
numerically.
\vsp

\noindent{\em Reduction to the Level of Pattern Overlaps.}
It is clear that (\ref{eq:attTDOU}) is again of the time-dependent
Ornstein-Uhlenbeck form, and will thus again have Gaussian solutions as
the natural ones:
\be
\rho_{t,\bEta}(u)=[2\pi\Sigma_{\bEta}^2(t)]^{-\frac{1}{2}}
e^{-\frac{1}{2}[u-\overline{u}_{\bEta}(t)]^2/\Sigma^2_{\bEta}(t)}
\label{eq:attTDOUsolution}
\ee
in which $\Sigma_{\bEta}(t)=[T\plus(\Sigma_{\bEta}^2(0)\minus T) e^{-2t}]^{\frac{1}{2}}$, and
with the $\overline{u}_{\bEta}(t)$ evolving in time according to
\be
\frac{d}{dt}\overline{u}_{\bEta}=
\sum_{\bEta^\prime}p_{\bEta^\prime}(\bEta\cdot\bA\bEta^\prime)
\int\!Dz ~\tanh[\gamma (\overline{u}_{\bEta^\prime}+\Sigma_{\bEta^\prime} z)]
-\overline{u}_{\bEta}
\label{eq:attTDOUaverage}
\ee
Our problem has thus been reduced successfully to the study of the
$2^p$ coupled scalar equations (\ref{eq:attTDOUaverage}). We can also
measure the correlation between the firing activities $s_i(u_i)=\frac{1}{2}[1\plus\tanh(\gamma u_i)]$
and the pattern components (similar to the overlaps in the case of
binary neurons). If the pattern bits are drawn at random, i.e. $\lim_{N\to\infty}|I_{\bEta}|/N=p_{\bEta}=2^{-p}$
for all $\bEta$, we can define a `graded response' equivalent
$m_\mu(\bu)=2N^{-1}\sum_i \xi_i^\mu s_i(u_i)\in[-1,1]$ of
the pattern overlaps:
\bd
m_\mu(\bu)=\frac{2}{N}\sum_i\xi_i^\mu s_i(\bu)
=\frac{1}{N}\sum_i\xi_i^\mu \tanh(\gamma u_i)
+\order(N^{-\frac{1}{2}})
\vspace*{-2mm}
\ed
\be
=\sum_{\bEta}p_{\bEta}~\eta_\mu \int\!du~\rho_{\bEta}(u;\bu)
\tanh(\gamma u)+\order(N^{-\frac{1}{2}})
\label{eq:attoverlaps}
\ee
Full recall of pattern $\mu$ implies $s_i(u_i)=\frac{1}{2}[\xi_i^\mu\plus
1]$, giving $m_\mu(\bu)=1$. Since the distributions $\rho_{\bEta}(u)$ obey deterministic laws for $N\to\infty$,
the same will be true for the overlaps $\bm=(m_1,\ldots,m_p)$.
For the Gaussian solutions
(\ref{eq:attTDOUaverage}) of (\ref{eq:attTDOU}) we can now proceed to replace the $2^p$ macroscopic laws
(\ref{eq:attTDOUaverage}), which reduce to
$\frac{d}{dt}\overline{u}_{\bEta}=\bEta\cdot\bA
\bm-\overline{u}_{\bEta}$ and give $\overline{u}_{\bEta}=\overline{u}_{\bEta}(0)e^{-t}+\bEta\cdot\bA\int_0^t\!ds
~e^{s-t}\bm(s)$, by $p$ integral equations
 in terms of overlaps only:
\be
m_\mu(t)
=\sum_{\bEta}p_{\bEta}~\eta_\mu \int\!Dz~\tanh\left[\gamma\left(\overline{u}_{\bEta}(0)e^{-t}+\bEta\cdot\bA\int_0^t\!ds
~e^{s-t}\bm(s)\plus
z\sqrt{T\plus(\Sigma_{\bEta}^2(0)\minus T) e^{-2t}}
\right)\right]
\label{eq:overlapsintermediate}
\ee
with $Dz=(2\pi)^{-\frac{1}{2}}e^{-\frac{1}{2}z^2}dz$.
Here the sub-lattices only come in via the initial
conditions.
\vsp

\noindent{\em Extracting the Physics from the Macroscopic Laws.}
The equations describing the asymptotic (stationary) state
can be written entirely without sub-lattices, by taking the $t\to\infty$ limit in (\ref{eq:overlapsintermediate}),
using $\overline{u}_{\bEta}\to\bEta\cdot\bA\bm$,
$\Sigma_{\bEta}\to\sqrt{T}$, and the familiar notation
$\bra g(\bxi)\ket_{\bxi}=\lim_{N\to\infty}\frac{1}{N}\sum_i g(\bxi_i)=2^{-p}\sum_{\bxi\in\{-1,1\}^p} g(\bxi)$:
\be
m_\mu
=\bra \xi_\mu \int\!Dz~\tanh[\gamma(
\bxi\cdot\bA\bm\plus z\sqrt{T})]\ket_{\bxi}
~~~~~~~~~~~~
\rho_{\bEta}(u)=[2\pi T]^{-\frac{1}{2}}
e^{-\frac{1}{2}[u-\bEta\cdot\bA\bm]^2/T}
\label{eq:attstationary}
\ee
\begin{figure}[t]
\begin{center}\vspace*{-9mm}
\epsfysize=67mm\epsfbox{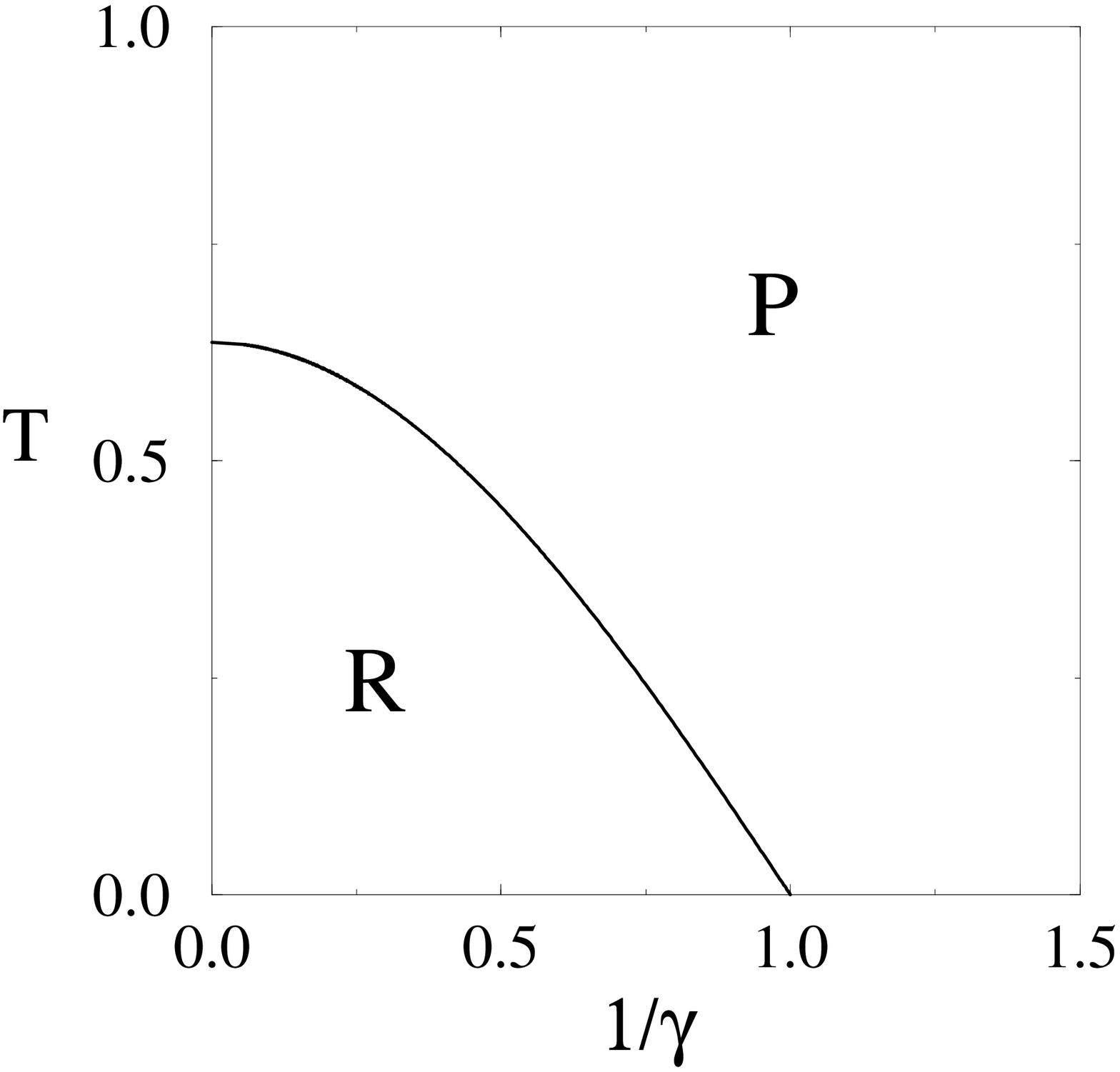}
\epsfysize=67mm\epsfbox{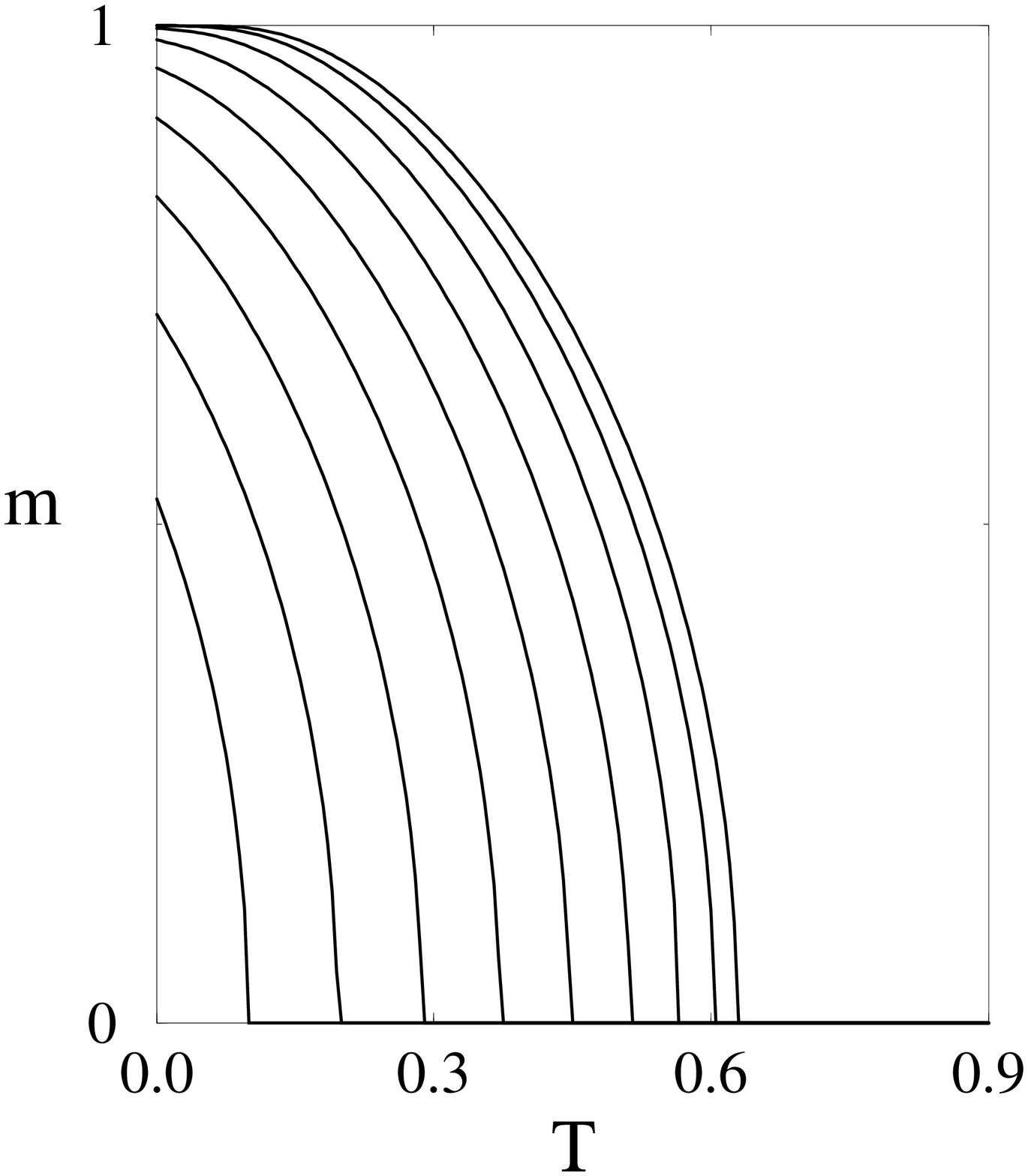}
\end{center}
\vspace*{-7mm}
\caption{Left: phase diagram of the Hopfield model with graded-response neurons
and $J_{ij}=(2/N)\sum_\mu\xi_i^\mu\xi_j^\mu$, away from saturation.
P: paramagnetic phase, no recall. R: pattern recall
phase.
Solid line: separation of the above phases, marked by a continuous
transition.
Right: asymptotic recall amplitudes $m=(2/N)\sum_i \xi_i^\mu s_i$
of pure states (defined such that full recall corresponds to $m=1$), as functions of the noise level $T$,
for $\gamma^{-1}\in\{0.1,0.2,\ldots,0.8,0.9\}$ (from top to bottom).}
\label{fig:gradedhopfieldphasediagram}
\end{figure}
Note the appealing similarity with previous results  on
networks with binary neurons in equilibrium \cite{part1}. For $T=0$ the overlap
equations (\ref{eq:attstationary}) become identical to those found for attractor networks with
binary neurons and finite $p$ (hence our choice to insert an extra factor 2 in defining the synapses),
with $\gamma$ replacing the inverse noise level $\beta$
in the former.

For the simplest non-trivial choice  $A_{\mu\nu}=\delta_{\mu\nu}$
(i.e. $J_{ij}=(2/N)\sum_{\mu}\xi_i^\mu\xi_j^\mu$, as in the Hopfield
\cite{Hopfield} model) equation (\ref{eq:attstationary})
 yields the familiar pure and mixture state solutions. For $T=0$ we find
a continuous phase transition from non-recall to pure states of the form $m_\mu=m\delta_{\mu\nu}$
(for some $\nu$) at $\gamma_c=1$. For $T>0$ we
have in (\ref{eq:attstationary}) an
additional Gaussian noise, absent in the models with binary neurons. Again the pure states are the first
non-trivial solutions to enter the stage. Substituting $m_{\mu}=m\delta_{\mu\nu}$
into (\ref{eq:attstationary}) gives
\be
m=\int\!Dz~\tanh[\gamma(m\plus z\sqrt{T})]
\label{eq:attgradedpure}
\ee
Writing (\ref{eq:attgradedpure}) as
$m^2=\gamma m\int_0^m dk[1\minus \int\!Dz~\tanh^2[\gamma(k+z\sqrt{T})]]\leq
\gamma m^2$,
reveals
that $m=0$ as soon as $\gamma<1$.
A continuous transition to an $m>0$ state occurs when
$\gamma^{-1}=1\minus \int\!Dz~\tanh^2[\gamma
z\sqrt{T}]$. A parametrisation of this transition line in the
$(\gamma,T)$-plane is given by
\be
\gamma^{-1}(x)=1\minus \int\!Dz~\tanh^2(zx),~~~~~~~~T(x)=x^2/\gamma^2(x),~~~~~~~~x\geq 0
\label{eq:parametrization}
\ee
Discontinuous transitions away from $m=0$ (for which there is no evidence) would have to be
calculated numerically.
For $\gamma=\infty$ we get the equation $m={\rm erf}[m/\sqrt{2 T}]$,
giving a continuous transition to $m>0$ at $T_c=2/\pi\approx 0.637$.
Alternatively the latter number can also be found by taking $\lim_{x\to\infty}T(x)$
in the above parametrisation:
\bd
T_c(\gamma=\infty)=\lim_{x\to\infty}
x^2[1\minus \int\!\!Dz~\tanh^2(zx)]^2
=\lim_{x\to\infty}[\int\!\!Dz~\frac{d}{dz}\tanh(zx)]^2
=[2\int\!\!Dz~\delta(z)]^2=2/\pi
\ed
The resulting picture of the network's stationary state properties
is illustrated in figure \ref{fig:gradedhopfieldphasediagram}, which shows the phase diagram
and the stationary recall overlaps of the pure states, obtained by numerical
calculation and solution of equations (\ref{eq:parametrization}) and
(\ref{eq:attgradedpure}).
\vsp

Let us now turn to dynamics. It follows from
(\ref{eq:attstationary}) that the `natural' initial conditions
for $\overline{u}_{\bEta}$ and $\Sigma_{\bEta}$ are of the
form: $\overline{u}_{\bEta}(0)=\bEta\cdot\bk_0$ and $\Sigma_{\bEta}(0)=\Sigma_0$
for all $\bEta$. Equivalently:
\bd
t=0:~~~~~~~~\rho_{\bEta}(u)=[2\pi \Sigma_0^2]^{-\frac{1}{2}}
e^{-\frac{1}{2}[u-\bEta\cdot\bk_0]^2/\Sigma^2_0},~~~~~~~~~\bk_0\in\Re^p,~\Sigma_0\in\Re
\ed
These would also be the typical and natural statistics if we were to
prepare an initial firing state $\{s_i\}$ by hand, via manipulation of the potentials
$\{u_i\}$.
For such initial conditions we can simplify the dynamical equation
(\ref{eq:overlapsintermediate}) to
\be
m_\mu(t)
=\bra~\xi_\mu \int\!Dz~\tanh\left[
\gamma\left(\bxi\cdot[\bk_0 e^{-t}\plus\bA\int_0^t\!ds
~e^{s-t}\bm(s)]+
 z\sqrt{T\plus(\Sigma_0^2\minus T) e^{-2t}}\right)
\right]\ket_{\bxi}
\label{eq:overlapsfinal}
\ee
For the special case of the Hopfield synapses, i.e.
$A_{\mu\nu}=\delta_{\mu\nu}$,
it follows from (\ref{eq:overlapsfinal}) that recall of a given pattern $\nu$ is triggered upon choosing
$k_{0,\mu}=k_0\delta_{\mu\nu}$
(with $k_0>0$), since then equation (\ref{eq:overlapsfinal})
generates $m_\mu(t)=m(t)\delta_{\mu\nu}$ at any time, with the
amplitude $m(t)$ following from
\be
m(t)
=\int\!Dz~\tanh\left[
\gamma[k_0 e^{-t}\plus \int_0^t\!ds
~e^{s-t}m(s)\plus z\sqrt{T\plus(\Sigma_0^2\minus T) e^{-2t}}]
\right]
\label{eq:attpureflow}
\ee
which is the dynamical counterpart of equation
(\ref{eq:attgradedpure}) (to which indeed it reduces for $t\to\infty$).

\begin{figure}[t]
\begin{center}\vspace*{-10mm}
\epsfysize=67mm\epsfbox{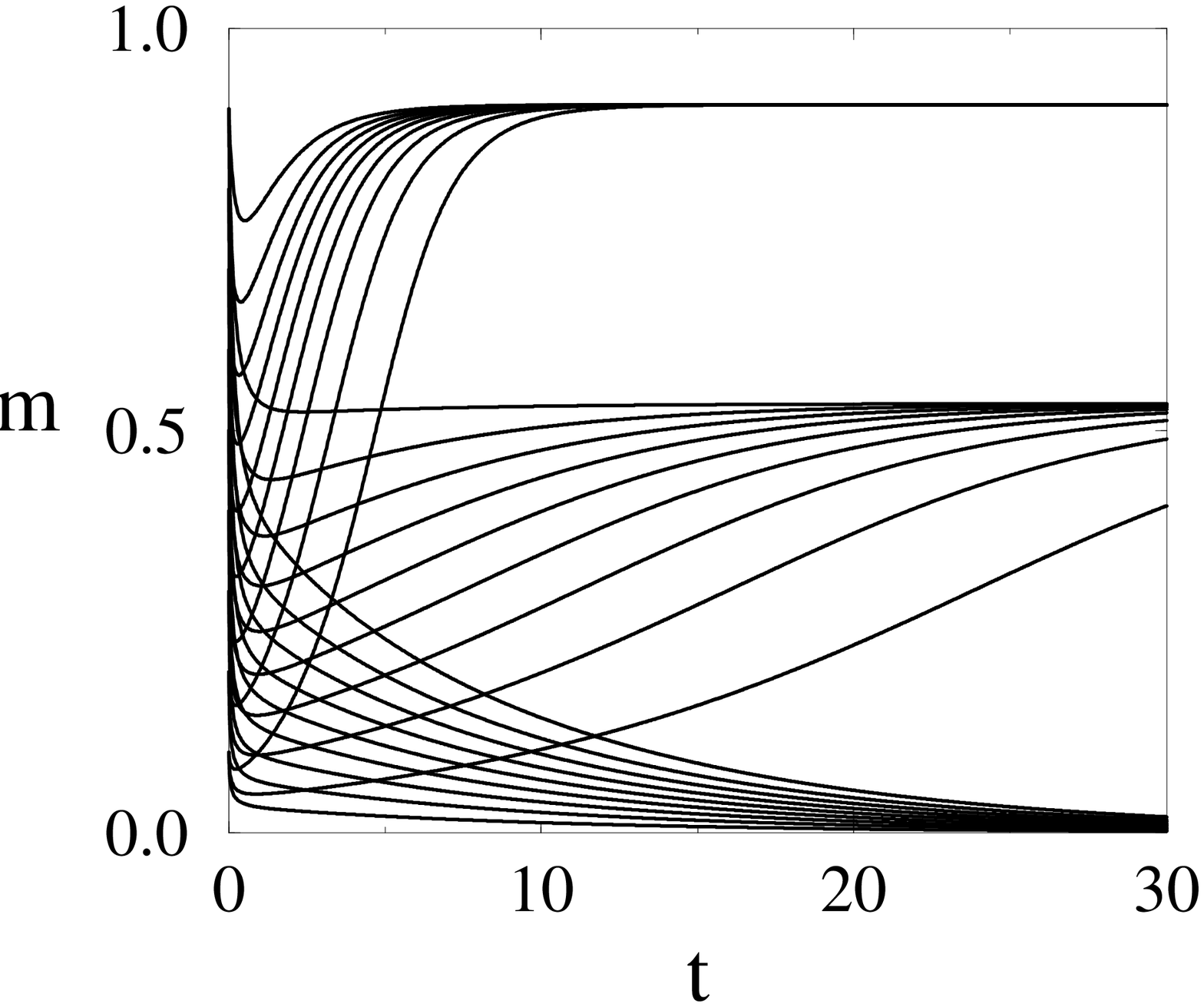}
\end{center}
\vspace*{-5mm}
\caption{Overlap evolution in the Hopfield model with graded-response
neurons
and $J_{ij}=(2/N)\sum_\mu\xi_i^\mu\xi_j^\mu$, away from saturation.
Gain parameter: $\gamma=4$. Initial conditions: $\rho_{\bEta}(u)=\delta[u-k_0\eta_\nu]$ (i.e.
triggering recall of pattern $\nu$, with uniform membrane
potentials within sub-lattices).
Lines:  recall amplitudes $m=(2/N)\sum_i \xi_i^\nu s_i$ of
pure state $\nu$ as functions of time,
for $T=0.25$ (upper set), $T=0.5$ (middle set) and $T=0.75$ (lower set), following
different initial overlaps $m_0\in\{0.1,0.2,\ldots,0.8,0.9\}$.}
\label{fig:gradedhopfieldoverlaps}
\end{figure}
We finally specialize further to the case where our Gaussian initial conditions
are not only chosen to trigger recall of a single pattern $\bxi^\nu$, but in addition
describe uniform membrane potentials within the
sub-lattices, i.e. $k_{0,\mu}=k_0\delta_{\mu\nu}$ and $\Sigma_0=0$,
so $\rho_{\bEta}(u)=\delta[u-k_0\eta_\nu]$.  Here we can derive from
(\ref{eq:attpureflow}) at $t=0$ the identity
$m_0=\tanh[ \gamma k_0]$, which enables us to express $k_0$ as
$k_0=(2\gamma)^{-1}\log[(1\plus m_0)/(1\minus m_0)]$, and find
(\ref{eq:attpureflow}) reducing to
\be
m(t)
=\int\!Dz~\tanh\left[
e^{-t}\log[\frac{1\plus m_0}{1\minus m_0}]^{\frac{1}{2}}
+\gamma[ \int_0^t\!ds
~e^{s-t}m(s)\plus z\sqrt{T(1\minus e^{-2t})}]
\right]
\label{eq:attpuresimplest}
\ee
Solving this equation numerically  leads to graphs such as those
shown in figure \ref{fig:gradedhopfieldoverlaps} for the choice
$\gamma=4$ and $T\in\{0.25,0.5,0.75\}$.
Compared to the overlap evolution in large networks of binary
networks (away from saturation) one immediately observes richer behaviour,
e.g. non-monotonicity.

The analysis and results described in this section, which can be done
and derived
in a similar fashion for other networks with continuous units (such as coupled
oscillators), are somewhat difficult to
find in research papers. There are two reasons for this. Firstly,
non-equilibrium statistical mechanical studies only started being
carried out around 1988, and obviously concentrated at first on
the (simpler) networks with binary variables. Secondly, due to the
absence of detailed balance in networks of graded response
networks, the latter appear to have been  suspected of consequently
having highly complicated dynamics, and analysis terminated with
pseudo-equilibrium studies \cite{Kuehn}. In retrospect that
turns out to have been too pessimistic a view on the
power of non-equilibrium statistical mechanics: one finds that dynamical tools
can be applied without serious technical problems (although the calculations are somewhat
more involved), and again yield
interesting and explicit results in the form of phase diagrams and
dynamical curves for macroscopic observables, with sensible physical
interpretations.


\section{Correlation- and Response-Functions}

We now turn to correlation functions $C_{ij}(t,t^{\prime})$ and
response functions $G_{ij}(t,t^{\prime})$. These will become the
language in which the generating functional methods are
formulated, which will enable us to solve the dynamics of
recurrent networks in the (complex) regime near saturation
(we take $t>t^{\prime}$):
\be
C_{ij}(t,t^{\prime})=\bra \sigma_i(t)\sigma_j(t^{\prime})\ket
~~~~~~~~~~
G_{ij}(t,t^{\prime})=\partial\bra\sigma_i(t)\ket/\partial\theta_j(t^{\prime})
\label{eq:correlation_response}
\ee
The $\{\sigma_i\}$ evolve in time according to
equations of the form (\ref{eq:sequentialmaster}) (binary neurons, sequential updates),
(\ref{eq:parallelmarkov}) (binary neurons, parallel updates) or
(\ref{eq:fokkerplanck}) (continuous neurons). The $\theta_i$
represent thresholds and/or external stimuli, which are added to the local
fields in the cases
(\ref{eq:sequentialmaster},\ref{eq:parallelmarkov}), or added to
the deterministic forces in the case of a Fokker-Planck equation
(\ref{eq:fokkerplanck}).
We retain $\theta_i(t)=\theta_i$, except for a perturbation
$\delta\theta_j(t^{\prime})$
applied at time $t^\prime$ in defining the response function.
Calculating averages such as (\ref{eq:correlation_response}) requires
determining joint probability distributions involving neuron
states at different times.

\subsection{Fluctuation-Dissipation Theorems}

\noindent{\em Networks of Binary Neurons.}
For networks of binary neurons with
discrete time dynamics of the form
$p_{\ell+1}(\bsigma)=\sum_{\bsigma^{\prime}}W\left[\bsigma;\bsigma^{\prime}\right]p_\ell(\bsigma^{\prime})$,
the probability of observing a given `path' $\bsigma(\ell^\prime)\to\bsigma(\ell^\prime\plus
1)\to\ldots\to
\bsigma(\ell\minus 1)\to\bsigma(\ell)$
of successive configurations between step $\ell^\prime$ and step $\ell$ is given by
the product of the corresponding transition matrix elements (without summation):
\bd
{\rm Prob}[\bsigma(\ell^\prime),\ldots,\bsigma(\ell)]=W[\bsigma(\ell);\bsigma(\ell\minus 1)]
W[\bsigma(\ell\minus 1);\bsigma(\ell\minus 2)]
\ldots W[\bsigma(\ell^\prime\plus
1);\bsigma(\ell^\prime)]p_{\ell^\prime}(\bsigma(\ell^\prime))
\ed
This allows us to write
\be
C_{ij}(\ell,\ell^{\prime})=
\sum_{\bsigma(\ell^\prime)}\!\cdots\!\sum_{\bsigma(\ell)}\!
{\rm Prob}[\bsigma(\ell^\prime),\ldots,\bsigma(\ell)]\sigma_i(\ell)\sigma_j(\ell^\prime)
=
\sum_{\bsigma\bsigma^\prime}\!\sigma_i\sigma_j^{\prime}
W^{\ell-\ell^{\prime}}[\bsigma;\bsigma^{\prime}]
 p_{\ell^\prime}(\bsigma^\prime)
\label{eq:eqcorrelation}
\ee
\be
G_{ij}(\ell,\ell^{\prime})=
\sum_{\bsigma\bsigma^\prime\bsigma^\pprime}
\sigma_i W^{\ell-\ell^{\prime}-1}[\bsigma;\bsigma^\pprime]
\left[\frac{\partial}{\partial \theta_j}
W[\bsigma^\pprime;\bsigma^\prime]\right]
p_{\ell^\prime}(\bsigma^\prime)
\label{eq:eqresponse}
\ee
From (\ref{eq:eqcorrelation}) and (\ref{eq:eqresponse})
it follows that both $C_{ij}(\ell,\ell^\prime)$  and $G_{ij}(\ell,\ell^\prime)$
will in the stationary state, i.e. upon substituting
$p_{\ell^\prime}(\bsigma^\prime)=p_\infty(\bsigma^\prime)$,
only depend on $\ell\minus \ell^{\prime}$:
$C_{ij}(\ell,\ell^{\prime})\rightarrow C_{ij}(\ell\minus \ell^{\prime})$ and
$G_{ij}(\ell,\ell^{\prime})\rightarrow G_{ij}(\ell\minus
\ell^{\prime})$.
For this we do not require
detailed balance. Detailed balance, however, leads to
a simple relation between the response function $G_{ij}(\tau)$ and the
temporal derivative of the correlation function $C_{ij}(\tau)$.

We now turn to equilibrium systems, i.e. networks with symmetric synapses
(and with all $J_{ii}=0$ in the case of sequential dynamics).
We  calculate the derivative of the transition matrix
that occurs in (\ref{eq:eqresponse})
by differentiating the equilibrium condition $p_{\rm
eq}(\bsigma)=\sum_{\bsigma^\prime}W[\bsigma;\bsigma^\prime]p_{\rm eq}(\bsigma^\prime)$
 with respect to external fields:
\bd
\frac{\partial}{\partial\theta_j}p_{\rm eq}(\bsigma)
=\sum_{\bsigma^{\prime}}\left\{
\frac{\partial
W[\bsigma;\bsigma^{\prime}]}{\partial\theta_j}
p_{\rm eq}(\bsigma^{\prime})
+W[\bsigma;\bsigma^{\prime}]\frac{\partial}{\partial\theta_j}p_{\rm
eq}(\bsigma^{\prime})\right\}
\ed
Detailed balance implies $p_{\rm
eq}(\bsigma)=Z^{-1}e^{-\beta H(\bsigma)}$ (in the parallel case we
simply substitute the appropriate Hamiltonian
$H\rightarrow\tilde{H}$), giving
$\partial p_{\rm eq}(\bsigma)/\partial\theta_j=-
[Z^{-1}\partial Z/\partial\theta_j
\plus \beta \partial H(\bsigma)/\partial\theta_j]p_{\rm
eq}(\bsigma)$,
so that
\bd
\sum_{\bsigma^{\prime}}
\frac{\partial W[\bsigma;\bsigma^{\prime}]}{\partial\theta_j}
p_{\rm eq}(\bsigma^{\prime})
=
\beta\left\{
\sum_{\bsigma^{\prime}}W\left[\bsigma;\bsigma^{\prime}\right]\frac{\partial
H(\bsigma^{\prime})}{\partial\theta_j}p_{\rm
eq}(\bsigma^{\prime})
-\frac{\partial H(\bsigma)}{\partial\theta_j} p_{\rm eq}(\bsigma)
\right\}
\ed
(the term containing $Z$ drops out). We now obtain for the
response function (\ref{eq:eqresponse}) in equilibrium :
\be
G_{ij}(\ell)=\beta
\sum_{\bsigma\bsigma^{\prime}}\sigma_i
W^{\ell-1}\left[\bsigma;\bsigma^{\prime}\right]
\left\{
\sum_{\bsigma^{\prime\prime}}W\left[\bsigma^{\prime};\bsigma^{\prime\prime}\right]\frac{\partial
H(\bsigma^{\prime\prime})}{\partial\theta_j}p_{\rm
eq}(\bsigma^{\prime\prime})
-\frac{\partial H(\bsigma^{\prime})}{\partial\theta_j} p_{\rm eq}(\bsigma^{\prime})
\right\}
\label{eq:neweqresponse}
\ee
The structure of (\ref{eq:neweqresponse}) is similar to what
follows upon calculating the evolution of the equilibrium correlation
function (\ref{eq:eqcorrelation}) in a single iteration step:
\be
C_{ij}(\ell)-C_{ij}(\ell\minus 1)=
\sum_{\bsigma\bsigma^{\prime}}\sigma_i
W^{\ell-1}\left[\bsigma;\bsigma^{\prime}\right]
\left\{
\sum_{\bsigma^{\prime\prime}}W\left[\bsigma^{\prime};\bsigma^{\prime\prime}\right]\sigma_j^{\prime\prime}
p_{\rm eq}(\bsigma^{\prime\prime})-\sigma_j^{\prime} p_{\rm eq}(\bsigma^{\prime})\right\}
\label{eq:neweqcorrelation}
\ee
Finally we calculate the relevant derivatives of the two Hamiltonians
$H(\bsigma)=\minus\sum_{i<j}J_{ij}\sigma_i\sigma_j\plus\sum_i\theta_i\sigma_i$
and $\tilde{H}(\bsigma)=\minus \sum_{i}\theta_i\sigma_i\minus \beta^{-1}\sum_{i}\log 2 \cosh[\beta
h_i(\bsigma)]$ (with $h_i(\bsigma)=\sum_j
J_{ij}\sigma_j\plus\theta_i$), see \cite{part1}:
\bd
\partial H(\bsigma)/\partial\theta_j=- \sigma_j~~~~~~~~~~~~~~
\partial\tilde{H}(\bsigma)/\partial\theta_j=-\sigma_j\minus \tanh[\beta
h_j(\bsigma)]
\ed
For sequential dynamics we hereby arrive directly at a
fluctuation-dissipation theorem. For parallel dynamics we need one
more identity (which follows from the definition of the transition matrix in (\ref{eq:parallelmarkov}) and the
detailed balance property) to transform the $tanh$ occurring in the derivative of
$\tilde{H}$:
\bd
\tanh[\beta h_j(\bsigma^{\prime})]p_{\rm eq}(\bsigma^{\prime})
=\sum_{\bsigma^{\prime\prime}}
\sigma_j^{\prime\prime}W\left[\bsigma^{\prime\prime};\bsigma^{\prime}\right]p_{\rm  eq}(\bsigma^{\prime})
=\sum_{\bsigma^{\prime\prime}}
W\left[\bsigma^{\prime};\bsigma^{\prime\prime}\right]
\sigma_j^{\prime\prime}p_{\rm eq}(\bsigma^{\prime\prime})
\ed
For parallel dynamics $\ell$ and $\ell^\prime$ are the real
time labels $t$ and $t^\prime$, and we obtain, with $\tau=t\minus t^\prime$:
\be
{\sl Binary~\&~Parallel:}~~~~~~~~~~G_{ij}(\tau>0)= -\beta[C_{ij}(\tau\plus
1)-C_{ij}(\tau\minus 1)],~~~~~~G_{ij}(\tau\leq 0)=0
\label{eq:FDT_Isingparallel}
\ee
For the continuous-time version (\ref{eq:sequentialmaster}) of sequential
dynamics the
time $t$ is defined as $t=\ell/N$, and
the difference equation (\ref{eq:neweqcorrelation})
becomes a differential equation. For perturbations at time $t^{\prime}$ in the definition  of the
response function (\ref{eq:eqresponse}) to retain a non-vanishing effect
at (re-scaled) time $t$ in the limit $N\rightarrow\infty$, they will
have to be re-scaled as well:  $\delta\theta_j(t^{\prime})\rightarrow
N\delta\theta_j(t^{\prime})$. As a result:
\be
{\sl Binary~\&~Sequential:}~~~~~~~~~~
G_{ij}(\tau)=-\beta\theta(\tau)\frac{d}{d\tau}C_{ij}(\tau)
\label{eq:FDT_Isingsequential}
\ee
The need to re-scale perturbations in making the transition from discrete to continuous
times has the same origin as the need to re-scale the
random forces in the derivation of the continuous-time Langevin
equation from a discrete-time process.
Going from ordinary derivatives to functional derivatives (which is
what happens in the continuous-time limit), implies
replacing Kronecker delta's $\delta_{t,t^\prime}$ by Dirac
delta-functions according to $\delta_{t,t^\prime}\to\Delta\delta(t\minus
t^\prime)$, where $\Delta$ is the average duration of an iteration step.
Equations (\ref{eq:FDT_Isingparallel}) and
(\ref{eq:FDT_Isingsequential}) are examples of so-called
fluctuation-dissipation theorems (FDT).
\vsp

\noindent{\em Networks with Continuous Neurons.} For systems described by a Fokker-Planck equation
(\ref{eq:fokkerplanck}) the simplest way to calculate correlation-
and response-functions is by first returning to
the underlying
discrete-time system and leaving
the continuous time limit $\Delta\to 0$ until the end. In \cite{part1} we saw that for
small but finite time-steps $\Delta$ the underlying discrete-time process is
described by
\bd
t=\ell\Delta,~~~~~~~~
p_{\ell\Delta\plus\Delta}(\bsigma)=[1\plus\Delta\cL_{\bsigma}\plus\order(\Delta^{\frac{3}{2}})]p_{\ell\Delta}(\bsigma)
\ed
with $\ell=0,1,2,\ldots$ and with the differential operator
\be
\cL_{\bsigma}=-\sum_i\frac{\partial}{\partial\sigma_i}[f_i(\bsigma)-T\frac{\partial}{\partial\sigma_i}]
\label{eq:operator}
\ee
From this it follows that the conditional probability density $p_{\ell\Delta}(\bsigma|\bsigma^\prime,\ell^\prime\Delta)$
for finding state $\sigma$ at time $\ell\Delta$, given the system was in state $\bsigma^\prime$ at time
$\ell^\prime\Delta$, must be
\be
p_{\ell\Delta}(\bsigma|\bsigma^\prime,\ell^\prime\Delta)=
[1\plus\Delta\cL_{\bsigma}\plus\order(\Delta^{\frac{3}{2}})]^{\ell-\ell^\prime}
\delta[\bsigma\minus\bsigma^\prime]
\label{eq:Langevin_propagator}
\ee
Equation (\ref{eq:Langevin_propagator}) will be our main building
block. Firstly, we will calculate the correlations:
\bd
C_{ij}(\ell\Delta,\ell^\prime\Delta)=\bra\sigma_i(\ell\Delta)\sigma_j(\ell^\prime\Delta)\ket
=\int\!d\bsigma
d\bsigma^\prime ~\sigma_i\sigma_j^\prime
~p_{\ell\Delta}(\bsigma|\bsigma^\prime,\ell^\prime\Delta)p_{\ell^\prime\Delta}(\bsigma^\prime)
\ed
\bd
=\int\!d\bsigma~\sigma_i
[1\plus\Delta\cL_{\bsigma}\plus\order(\Delta^{\frac{3}{2}})]^{\ell-\ell^\prime}
\int\! d\bsigma^\prime ~\sigma_j^\prime\delta[\bsigma\minus\bsigma^\prime]
p_{\ell^\prime\Delta}(\bsigma^\prime)
\ed
\bd
=\int\!d\bsigma~\sigma_i
[1\plus\Delta\cL_{\bsigma}\plus\order(\Delta^{\frac{3}{2}})]^{\ell-\ell^\prime}
\left[\sigma_j ~p_{\ell^\prime\Delta}(\bsigma)\right]
\ed
At this stage we can take the limits $\Delta \to 0$ and $\ell,\ell^\prime\to\infty$, with $t=\ell\Delta$ and
$t^\prime=\ell^\prime\Delta$ finite, using $\lim_{\Delta\to 0}[1\plus \Delta A]^{k/\Delta}=e^{kA}$:
\be
C_{ij}(t,t^\prime)=\int\!d\bsigma ~\sigma_i ~e^{(t-t^\prime)\cL_{\bsigma}}\left[
\sigma_j~ p_{t^\prime}(\bsigma)\right]
\label{eq:Langevin_correlations}
\ee
Next we turn to the response function. A perturbation applied at time $t^\prime=\ell^\prime\Delta$
to the Langevin forces $f_i(\bsigma)$
comes in at the transition $\bsigma(\ell^\prime\Delta)\to\bsigma(\ell^\prime\Delta\plus\Delta)$.
As with sequential dynamics binary networks,
the perturbation is re-scaled with the step size $\Delta$ to retain significance as $\Delta\to 0$:
\bd
G_{ij}(\ell\Delta,\ell^\prime\Delta)=
\frac{\partial\bra\sigma_i(\ell\Delta)\ket}{\Delta\partial\theta_j(\ell^\prime\Delta)}
=\frac{\partial}{\Delta\partial\theta_j(\ell^\prime\Delta)}
\int\!d\bsigma
d\bsigma^\prime ~\sigma_i
~p_{\ell\Delta}(\bsigma|\bsigma^\prime,\ell^\prime\Delta)p_{\ell^\prime\Delta}(\bsigma^\prime)
\ed
\bd
=
\int\!d\bsigma
d\bsigma^\prime d\bsigma^{\pprime} ~\sigma_i
~p_{\ell\Delta}(\bsigma|\bsigma^\pprime,\ell^\prime\Delta\plus\Delta)
\left[\frac{\partial p_{\ell^\pprime\Delta\plus\Delta}(\bsigma|\bsigma^\prime,\ell^\prime\Delta)}
{\Delta\partial\theta_j}\right]
p_{\ell^\prime\Delta}(\bsigma^\prime)
\ed
\bd
=\int\!d\bsigma
d\bsigma^\prime d\bsigma^{\pprime} ~\sigma_i
[1\plus\Delta\cL_{\bsigma}\plus\order(\Delta^{\frac{3}{2}})]^{\ell-\ell^\prime-1}
\delta[\bsigma\minus\bsigma^\pprime]
\left[\frac{1}{\Delta}\frac{\partial}{\partial\theta_j}
[1\plus\Delta\cL_{\bsigma^\pprime}\plus\order(\Delta^{\frac{3}{2}})]
\delta[\bsigma^\pprime\minus\bsigma^\prime]\right]
p_{\ell^\prime\Delta}(\bsigma^\prime)
\ed
\bd
=-\int\!d\bsigma
d\bsigma^\prime d\bsigma^{\pprime} ~\sigma_i
[1\plus\Delta\cL_{\bsigma}\plus\order(\Delta^{\frac{3}{2}})]^{\ell-\ell^\prime-1}
\delta[\bsigma\minus\bsigma^\pprime]
\delta[\bsigma^\pprime\minus\bsigma^\prime]
[\frac{\partial}{\partial\sigma^\prime_j}\plus\order(\Delta^{\frac{1}{2}})]~p_{\ell^\prime\Delta}(\bsigma^\prime)
\ed
\bd
=-\int\!d\bsigma~\sigma_i
[1\plus\Delta\cL_{\bsigma}\plus\order(\Delta^{\frac{3}{2}})]^{\ell-\ell^\prime-1}
[\frac{\partial}{\partial\sigma_j}\plus\order(\Delta^{\frac{1}{2}})]~p_{\ell^\prime\Delta}(\bsigma)
\ed
We take the limits $\Delta \to 0$ and $\ell,\ell^\prime\to\infty$, with $t=\ell\Delta$ and
$t^\prime=\ell^\prime\Delta$ finite:
\be
G_{ij}(t,t^\prime)=-
\int\!d\bsigma ~\sigma_i ~e^{(t-t^\prime)\cL_{\bsigma}}\frac{\partial}{\partial\sigma_j}p_{t^\prime}(\bsigma)
\label{eq:Langevin_response}
\ee
Equations (\ref{eq:Langevin_correlations}) and
(\ref{eq:Langevin_response}) apply to arbitrary systems described
by Fokker-Planck equations. In the case of conservative forces,
i.e. $f_i(\bsigma)=-\partial H(\bsigma)/\partial\sigma_i$, and when the system is in an equilibrium state
at time $t^\prime$ so that $C_{ij}(t,t^\prime)=C_{ij}(t\minus t^\prime)$ and
$G_{ij}(t,t^\prime)=G_{ij}(t\minus t^\prime)$, we can take a further step
using  $p_{t^\prime}(\bsigma)=p_{\rm
eq}(\bsigma)=Z^{-1}e^{-\beta H(\bsigma)}$. In that case, taking the time
derivative of expression (\ref{eq:Langevin_correlations}) gives
\bd
\frac{\partial}{\partial \tau}C_{ij}(\tau)=\int\!d\bsigma ~\sigma_i ~e^{\tau\cL_{\bsigma}}\cL_{\bsigma}\left[
\sigma_j~ p_{\rm eq}(\bsigma)\right]
\ed
Working out the key term in this expression gives
\bd
\cL_{\bsigma}[\sigma_j~ p_{\rm eq}(\bsigma)]
=-\sum_i\frac{\partial}{\partial\sigma_i}[f_i(\bsigma)\minus
T\frac{\partial}{\partial\sigma_i}][\sigma_j~ p_{\rm eq}(\bsigma)]
=T\frac{\partial}{\partial\sigma_j}p_{\rm eq}(\bsigma)-
\sum_i\frac{\partial}{\partial\sigma_i}[\sigma_j J_i(\bsigma)]
\ed
with the components of the probability current density  $J_i(\bsigma)=[f_i(\bsigma)\minus
T\frac{\partial}{\partial\sigma_i}]p_{\rm eq}(\bsigma)$. In
equilibrium, however, the current is zero by definition, so only the
first term in the above expression survives. Insertion into our
previous equation for $\partial C_{ij}(\tau)/\partial\tau$, and comparison with
(\ref{eq:Langevin_response}) leads
to the FDT for continuous systems:
\be
{\sl Continuous:}~~~~~~~~~~
G_{ij}(\tau)=-\beta\theta(\tau)\frac{d}{d\tau}C_{ij}(\tau)
\label{eq:FDT_continuous}
\ee
We will now calculate the correlation and response functions
explicitly, and verify the validity or otherwise of the FDT relations,
 for attractor networks away from saturation.

\subsection{Example: Simple Attractor Networks with Binary Neurons}

\noindent{\em Correlation- and Response Functions for Sequential
Dynamics}.
We will consider the continuous time version
(\ref{eq:sequentialmaster}) of the sequential
dynamics, with the
local fields $h_i(\bsigma)=\sum_j J_{ij}\sigma_j+\theta_i$, and the separable interaction matrix
(\ref{eq:coolenmatrix}). We already solved the dynamics of
this model for the case with zero external fields and away from
saturation (i.e. $p\ll \sqrt{N}$). Having non-zero, or
even time-dependent, external fields does not affect the calculation
much; one adds the
external fields to the internal ones and finds the macroscopic laws
(\ref{eq:overlapflow}) for the overlaps with the stored patterns
being replaced by
\be
\frac{d}{dt}\bm(t)=\lim_{N\to\infty}\frac{1}{N}\sum_i
\bxi_i\tanh\left[\beta\bxi_i\cdot
\bA\bm(t)\plus \theta_i(t)\right] -\bm(t)~~~~~~~~
\label{eq:overlapflowwithfields}
\ee
Fluctuations in the local fields are of vanishing
order in $N$ (since the fluctuations in $\bm$ are),
so that one can easily
derive from the master equation (\ref{eq:sequentialmaster}) the
following expressions for spin averages:
\be
\frac{d}{dt}\bra \sigma_i(t)\ket=
\tanh\beta[\bxi_i\!\cdot\! \bA\bm(t)\plus\theta_i(t)]
- \bra \sigma_i(t)\ket
\label{eq:singlespinaverage}
\ee
\be
i\neq j:~~\frac{d}{dt}\bra \sigma_i(t)\sigma_j(t)\ket=
\tanh\beta[\bxi_i\!\cdot\! \bA\bm(t)\plus\theta_i(t)] \bra \sigma_j(t)\ket
+\tanh\beta[\bxi_j\!\cdot\! \bA\bm(t)\plus\theta_j(t)] \bra \sigma_i(t)\ket
-2\bra \sigma_i(t)\sigma_j(t)\ket
\label{eq:doublespinaverage}
\ee
Correlations at different times are calculated by applying
(\ref{eq:singlespinaverage}) to situations where the
microscopic state at time $t^\prime$ is known exactly, i.e. where
$p_{t^\prime}(\bsigma)=\delta_{\bsigma,\bsigma^\prime}$ for some
$\bsigma^\prime$:
\be
\bra \sigma_i(t)\ket|_{\bsigma(t^\prime)=\bsigma^\prime}=
\sigma_i^\prime e^{-(t-t^\prime)}+\int_{t^\prime}^t\!ds~e^{s-t}
\tanh\beta[\bxi_i\cdot \bA\bm(s;\bsigma^\prime,t^\prime)\plus\theta_i(s)]
\label{eq:conditionalspinaverage}
\ee
with $\bm(s;\bsigma^\prime,t^\prime)$ denoting the solution of
(\ref{eq:overlapflowwithfields}) following initial condition
$\bm(t^\prime)=\frac{1}{N}\sum_i \sigma_i^\prime \bxi_i$.  If we
multiply both sides of (\ref{eq:conditionalspinaverage}) by
$\sigma_j^\prime$ and average over all possible states
$\bsigma^\prime$ at time $t^\prime$ we obtain in leading order in $N$:
\bd
\bra \sigma_i(t)\sigma_j(t^\prime)\ket=
\bra \sigma_i(t^\prime) \sigma_j(t^\prime)\ket  e^{-(t-t^\prime)}+\int_{t^\prime}^t\!ds~e^{s-t}
\bra \tanh\beta[\bxi_i\cdot \bA\bm(s;\bsigma(t^\prime),t^\prime)
\plus\theta_i(s)]\sigma_j(t^\prime)\ket
\ed
Because of the existence of deterministic laws for the overlaps $\bm$
in the $N\to\infty$ limit,
 we know with probability one that during the stochastic
process the actual value $\bm(\bsigma(t^\prime))$ must be given by the
solution of  (\ref{eq:overlapflowwithfields}), evaluated at time
$t^\prime$. As a result we obtain, with $C_{ij}(t,t^\prime)=\bra\sigma_i(t)\sigma_j(t^\prime)\ket$:
\be
C_{ij}(t,t^\prime)=
C_{ij}(t^\prime,t^\prime) e^{-(t-t^\prime)}+
\int_{t^\prime}^t\!ds~e^{s-t}
\tanh\beta[\bxi_i\cdot \bA\bm(s)\plus\theta_i(s)]\bra\sigma_j(t^\prime)\ket
\label{eq:exactcorrelations}
\ee
Similarly we obtain from the solution of (\ref{eq:singlespinaverage})
an equation for
the leading
order in $N$ of the response
functions, by derivation with respect to external fields:
\bd
\frac{\partial\bra \sigma_i(t)\ket}{\partial\theta_j(t^\prime)}
=
\beta\theta(t\minus t^\prime)\int_{-\infty}^t\!ds~e^{s-t}
\left[1\minus \tanh^2\beta[\bxi_i\cdot \bA\bm(s)\plus\theta_i(s)]\right]
\left[ \frac{1}{N}\sum_k (\bxi_i\cdot
\bA\bxi_k)\frac{\partial\bra\sigma_k(s)\ket}{\partial\theta_j(t^\prime)}
\plus
\delta_{ij}\delta(s\minus t^\prime)\right]
\ed
or
\bd
G_{ij}(t,t^\prime)=
\beta\delta_{ij}\theta(t\minus t^\prime) e^{-(t-t^\prime)}
\left[1\minus \tanh^2\beta[\bxi_i\cdot
\bA\bm(t^\prime)\plus\theta_i(t^\prime)]\right]
~~~~~~~~~~~~~~~~~~~~
\ed
\be
~~~~~~~~~~~~~~~~~~~~+
\beta\theta(t\minus t^\prime)\int_{t^\prime}^t\!ds~e^{s-t}
\left[1\minus \tanh^2\beta[\bxi_i\cdot \bA\bm(s)\plus\theta_i(s)]\right]
\frac{1}{N}\sum_k (\bxi_i\cdot \bA\bxi_k) G_{kj}(s,t^\prime)
\label{eq:exactresponse}
\ee
For $t=t^\prime$ we retain in leading order in
$N$ only the instantaneous single-site contribution
\be
\lim_{t^\prime\uparrow t} G_{ij}(t,t^\prime)=
\beta\delta_{ij}
\left[1\minus \tanh^2\beta[\bxi_i\cdot \bA\bm(t)\plus\theta_i(t)]\right]
\label{eq:Gtauzero}
\ee
This leads to the following ansatz for the scaling with $N$ of the
 $G_{ij}(t,t^\prime)$, which can be shown to be correct by
insertion into (\ref{eq:exactresponse}), in combination with the
 correctness at $t=t^\prime$ following from
(\ref{eq:Gtauzero}):
\bd
i=j:~~~G_{ii}(t,t^\prime)=\order(1),
~~~~~~~~~~~~~~~~~~
i\neq j:~~~ G_{ij}(t,t^\prime)=\order(N^{-1})
\ed
Note that this implies $\frac{1}{N}\sum_k (\bxi_i\cdot \bA\bxi_k)
G_{kj}(s,t^\prime)=\order(\frac{1}{N})$. In leading order in $N$ we now find
\be
G_{ij}(t,t^\prime)=
\beta\delta_{ij}\theta(t\minus t^\prime) e^{-(t-t^\prime)}
\left[1\minus \tanh^2\beta[\bxi_i\cdot
\bA\bm(t^\prime)\plus\theta_i(t^\prime)]\right]
\label{eq:leadingresponse}
\ee
For those cases where the macroscopic laws
(\ref{eq:overlapflowwithfields}) describe evolution to a
stationary state $\bm$, obviously requiring stationary external fields
$\theta_i(t)=\theta_i$, we can take the limit $t\to \infty$, with
$t\minus t^\prime=\tau$ fixed, in the two results
(\ref{eq:exactcorrelations},\ref{eq:leadingresponse}).
Using the $t\to\infty$ limits of
(\ref{eq:singlespinaverage},\ref{eq:doublespinaverage}) we subsequently find
time translation invariant expressions:
$\lim_{t\to\infty}C_{ij}(t,t\minus\tau)= C_{ij}(\tau)$ and
$\lim_{t\to\infty}G_{ij}(t,t\minus\tau)= G_{ij}(\tau)$, with in
leading order in $N$
\be
C_{ij}(\tau)=
\tanh\beta[\bxi_i\cdot \bA\bm \plus\theta_i]\tanh\beta[\bxi_j\cdot A\bm
\plus\theta_j]
+\delta_{ij}e^{-\tau}
\left[
1\minus \tanh^2\beta[\bxi_i\cdot \bA\bm \plus\theta_i]
\right]
\label{eq:statcorrelations}
\ee
\be
G_{ij}(\tau)=
\beta\delta_{ij}\theta(\tau) e^{-\tau}
\left[ 1\minus \tanh^2\beta[\bxi_i\cdot
\bA\bm\plus\theta_i]\right]
\label{eq:statresponse}
\ee
for which indeed the Fluctuation-Dissipation Theorem (\ref{eq:FDT_Isingsequential})
holds:
$G_{ij}(\tau)=-\beta\theta(\tau) \frac{d}{d\tau}C_{ij}(\tau)$.
\vsp

\noindent{\em Correlation- and Response Functions for Parallel
Dynamics}.
We now turn to the parallel dynamical rules
(\ref{eq:parallelmarkov}),
with the local fields $h_i(\bsigma)=\sum_j J_{ij}\sigma_j\plus\theta_i$, and the interaction matrix
(\ref{eq:coolenmatrix}). As before, having time-dependent external
fields amounts simply to  adding these fields to the internal ones,
and the dynamic laws
(\ref{eq:overlapmap}) are found to be replaced by
\be
\bm(t+1)=\lim_{N\to\infty}\frac{1}{N}\sum_i
\bxi_i\tanh\left[\beta\bxi_i\cdot
\bA\bm(t)\plus\theta_i(t)\right]
\label{eq:overlapmapwithfields}
\ee
Fluctuations in the local fields are again of vanishing
order in $N$, and the parallel dynamics versions of equations
(\ref{eq:singlespinaverage},\ref{eq:doublespinaverage}), to be derived
from (\ref{eq:parallelmarkov}), are found to be
\be
\bra \sigma_i(t\plus 1)\ket= \tanh\beta[\bxi_i\cdot A\bm(t)\plus\theta_i(t)]
\label{eq:parsinglespinaverage}
\ee
\be
i\neq j:~~~~~~\bra \sigma_i(t\plus 1)\sigma_j(t\plus 1)\ket=
\tanh\beta[\bxi_i\cdot \bA\bm(t)\plus\theta_i(t)]
\tanh\beta[\bxi_j\cdot \bA\bm(t)\plus\theta_j(t)]
\label{eq:pardoublespinaverage}
\ee
With $\bm(t;\bsigma^\prime,t^\prime)$ denoting the solution of
the map (\ref{eq:overlapmapwithfields}) following initial condition
$\bm(t^\prime)=\frac{1}{N}\sum_i \sigma_i^\prime \bxi_i$, we
immediately obtain from equations
(\ref{eq:parsinglespinaverage},\ref{eq:pardoublespinaverage})
the correlation functions:
\be
C_{ij}(t,t)=\delta_{ij}+[1\minus\delta_{ij}]
\tanh\beta[\bxi_i\cdot \bA\bm(t\minus 1)\plus\theta_i(t\minus 1)]
\tanh\beta[\bxi_j\cdot \bA\bm(t\minus 1)\plus\theta_j(t\minus 1)]
\label{eq:parexactcorrelations1}
\ee
\bd
t>t^\prime:~~~~C_{ij}(t,t^\prime)=
 \bra \tanh\beta[\bxi_i\cdot
\bA\bm(t\minus 1;\bsigma(t^\prime),t^\prime)\plus\theta_i(t\minus
1)]\sigma_j(t^\prime)\ket
~~~~~~~~~~~~~~~~~~~~~~~~~~~~~~~~~~~~~~~
\ed
\be
~~~~~~~~
= \tanh\beta[\bxi_i\cdot \bA\bm(t\minus 1)\plus\theta_i(t\minus 1)]
\tanh\beta[\bxi_j\cdot \bA\bm(t^\prime\minus 1)\plus\theta_j(t^\prime\minus 1)]
\label{eq:parexactcorrelations2}
\ee
From (\ref{eq:parsinglespinaverage}) also follow equations determining
the leading
order in $N$ of the response
functions $G_{ij}(t,t^\prime)$, by derivation with respect to the external fields $\theta_j(t^\prime)$:
\be
\begin{array}{lll}
t^\prime> t\minus 1: & & G_{ij}(t,t^\prime)=0 \\[1mm]
t^\prime =t\minus 1: & & G_{ij}(t,t^\prime)= \beta
\delta_{ij}\left[1\minus \tanh^2\beta[\bxi_i\cdot \bA\bm(t\minus 1)\plus\theta_i(t\minus 1)]\right]\\
t^\prime< t\minus 1: & & G_{ij}(t,t^\prime)= \beta
\left[1\minus \tanh^2\beta[\bxi_i\cdot \bA\bm(t\minus
1)\plus\theta_i(t\minus 1)]\right]
\frac{1}{N}\sum_k(\bxi_i\cdot \bA\bxi_k)G_{kj}(t\minus 1,t^\prime)
\end{array}
\label{eq:parexactresponse}
\ee
It now follows iteratively that all off-diagonal elements must be of
vanishing order in $N$: $G_{ij}(t,t\minus 1)=\delta_{ij}G_{ii}(t,t\minus 1)
~\to~
G_{ij}(t,t\minus 2)=\delta_{ij}G_{ii}(t,t\minus 2)
~\to~\ldots$, so that in leading order
\be
G_{ij}(t,t^\prime)= \beta
\delta_{ij}\delta_{t,t^\prime+1}
\left[1\minus \tanh^2\beta[\bxi_i\cdot \bA\bm(t^\prime)\plus\theta_i(t^\prime)]\right]
\label{eq:parleadingresponse}
\ee
For those cases where the macroscopic laws
(\ref{eq:overlapmapwithfields}) describe evolution to a stationary
state $\bm$, with stationary external fields,
we can take the limit $t\to \infty$, with
$t\minus t^\prime=\tau$ fixed, in
(\ref{eq:parexactcorrelations1},\ref{eq:parexactcorrelations2},\ref{eq:parleadingresponse}).
We find
time translation invariant expressions:
$\lim_{t\to\infty}C_{ij}(t,t\minus\tau)= C_{ij}(\tau)$ and
$\lim_{t\to\infty}G_{ij}(t,t\minus\tau)= G_{ij}(\tau)$, with in
leading order in $N$:
\be
C_{ij}(\tau)=
\tanh\beta[\bxi_i\cdot \bA\bm\plus\theta_i]
\tanh\beta[\bxi_j\cdot \bA\bm\plus\theta_j]
+\delta_{ij}\delta_{\tau,0}\left[1\minus
\tanh^2\beta[\bxi_i\cdot \bA\bm\plus\theta_i]\right]
\label{eq:parstatcorrelations}
\ee
\be
G_{ij}(\tau)= \beta
\delta_{ij}\delta_{\tau,1}
\left[1\minus \tanh^2\beta[\bxi_i\cdot \bA\bm\plus\theta_i]\right]
\label{eq:parstatresponse}
\ee
obeying the Fluctuation-Dissipation Theorem
(\ref{eq:FDT_Isingparallel}):
$G_{ij}(\tau>0)=- \beta[C_{ij}(\tau\plus 1)\minus C_{ij}(\tau\minus
1)]$.

\subsection{Example: Graded Response Neurons with
Uniform Synapses}

Let us finally find out how to calculate correlation and response function
for the simple network (\ref{eq:graded_response}) of graded response neurons,
 with (possibly
time-dependent)
external forces $\theta_i(t)$, and with uniform synapses
$J_{ij}=J/N$:
\be
\frac{d}{dt}u_i(t)
=\frac{J}{N}\sum_{j} g[\gamma u_j(t)]-u_i(t) +\theta_i(t)+\eta_i(t)
\label{eq:gradedtoy}
\ee
For a given realisation of the external forces and the Gaussian noise variables $\{\eta_i(t)\}$
we can formally integrate (\ref{eq:gradedtoy}) and find
\be
u_i(t)=u_i(0)e^{-t}+\int_0^t\!ds ~e^{s-t}\left[J \int\!du~\rho(u;\bu(s))~g[\gamma u]+
\theta_i(s)+\eta_i(s)\right]
\label{eq:formaltoysolution}
\ee
with the distribution of membrane potentials $\rho(u;\bu)=N^{-1}\sum_i\delta[u\minus
u_i]$. The correlation function $C_{ij}(t,t^\prime)=\bra u_i(t) u_j(t^\prime)\ket$
immediately follows from (\ref{eq:formaltoysolution}). Without loss of generality we can define $t\geq t^\prime$.
For absent
external forces (which were only needed in order to define the response
function), and upon using $\bra \eta_i(s)\ket=0$ and $\bra\eta_i(s)\eta_j(s^\prime)\ket=2T\delta_{ij}
\delta(s\minus s^\prime)$, we arrive at
\bd
C_{ij}(t,t^\prime)=  T\delta_{ij}(e^{t^\prime-t}-e^{-t^\prime-t})
~~~~~~~~~~~~~~~~~~~~~~~~~~~~~~~~~~~~~~~~~~~~~~~~~~~~~~~~~~~~~~~~~~~~~~~~~~~~~~~~~~~~~~~~~~~~~~
\vspace*{-2mm}
\ed
\bd
+~\bra
\left[u_i(0)e^{-t}\plus J \!\int\!du~g[\gamma u]\int_0^t\!ds~e^{s-t}\rho(u;\bu(s))\right]
\left[u_j(0)e^{-t^\prime}\plus J\! \int\!du~g[\gamma
u]\int_0^{t^\prime}\!ds^\prime
~e^{s^\prime-t^\prime}\rho(u;\bu(s^\prime))\right]\ket
\ed
For $N\to\infty$, however, we know the distribution of potentials to
evolve deterministically: $\rho(u;\bu(s))\to \rho_s(u)$
where $\rho_s(u)$ is the solution of (\ref{eq:TDOU}). This allows
us to simplify the above expression to
\bd
N\to\infty:~~~~C_{ij}(t,t^\prime)=  T\delta_{ij}(e^{t^\prime-t}-e^{-t^\prime-t})
~~~~~~~~~~~~~~~~~~~~~~~~~~~~~~~~~~~~~~~~~~~~~~~~~~~~~~~~~~~~~~~~~~~~~~~~~~~~~~~~~~~~~~~~~~~~~~
\vspace*{-2mm}
\ed
\be
~~+~\bra
\left[u_i(0)e^{-t}\plus J \!\int\!du~g[\gamma u]\int_0^t\!ds~e^{s-t}\rho_s(u)\right]
\left[u_j(0)e^{-t^\prime}\plus J\! \int\!du~g[\gamma
u]\int_0^{t^\prime}\!ds^\prime
~e^{s^\prime-t^\prime}\rho_{s^\prime}(u)\right]\ket
\label{eq:toycorrelations}
\ee
Next we turn to the response function $G_{ij}(t,t^\prime)=\delta\bra
u_i(t)\ket/\delta\xi_j(t^\prime)$ (its definition involves
functional rather than scalar differentiation, since time is
continuous). After this differentiation the forces $\{\theta_i(s)\}$
can be put to zero. Functional differentiation of
(\ref{eq:formaltoysolution}), followed by averaging, then leads us to
\bd
G_{ij}(t,t^\prime)=\theta(t\minus
t^\prime)~\delta_{ij}~e^{t^\prime-t}-
J\int\!du~g[\gamma u]\frac{\partial}{\partial u}
 \int_0^t\!ds ~e^{s-t}~ \frac{1}{N}\sum_k \lim_{\btheta\to{\bf 0}}\bra ~\delta[u\minus u_k(s)]\frac{\delta
 u_k(s)}{\delta \theta_j(t^\prime)}~\ket
\ed
In view of (\ref{eq:formaltoysolution}) we make the self-consistent ansatz
$\delta u_k(s)/\delta\xi_j(s^\prime)=\order(N^{-1})$ for $k\neq j$. This produces
\be
N\!\to\!\infty:~~~~G_{ij}(t,t^\prime)=
\theta(t\minus t^\prime)~\delta_{ij}~e^{t^\prime-t}
\label{eq:toyresponse}
\ee
Since equation (\ref{eq:TDOU}) evolves towards a stationary state,
we can also take the limit $t\to \infty$, with
$t\minus t^\prime=\tau$ fixed, in (\ref{eq:toycorrelations}).
Assuming non-pathological decay of the distribution of potentials
allows us to put $\lim_{t\to\infty}\int_0^t\!ds~e^{s-t}\rho_s(u)=\rho(u)$ (the stationary solution of
(\ref{eq:TDOU})), with which we find not only (\ref{eq:toyresponse}) but also (\ref{eq:toycorrelations})
reducing to
time translation invariant expressions for $N\to\infty$,
$\lim_{t\to\infty}C_{ij}(t,t\minus\tau)= C_{ij}(\tau)$ and
$\lim_{t\to\infty}G_{ij}(t,t\minus\tau)= G_{ij}(\tau)$, in which
\be
\vspace*{-1mm}
C_{ij}(\tau)=  T\delta_{ij}e^{-\tau}
+ J^2 \left\{\int\!du~\rho(u) g[\gamma u]\right\}^2
~~~~~~~~~~~~
G_{ij}(\tau)=
\theta(\tau)\delta_{ij} e^{-\tau}
\vspace*{-1mm}
\ee
Clearly the leading orders in $N$ of these two functions obey the
fluctuation-dissipation theorem (\ref{eq:FDT_continuous}): $G_{ij}(\tau)=-\beta
\theta(\tau)\frac{d}{d\tau}C_{ij}(\tau)$.
As with the binary neuron attractor networks for which
we calculated the correlation and response functions earlier, the
impact of detailed balance violation (occurring when $A_{\mu\nu}\neq A_{\nu\mu}$
in networks with binary neurons and synapses
(\ref{eq:coolenmatrix}), and in all networks with graded
response neurons \cite{part1}) on the validity of the fluctuation-dissipation
theorems, vanishes for $N\to\infty$, provided our networks are relatively simple
and evolve to a stationary
state in terms of the macroscopic observables (the latter need not necessarily happen,
see e.g. figures \ref{fig:ruijgrokflows} and
\ref{fig:cycles}). Detailed balance violation, however, would be
noticed in the finite size effects \cite{Castellanos}.


\section{Dynamics in the Complex Regime}

The approach we followed so far to derive closed macroscopic laws
from the microscopic equations fails when the number of
attractors is no longer small compared to the number $N$ of
microscopic neuronal variables. In statics we have seen \cite{part1} that,
at the work floor level,
the fingerprint of complexity is the need to use replica theory, rather than
the relatively
simple and straightforward methods based on (or equivalent to)
calculating the density of states for given
realisations of the macroscopic observables. This is caused by
the presence of a number of `disorder' variables per degree of
freedom which is proportional to $N$, over which we are forced to
average the macroscopic laws.
One finds that in dynamics this situation is  reflected in
the inability to find an exact set of closed equations
for a finite number of observables (or densities). We will see that the natural dynamical
counterpart of equilibrium replica theory is generating
functional analysis.

\subsection{Overview of Methods and Theories}

Let us return to the simplest setting in which to study the problem:
single pattern recall in an attractor neural network with $N$ binary neurons and  $p=\alpha N$
stored patterns in the non-trivial regime, where $\alpha>0$.
We choose parallel dynamics, i.e. (\ref{eq:parallelmarkov}), with
Hebbian-type
synapses of the form (\ref{eq:coolenmatrix}) with $A_{\mu\nu}=\delta_{\mu\nu}$, i.e.
 $J_{ij}=N^{-1}\sum_{\mu}^p\xi_i^\mu \xi_j^\mu$, giving us the parallel dynamics
version of the Hopfield model \cite{Hopfield}. Our interest is in
the recall overlap $m(\bsigma)=N^{-1}\sum_i\sigma_i\xi_i^1$ between
system state and pattern one.
We saw in \cite{part1} that
for $N\to\infty$ the fluctuations in the
values of the recall overlap $m$ will vanish, and that
for initial states where all $\sigma_i(0)$ are drawn independently
the overlap $m$ will obey
\be
m(t\plus 1)=\int\!dz~P_t(z) \tanh[\beta(m(t)\plus z)]~~~~~~~~~~
P_t(z)=\lim_{N\to \infty} \frac{1}{N}\sum_i \bra \delta[z-
\frac{1}{N}\sum_{\mu>1}\xi_i^1\xi_i^\mu \sum_{j\neq i} \xi_j^\mu
\sigma_j(t)]\ket
~~
\label{eq:equation_m}
\ee
and that all complications in a dynamical
analysis of the $\alpha>0$ regime are concentrated in the
calculation of the distribution $P_t(z)$ of the (generally
non-trivial) interference noise.
\vsp

\noindent{\em Gaussian Approximations.}
As a simple approximation one could just assume \cite{Amari} that the $\sigma_i$
remain uncorrelated at all times, i.e. ${\rm Prob}[\sigma_i(t)\!=\!\pm
\xi_i^1]\!=\!\frac{1}{2}[1\pm m(t)]$ for all $t\geq 0$,
such that the argument given in \cite{part1} for $t=0$ (leading to a Gaussian $P(z)$)
would hold generally, and where
the mapping (\ref{eq:equation_m}) would describe
the overlap evolution at {\rm all} times:
\be
P_t(z)=[2\pi\alpha]^{-\frac{1}{2}}e^{-\frac{1}{2}z^2/\alpha}:~~~~~~~~~~
m(t+1)=\int\!Dz~ \tanh[\beta (m(t)+z\sqrt{\alpha})]
\label{eq:gaussian_simple}
\ee
with the Gaussian measure $Dz=(2\pi)^{-\frac{1}{2}}e^{-\frac{1}{2}z^2}dz$.
This equation, however, must be generally incorrect. Firstly, figure 5 in
\cite{part1} shows that knowledge of $m(t)$ {\em only} does
not permit prediction of $m(t\plus 1)$.
Secondly, expansion of the right-hand side of (\ref{eq:gaussian_simple}) for small $m(t)$
shows that (\ref{eq:gaussian_simple}) predicts a critical noise level (at $\alpha=0$)
of $T_c=\beta_c^{-1}=1$, and a storage capacity
(at $T=0$) of $\alpha_c=2/\pi\approx 0.637$, whereas both numerical simulations and
equilibrium statistical mechanical calculations \cite{part1} point to $\alpha_c\approx
0.139$.
Rather than taking all $\sigma_i$ to be independent, a weaker
assumption would be to just assume the interference noise distribution $P_t(z)$ to
be a zero-average Gaussian one, at any time, with statistically independent noise variables
$z$ at different times.
One can then derive (for $N\to\infty$ and fully connected networks) an evolution equation for the
width $\Sigma(t)$, giving \cite{AmariMaginu,NishimoriOzeki}:
\bd
P_t(z)=[2\pi\Sigma^2(t)]^{-\frac{1}{2}} e^{-\frac{1}{2}z^2/\Sigma^2(t)}:~~~~~~~~~~
m(t\plus 1)=\int\!Dz ~\tanh[\beta(m(t)+z\Sigma(t))]
\ed
\bd
\Sigma^2(t\plus 1)=\alpha + 2\alpha
m(t\plus 1)m(t)h[m(t),\Sigma(t)]+\Sigma^2(t)h^2[m(t),\Sigma(t)]
\ed
with
$h[m,\Sigma]=\beta\left[1\minus
\int\!Dz~\tanh^2[\beta(m+z\Sigma)]\right]$.
These equations describe correctly the qualitative features of recall dynamics,
and  are found to work well when retrieval
actually occurs. For non-retrieval trajectories, however, they
appear to underestimate the impact of interference noise:
they predict $T_c=1$ (at $\alpha=0$) and a storage capacity
(at $T=0$) of $\alpha_c\approx 0.1597$ (which should have been about 0.139).
A final refinement of the Gaussian approach \cite{Okada} consisted
in allowing for correlations between the noise variables $z$ at different
times (while still describing them by Gaussian distributions).
This results in a hierarchy of macroscopic equations, which improve upon the previous
Gaussian theories
and even predict the correct stationary state and phase diagrams, but still
fail to be correct at intermediate times.
The fundamental problem with all Gaussian theories, however sophisticated, is
clearly illustrated in figure 6 of \cite{part1}:
the interference noise distribution is generally {\em not} of a Gaussian
shape.
$P_t(z)$ is only approximately Gaussian when pattern recall occurs.
Hence the successes of Gaussian theories in describing recall
trajectories, and their perpetual problems in describing the non-recall ones.
\vsp

\noindent{\em Non-Gaussian Approximations.}
In view of the non-Gaussian shape of the interference noise
distribution, several attempts have been made at constructing
non-Gaussian approximations. In all cases the aim is to arrive at a
theory involving only macroscopic observables with a {\em single}
time-argument. Figure 6 of \cite{part1}
suggests that for a fully connected network with binary neurons and parallel dynamics
a more accurate ansatz for $P_t(z)$ would be the sum
of two Gaussians. In \cite{HenkelOpper} the following choice was
proposed, guided by the structure of the exact formalism to be
described later:
\bd
P_t(z)=P_{t}^{+}(z)+P_t^{-}(z),~~~~~~~~~~
P_t^{\pm}(z)=\lim_{N\to \infty} \frac{1}{N}\sum_i \delta_{\sigma_i(t),\pm \xi_i^1}\bra \delta[z-
\frac{1}{N}\sum_{\mu>1}\xi_i^1\xi_i^\mu \sum_{j\neq i} \xi_j^\mu
\sigma_j(t)]\ket
\ed
\bd
P_t^{\pm}(z)=\frac{1\pm m(t)}{2\Sigma(t)\sqrt{2\pi
}}e^{-\frac{1}{2}[z\mp d(t)]^2/\Sigma^2(t)}
\ed
followed by a self-consistent calculation of
$d(t)$ (representing an effective `retarded self-interaction', since it has an effect
equivalent to adding $h_i(\bsigma(t))\to
h_i(\bsigma(t))+d(t)\sigma_i(t)$),
and of the width $\Sigma(t)$ of the two distributions
$P_t^{\pm}(z)$, together with
\bd
m(t\plus 1)=
\frac{1}{2}[1\plus m(t)] \int\!Dz~\tanh[\beta(m(t)\plus d(t)\plus z\Sigma(t))]+
\frac{1}{2}[1\minus m(t)]\int\!Dz~\tanh[\beta(m(t)\minus d(t)\plus z\Sigma(t))]
\ed
The resulting three-parameter theory, in the form of closed dynamic equations for $\{m,d,\Sigma\}$,
is found to give a nice (but not perfect)
agreement with numerical simulations.

A different philosophy was followed in \cite{CoolenSherrington} (for sequential dynamics).
First (as yet exact) equations are derived for the evolution of the two
macroscopic observables
$m(\bsigma)=m_1(\bsigma)$ and
$r(\bsigma)=\alpha^{-1}\sum_{\mu>1}m_\mu^2(\bsigma)$, with
$m_\mu(\bsigma)=N^{-1}\sum_i\xi_i^1\sigma_i$,
which are both found to involve $P_t(z)$:
\bd
\frac{d}{dt}m=\int\!dz~P_t(z) \tanh[\beta(m\plus z)]~~~~~~~~~~
\frac{d}{dt}r=\frac{1}{\alpha}\int\!dz~P_t(z) z \tanh[\beta(m\plus z)]+1-r
\ed
Next one closes these equations {\em by hand}, using a maximum-entropy
(or `Occam's Razor') argument: instead of calculating $P_t(z)$ from (\ref{eq:equation_m}) with
the real (unknown) microscopic distribution $p_t(\bsigma)$, it is calculated upon
assigning equal probabilities to all states $\bsigma$ with $m(\bsigma)=m$ and
$r(\bsigma)=r$, followed by averaging over all realisations of the
stored patterns with $\mu>1$. In order words: one assumes (i) that the microscopic
states visited by the system are `typical' within the appropriate $(m,r)$
sub-shells of state space, and (ii) that one can average over the
disorder. Assumption (ii) is harmless, the most important step is
(i). This procedure results in an  explicit
(non-Gaussian) expression for the noise
distribution in terms of $(m,r)$ only, a closed two-parameter theory which is exact for
short times and in equilibrium, accurate predictions
of the macroscopic flow in the $(m,r)$-plane (such as that shown in figure
5 of \cite{part1}), but (again) deviations
in predicted time-dependencies at intermediate times.
This theory, and its performance, was later improved by applying
the same ideas to a derivation of a dynamic equation for the function
$P_t(z)$ itself (rather than for $m$ and $r$ only)
\cite{Laughtonetal}; research is still under way with the aim to
construct a theory along these lines which is fully exact.
\vsp

\noindent{\em Exact Results: Generating Functional Analysis.}
The only fully exact procedure available at present is known under various names,
such as `generating functional analysis',
`path integral formalism' or `dynamic mean-field theory', and is based on a
philosophy different from those described so far. Rather than
working with the probability $p_t(\bsigma)$ of finding a microscopic
state $\bsigma$ at time $t$ in order to calculate the statistics of a set of macroscopic
observables $\bOmega(\bsigma)$ at time $t$,
one here turns to the probability ${\rm Prob}[\bsigma(0),\ldots,\bsigma(t_m)]$ of
finding a microscopic {\em path}
$\bsigma(0)\to\bsigma(1)\to\ldots\to\bsigma(t_m)$.
One also adds time-dependent external sources to the local fields, $h_i(\bsigma)\to
h_i(\bsigma)+\theta_i(t)$, in order to probe the networks via perturbations and define a response function.
The idea is to concentrate on the moment generating
function $Z[\bpsi]$, which, like ${\rm Prob}[\bsigma(0),\ldots,\bsigma(t_m)]$,
fully captures the statistics of paths:
\be
Z[\bpsi]=\bra e^{-i\sum_{i}\sum_{t=0}^{t_m} \psi_i(t)\sigma_i(t)}\ket
\label{eq:generator}
\ee
It generates averages of the relevant observables, including those involving neuron
states at different times, such as correlation functions $C_{ij}(t,t^\prime)=\bra \sigma_i(t)
\sigma_j(t^\prime)\ket$
and response
functions $G_{ij}(t,t^\prime)=\partial\bra \sigma_i(t)\ket/\partial \theta_j(t^\prime)$, upon differentiation
with respect to the dummy variables $\{\psi_i(t)\}$:
\be
\bra \sigma_i(t)\ket=i\lim_{\bpsi\rightarrow\bnul}\frac{\partial
Z[\bpsi]}{\partial\psi_i(t)}
~~~~~~~~
C_{ij}(t,t^\prime)=
-\lim_{\bpsi\rightarrow\bnul}\frac{\partial^2
Z[\bpsi]}{\partial\psi_i(t)\partial\psi_j(t^\prime)}
~~~~~~~~
G_{ij}(t,t^\prime)=
i\lim_{\bpsi\rightarrow\bnul}\frac{\partial^2
Z[\bpsi]}{\partial\psi_i(t)\partial\theta_j(t^\prime)}
\label{eq:generated}
\ee
Next one assumes (correctly) that for $N\to\infty$ only the statistical
properties of the stored patterns will influence the macroscopic
quantities, so that the generating function $Z[\bpsi]$ can be
averaged over all pattern realisations, i.e.
$Z[\bpsi]\to\overline{Z[\bpsi]}$.
As in replica theories (the canonical tool to deal with complexity
in equilibrium) one carries out the disorder average {\em
before} the average over the statistics of the neuron states,
resulting for $N\to\infty$ in what can be interpreted as a theory describing a single
`effective' binary neuron $\sigma(t)$, with an effective local
field $h(t)$ and the dynamics $Prob[\sigma(t\plus 1)=\pm 1]=\frac{1}{2}[1\pm \tanh[\beta h(t)]]$.
 However, this effective local field is found to generally depend
on past states of the neuron, and on zero-average but temporally correlated Gaussian
noise contributions $\phi(t)$:
\be
h(t|\{\sigma\},\{\phi\})=m(t)+\theta(t)+\alpha\sum_{t^\prime<t}R(t,t^\prime)\sigma(t^\prime)+\sqrt{\alpha}\phi(t)
\label{eq:effective_field}
\ee
The first comprehensive neural network studies along these
lines, dealing with fully connected networks, were carried out in
\cite{Riegeretal,Horneretal},
followed by applications to a-symmetrically and symmetrically
extremely diluted networks \cite{KreeZippelius,WatkinSherr}
(we will come back to those later).
More recent applications include sequence processing networks \cite{Duringetal}\footnote{In the case of
sequence recall the overlap $m$
is defined with respect to the `moving' target, i.e.
$m(t)=\frac{1}{N}\sum_i \sigma_i(t)\xi_i^t$}.
For $N\to\infty$ the
differences between different models are found to show up only
in the actual form taken by the effective local
field (\ref{eq:effective_field}),
i.e. in the dependence of the `retarded self-interaction' kernel
$R(t,t^\prime)$ and the covariance matrix $\bra
\phi(t)\phi(t^\prime)\ket$ of the interference-induced Gaussian
noise on the macroscopic objects
$\bC=\{C(s,s^\prime)=\lim_{N\to\infty}\frac{1}{N}
\sum_i C_{ii}(s,s^\prime)\}$ and
$\bG=\{G(s,s^\prime)=\lim_{N\to\infty}\frac{1}{N}\sum_i G_{ii}(s,s^\prime)\}$.
For instance\footnote{In the case of extremely diluted models the structure variables are also treated
as disorder, and thus averaged out.}:
\bd
\vspace*{2mm}
\begin{array}{ccccccc}
{\sl model} && {\sl synapses~} J_{ij} && R(t,t^\prime) && \bra\phi(t)\phi(t^\prime)\ket
\\[1mm]
\hline
\\[-1mm]
{\sl fully~connected,} && \frac{1}{N}\sum_{\mu=1}^{\alpha N}\xi_i^\mu\xi_j^\mu
&&
[(\one\minus \bG)^{-1}\bG](t,t^\prime) &&
[(\one\minus\bG)^{-1}\bC(\one\minus\bG^\dag)^{-1}](t,t^\prime)
\\[-1mm]
{\sl static~ patterns}\\[1mm]
{\sl fully~connected,} && \frac{1}{N}\sum_{\mu=1}^{\alpha N}\xi_i^{\mu+1}\xi_j^\mu
&& 0 && \sum_{n\geq 0}[(\bG^\dag)^n \bC \bG^n](t,t^\prime)
\\[-1mm]
{\sl pattern~sequence}
\\[1mm]
{\sl symm~extr~diluted,} && \frac{c_{ij}}{c}\sum_{\mu=1}^{\alpha c}\xi_i^\mu\xi_j^\mu
&& G(t,t^\prime) && C(t,t^\prime)
\\[-1mm]
{\sl static ~patterns}
\\[1mm]
{\sl asymm ~extr~diluted,} && \frac{c_{ij}}{c}\sum_{\mu=1}^{\alpha c}\xi_i^\mu\xi_j^\mu
&& 0 && C(t,t^\prime)
\\[-1mm]
{\sl static ~patterns}
\\[1mm]
\hline
\end{array}
\vspace*{2mm}
\ed
with the $c_{ij}$ drawn at random according to
$P(c_{ij})=\frac{c}{N}\delta_{c_{ij},1}+(1\minus \frac{c}{N})\delta_{c_{ij},0}$
(either symmetrically, i.e. $c_{ij}=c_{ji}$, or
independently) and where $c_{ii}=0$,
$\lim_{N\to\infty}c/N=0$, and $c\to\infty$.
In all cases the observables (overlaps and correlation- and response-functions) are to be solved
from the following closed equations, involving
the statistics of the single effective
neuron experiencing the field (\ref{eq:effective_field}):
\be
m(t)=\bra \sigma(t)\ket~~~~~~~~~~~~
C(t,t^\prime)=\bra \sigma(t)\sigma(t^\prime)\ket~~~~~~~~~~~~
G(t,t^\prime)=\partial\bra \sigma(t)\ket/ \partial \theta(t^\prime)
\label{eq:mandCandG}
\ee
It is now clear that Gaussian theories can at most produce exact results for asymmetric networks.
Any degree of symmetry in the synapses
is found to induce a
non-zero retarded self-interaction, via the kernel
$K(t,t^\prime)$, which constitutes a non-Gaussian contribution to
the local fields.
Exact closed macroscopic theories
apparently require a number of macroscopic observables which grows as $\order(t^2)$
in order to predict the dynamics up to time $t$. In the case of
sequential dynamics the picture is found to be very similar to the
one above; instead of discrete time labels $t\in\{0,1,\ldots,t_m\}$,
path summations  and matrices,
there one has a real time variable $t\in[0,t_m]$, path-integrals and integral operators.
The remainder of this paper is devoted to the derivation of the
above results and their implications.

\subsection{Generating Functional Analysis for Binary Neurons}

\noindent{\em General Definitions.}
I will now show more explicitly how the generating functional
formalism works for networks of binary neurons.
We define parallel dynamics, i.e. (\ref{eq:parallelmarkov}),
driven as usual by local fields of the form $h_i(\bsigma;t)=\sum_j
J_{ij}\sigma_j+\theta_i(t)$, but with a more general choice of
Hebbian-type synapses, in which we allow for a possible random dilution
(to reduce repetition in our subsequent derivations):
\be
J_{ij}=\frac{c_{ij}}{c} \sum_{\mu=1}^{p}\xi_i^\mu \xi_j^\mu
~~~~~~~~~~p=\alpha c
\label{eq:diluted_synapses}
\ee
Architectural
properties are reflected in the variables $c_{ij}\in\{0,1\}$,
whereas information storage is to be effected by the remainder in (\ref{eq:diluted_synapses}),
involving $p$ randomly and independently drawn patterns
$\bxi^\mu=(\xi_1^\mu,\ldots,\xi_N^\mu)\in\{-1,1\}^N$. I will deal both with symmetric and with
asymmetric architectures (always putting $c_{ii}=0$),
in which the variables $c_{ij}$ are
drawn randomly according to
\be
{\sl symmetric:}~~~~c_{ij}=c_{ji},~~~\forall i<j~~P(c_{ij})=\frac{c}{N}\delta_{c_{ij},1}+(1-\frac{c}{N})\delta_{c_{ij},0}
\vspace*{-1mm}
\label{eq:symm_dil}
\ee
\be
{\sl asymmetric:}~~~~~~~~~~~~~~~~~\forall i\neq j~~P(c_{ij})=\frac{c}{N}\delta_{c_{ij},1}+(1-\frac{c}{N})\delta_{c_{ij},0}
\label{eq:asymm_dil}
\ee
(one could also study intermediate degrees of symmetry; this would involve only simple adaptations).
Thus $c_{kl}$ is statistically independent of $c_{ij}$ as soon as $(k,l)\notin
\{(i,j),(j,i)\}$. In leading order in $N$ one has $\bra \sum_j c_{ij}\ket=c$ for all
$i$, so $c$ gives the average number of neurons contributing to the
field of any given neuron. In view of this, the number $p$ of
patterns to be stored can be expected to scale as $p=\alpha c$.
The connectivity parameter  $c$ is chosen to diverge with $N$, i.e.
$\lim_{N\to\infty}c^{-1}=0$.
If $c=N$ we obtain the fully connected (parallel dynamics) Hopfield model. Extremely
diluted networks are obtained when $\lim_{N\to\infty}c/N=0$.

For simplicity we make the so-called `condensed ansatz': we assume that the system state has
an $\order(N^0)$ overlap only with a single pattern, say $\mu=1$.
This situation is induced by initial conditions: we take
a randomly drawn $\bsigma(0)$, generated by
\be
p(\bsigma(0))=\prod_i \left\{
\frac{1}{2}[1\plus m_0]\delta_{\sigma_i(0),\xi_i^1}+\frac{1}{2}[1\minus m_0]\delta_{\sigma_i(0),-\xi_i^1}
\right\}
~~~~~~{\rm so}~~~~~~\frac{1}{N}\sum_i\xi_i^1\bra\sigma_i(0) \ket=m_0
\label{eq:initial_cond}
\ee
The patterns $\mu>1$, as well as  the architecture  variables
$c_{ij}$, are viewed as disorder. One assumes that for $N\to\infty$ the macroscopic behaviour of the system
is `self-averaging', i.e. only dependent on the statistical properties of the disorder (rather than on
its microscopic realisation). Averages over
the disorder are written as $\overline{\cdots}$.
We next define the disorder-averaged generating function:
\be
\overline{Z[\bpsi]}=\overline{\bra e^{-i\sum_i \sum_t
\psi_i(t)\sigma_i(t)}\ket }
\label{eq:disav_generator}
\ee
in which the time  $t$ runs from $t=0$ to some
(finite) upper limit $t_{m}$. Note that $\overline{Z[\bnul]}=1$. With a modest
amount of foresight we define the macroscopic
site-averaged and disorder-averaged objects $m(t)\!=\!N^{-1}\sum_i\xi_i^1\overline{\bra\sigma_i(t)\ket}$,
$C(t,t^\prime)\!=\!N^{-1}\sum_i\overline{\bra\sigma_i(t)\sigma_i(t^\prime)\ket}$ and
$G(t,t^\prime)\!=\!N^{-1}\sum_i\partial \overline{\bra\sigma_i(t)\ket}/\partial
\theta_i(t^\prime)$. According to (\ref{eq:generated}) they can be
obtained from (\ref{eq:disav_generator}) as follows:
\be
m(t)=\lim_{\bpsi\rightarrow\bnul}\frac{i}{N}\sum_{j}\xi_j^1\frac{\partial
\overline{Z[\bpsi]}}{\partial\psi_j(t)}
\label{eq:disav_m}
\ee
\be
C(t,t^\prime)=
-\lim_{\bpsi\rightarrow\bnul}\frac{1}{N}\sum_j\frac{\partial^2
\overline{Z[\bpsi]}}{\partial\psi_j(t)\partial\psi_j(t^\prime)}
~~~~~~~~
G(t,t^\prime)=
\lim_{\bpsi\rightarrow\bnul}\frac{i}{N}\sum_j \frac{\partial^2
\overline{Z[\bpsi]}}{\partial\psi_j(t)\partial\theta_j(t^\prime)}
\label{eq:disav_CG}
\ee
So far we have only
reduced our problem to the calculation of the function $\overline{Z[\bpsi]}$
in (\ref{eq:disav_generator}), which will play a part similar to that of the
disorder-averaged free
energy in equilibrium calculations (see \cite{part1}).

\clearpage
\noindent{\em Evaluation of the Disorder-Averaged Generating Function.}
As in equilibrium replica
calculations, the hope is that progress can be made
by carrying out the disorder averages first. In equilibrium
calculations we use the replica trick to convert our disorder
averages
into feasible ones; here the idea is to isolate the
local fields at different times and different sites
by inserting appropriate $\delta$-distributions:
\bd
1=\prod_{it}\int\!dh_i(t)\delta[h_i(t)\minus \sum_{j}J_{ij}\sigma_j(t)\minus\theta_i(t)]=\int\!\{d\bh d\hbh\} e^{i\sum_{it} \hh_i(t)[h_i(t)-\sum_j
J_{ij}\sigma_j(t)-\theta_i(t)]}
\vspace*{-1mm}
\ed
with $\{d\bh d\hbh\}=\prod_{it}[d\hh_i(t) dh_i(t)/2\pi]$, giving
\bd
\overline{Z[\bpsi]}=\int\!\{d\bh d\hbh\}
e^{i\sum_{it}\hh_i(t)[h_i(t)-\theta_i(t)]}
\bra e^{-i\sum_{it}\psi_i(t)\sigma_i(t)}
\overline{ \left[e^{-i\sum_{it}\hh_i(t)\sum_j
J_{ij}\sigma_j(t)}\right]}
\ket_{\rm pf}
\ed
in which $\bra \ldots\ket_{\rm pf}$ refers to averages over a
constrained stochastic process of the type (\ref{eq:parallelmarkov}),
but with prescribed fields $\{h_i(t)\}$ at all sites and at all times.
Note that with such prescribed  fields the probability of generating
a path $\{\bsigma(0),\ldots,\bsigma(t_{m})\}$ is given by
\bd
{\rm Prob}[\bsigma(0),\ldots,\bsigma(t_{m})|\{h_i(t)\}]=
p(\bsigma(0))e^{ \sum_{it}[\beta \sigma_i(t+1) h_i(t)-\log 2\cosh[\beta
h_i(t)]]}
\vspace*{-2mm}
\ed
so
\be
\overline{Z[\bpsi]}=\int\!\{d\bh
d\hbh\}\sum_{\bsigma(0)}\!\cdots\!\!\sum_{\bsigma(t_{m})}\!
p(\bsigma(0))e^{N{\cal F}[\{\bsigma\},\{\hbh\}]}
\prod_{it}
e^{i\hh_i(t)[h_i(t)-\theta_i(t)]-i\psi_i(t)\sigma_i(t)+ \beta \sigma_i(t+1) h_i(t)-\log 2\cosh[\beta
h_i(t)]}
\vspace*{-1mm}
\label{eq:Zav_stage1}
\ee
\noindent with
\vspace*{-2mm}
\be
{\cal F}[\{\bsigma\},\{\hbh\}]=\frac{1}{N}\log\overline{ \left[e^{-i\sum_{it}\hh_i(t)\sum_j
J_{ij}\sigma_j(t)}\right]}
\label{eq:Zav_disorder}
\ee
We concentrate on the term ${\cal F}[\ldots]$ (with the disorder), of which we need only
know the limit $N\to\infty$, since only terms inside $\overline{Z[\bpsi]}$ which are exponential in $N$ will retain
statistical relevance. In the
disorder-average of (\ref{eq:Zav_disorder})
every site $i$ plays an equivalent role, so the leading order in $N$
 of (\ref{eq:Zav_disorder}) should
depend only on site-averaged functions of the $\{\sigma_i(t),\hh_i(t)\}$, with no reference to any special direction
except the one defined by pattern $\bxi^1$. The
simplest such functions with a single time variable are
\be
\vspace*{-1mm}
a(t;\{\bsigma\})=\frac{1}{N}\sum_i\xi_i^1\sigma_i(t)
~~~~~~~~~~~~
k(t;\{\hbh\})=\frac{1}{N}\sum_i\xi_i^1\hh_i(t)
\label{eq:onetime}
\ee
whereas the
simplest ones with two time variables would appear to be
\be
q(t,t^\prime;\{\bsigma\})=\frac{1}{N}\sum_i \sigma_i(t)\sigma_i(t^\prime)
~~~~~~~~~~~~
Q(t,t^\prime;\{\hbh\})=\frac{1}{N}\sum_i \hh_i(t)\hh_i(t^\prime)
\vspace*{-2mm}
\label{eq:qQ}
\ee
\be
K(t,t^\prime;\{\bsigma,\hbh\})=\frac{1}{N}\sum_i \hh_i(t)\sigma_i(t^\prime)
\label{eq:K}
\ee
It will turn out that all models of the
type (\ref{eq:diluted_synapses}), with either
(\ref{eq:symm_dil}) or (\ref{eq:asymm_dil}), have the crucial property
that (\ref{eq:onetime},\ref{eq:qQ},\ref{eq:K}) are in fact
the {\em only} functions to appear in the leading order of
(\ref{eq:Zav_disorder}):
\be
{\cal F}[\ldots]=\Phi[\{a(t;\ldots),k(t;\ldots),q(t,t^\prime;\ldots),
Q(t,t^\prime;\ldots),K(t,t^\prime;\ldots)\}]
+ \ldots ~~~~~(N\to\infty)
\label{eq:dyn_meanfield}
\ee
for some as yet unknown function $\Phi[\ldots]$.
This allows us to proceed with the evaluation of
(\ref{eq:Zav_stage1}).
We can achieve site factorisation in (\ref{eq:Zav_stage1}) if we
isolate the macroscopic objects (\ref{eq:onetime},\ref{eq:qQ},\ref{eq:K})
by introducing suitable
$\delta$-distributions (taking care that all exponents scale linearly with $N$, to secure statistical relevance).
Thus we insert
\bd
1=\prod_{t=0}^{t_m}\int\!da(t)~\delta[a(t)\minus a(t;\{\bsigma\})]
=\left[\frac{N}{2\pi}\right]^{t_m+1}\!\!\int\!d\ba d\hba
~e^{iN\sum_{t}\ha(t)[a(t)-\frac{1}{N}\sum_j\xi_j^1 \sigma_j(t)]}
\vspace*{-2mm}
\ed
\bd
1=\prod_{t=0}^{t_m}\int\!dk(t)~\delta[k(t)\minus k(t;\{\hbh\})]
=\left[\frac{N}{2\pi}\right]^{t_m+1}\!\!\int\!d\bk d\hbk
~e^{iN\sum_{t}\hk(t)[k(t)-\frac{1}{N}\sum_j\xi_j^1 \hh_j(t)]}
\vspace*{-2mm}
\ed
\bd
1=\!\prod_{t,t^\prime=0}^{t_m}\int\!dq(t,t^\prime)~\delta[q(t,t^\prime)\minus
q(t,t^\prime;\{\bsigma\})]
=\left[\frac{N}{2\pi}\right]^{(t_m+1)^2}\!\!\!\int\!d\bq d\hbq
~e^{iN\sum_{t,t^\prime}\hq(t,t^\prime)[q(t,t^\prime)-\frac{1}{N}\sum_j\sigma_j(t) \sigma_j(t^\prime)]}
\vspace*{-2mm}
\ed
\bd
1=\!\prod_{t,t^\prime=0}^{t_m}\int\!dQ(t,t^\prime)~\delta[Q(t,t^\prime)\minus
Q(t,t^\prime;\{\hbh\})]
=\left[\frac{N}{2\pi}\right]^{(t_m+1)^2}\!\!\!\int\!d\bQ d\hbQ
~e^{iN\sum_{t,t^\prime}\hQ(t,t^\prime)[Q(t,t^\prime)-\frac{1}{N}\sum_j\hh_j(t) \hh_j(t^\prime)]}
\vspace*{-2mm}
\ed
\bd
1=\!\prod_{t,t^\prime=0}^{t_m}\int\!dK(t,t^\prime)~\delta[K(t,t^\prime)\minus
K(t,t^\prime;\{\bsigma,\hbh\})]
=\left[\frac{N}{2\pi}\right]^{(t_m+1)^2}\!\!\!\int\!d\bK d\hbK
~e^{iN\sum_{t,t^\prime}\hK(t,t^\prime)[K(t,t^\prime)-\frac{1}{N}\sum_j\hh_j(t)\sigma_j(t^\prime) ]}
\ed
Insertion of these integrals into (\ref{eq:Zav_stage1}), followed by
insertion of
(\ref{eq:dyn_meanfield}) and usage  of the short-hand
\bd
\Psi[\ba,\hba,\bk,\hbk,\bq,\hbq,\bQ,\hbQ,\bK,\hbK]
=i\sum_{t}[\ha(t)a(t)+\hk(t)k(t)]
~~~~~~~~~~~~~~~~~~~~~~~~~~~~~~~~~~~~~~~~~~~~~~~~~~~
\vspace*{-2mm}
\ed
\be
~~~~~~~~~~~~~~~~~~~~~~~~~~~~~~~~~
+i\sum_{t,t^\prime}[\hq(t,t^\prime)q(t,t^\prime)+\hQ(t,t^\prime)Q(t,t^\prime)+\hK(t,t^\prime)K(t,t^\prime)]
\vspace*{-3mm}
\label{eq:bookkeeping}
\ee
then leads us to
\bd
\overline{Z[\bpsi]}=
\int\!d\ba d\hba d\bk d\hbk d\bq d\hbq d\bQ d\hbQ d\bK d\hbK~
e^{N\Psi[\ba,\hba,\bk,\hbk,\bq,\hbq,\bQ,\hbQ,\bK,\hbK]
+N\Phi[\ba,\bk,\bq,\bQ,\bK]+\order(\ldots)}
\ed
\bd
\times
\int\!\{d\bh
d\hbh\}\sum_{\bsigma(0)}\!\cdots\!\!\sum_{\bsigma(t_{m})}\!
p(\bsigma(0))
\prod_{it}
e^{i\hh_i(t)[h_i(t)-\theta_i(t)]-i\psi_i(t)\sigma_i(t)+ \beta \sigma_i(t+1) h_i(t)-\log 2\cosh[\beta
h_i(t)]}
\ed
\be
\times\prod_i
e^{-i\xi_i^1 \sum_t[
\ha(t)\sigma_i(t)+\hk(t)\hh_i(t)]-i\sum_{t,t^\prime}[
\hq(t,t^\prime)\sigma_i(t)\sigma_i(t^\prime)+\hQ(t,t^\prime)\hh_i(t)\hh_i(t^\prime)
+\hK(t,t^\prime)\hh_i(t)\sigma_i(t^\prime)]}
\label{eq:Zav_stage2}
\ee
in which the term denoted as $\order(\ldots)$ covers both the
non-dominant orders in (\ref{eq:Zav_disorder}) and the $\order(\log N)$
relics of the various pre-factors $[N/2\pi]$ in the above integral
representations of the $\delta$-distributions (note: $t_m$ was
assumed fixed).
We now see explicitly in (\ref{eq:Zav_stage2}) that the summations
and integrations over neuron states and local fields fully
factorise over the $N$ sites.
A simple transformation $\{\sigma_i(t),h_i(t),\hh_i(t)\}\to \{\xi_i^1\sigma_i(t),\xi_i^1
h_i(t),\xi_i^1\hh_i(t)\}$
brings the result into the form
\bd
\int\!\{d\bh
d\hbh\}\sum_{\bsigma(0)}\!\cdots\!\!\sum_{\bsigma(t_{m})}\!
p(\bsigma(0))
\prod_{it}
e^{i\hh_i(t)[h_i(t)-\xi_i^1\theta_i(t)]-i\xi_i^1\psi_i(t)\sigma_i(t)
+\beta \sigma_i(t+1) h_i(t)-\log 2\cosh[\beta
h_i(t)]}
~~~~~~~~~~~~~~~~~~~~
\vspace*{-1mm}
\ed
\bd
~~~~~~~~~~~~~~~~~~~~
\times\prod_i
e^{-i\xi_i^1\sum_t[
\ha(t)\sigma_i(t)+\hk(t)\hh_i(t)]-i\sum_{t,t^\prime}[
\hq(t,t^\prime)\sigma_i(t)\sigma_i(t^\prime)+\hQ(t,t^\prime)\hh_i(t)\hh_i(t^\prime)
+\hK(t,t^\prime)\hh_i(t)\sigma_i(t^\prime)]}
\vspace*{-3mm}
\ed
\bd
=e^{N~\Xi[\hba,\hbk,\hbq,\hbQ,\hbK]}
\ed
with
\bd
\Xi[\hba,\hbk,\hbq,\hbQ,\hbK]=\frac{1}{N}\sum_i\log
\int\!\{dh
d\hh\}\!\!\!\sum_{\sigma(0)\cdots\sigma(t_{m})}\!\!\!\pi_0(\sigma(0))~
e^{\sum_t\{i\hh(t)[h(t)-\xi_i^1\theta_i(t)]-i\xi_i^1\psi_i(t)\sigma(t)\}}
~~~~~~~~~~
\ed
\be
~~~~~~~~~~
\times
e^{\sum_t\{\beta \sigma(t+1) h(t)-\log 2\cosh[\beta
h(t)]\}-i\sum_t[
\ha(t)\sigma(t)+\hk(t)\hh(t)]-i\sum_{t,t^\prime}[
\hq(t,t^\prime)\sigma(t)\sigma(t^\prime)+\hQ(t,t^\prime)\hh(t)\hh(t^\prime)
+\hK(t,t^\prime)\hh(t)\sigma(t^\prime)]}
\label{eq:define_Xi}
\ee
in which $\{dh d\hh\}=\prod_t[dh(t)d\hh(t)/2\pi]$
and $\pi_0(\sigma)=\frac{1}{2}[1\plus m_0]\delta_{\sigma,1}\plus \frac{1}{2}[1\minus
m_0]\delta_{\sigma,-1}$. At this stage (\ref{eq:Zav_stage2})
acquires the form of an integral to be evaluated via the
saddle-point (or `steepest descent') method:
\be
\overline{Z[\{\psi(t)\}]}=
\int\!d\ba d\hba d\bk d\hbk d\bq d\hbq d\bQ d\hbQ d\bK d\hbK~
e^{N\left\{\Psi[\ldots]+\Phi[\ldots]+\Xi[\ldots]\right\}+\order(\ldots)}
\label{eq:Zav_stage3}
\ee
in which the functions $\Psi[\ldots]$, $\Phi[\ldots]$ and $\Xi[\ldots]$
are defined by
(\ref{eq:dyn_meanfield},\ref{eq:bookkeeping},\ref{eq:define_Xi}).

\clearpage
\noindent{\em The Saddle-Point Problem.}
The disorder-averaged generating function (\ref{eq:Zav_stage3}) is
for $N\to\infty$ dominated by the physical saddle-point of the
macroscopic surface
\be
\Psi[\ba,\hba,\bk,\hbk,\bq,\hbq,\bQ,\hbQ,\bK,\hbK]+\Phi[\ba,\bk,\bq,\bQ,\bK]+\Xi[\hba,\hbk,\hbq,\hbQ,\hbK]
\label{eq:surface}
\ee
with the three contributions defined in
(\ref{eq:dyn_meanfield},\ref{eq:bookkeeping},\ref{eq:define_Xi}).
It will be advantageous at this stage to define the following effective
measure (which will be further simplified later):
\be
\bra f[\{\sigma\},\{h\},\{\hh\}]\ket_\star=\frac{1}{N}\sum_i\left\{
\frac{\int\!\{dh
d\hh\}\!\!\sum_{\sigma(0)\cdots\sigma(t_{m})}\!M_i[\{\sigma\},\{h\},\{\hh\}]~f[\{\sigma\},\{h\},\{\hh\}]}
{\int\!\{dh
d\hh\}\!\!\sum_{\sigma(0)\cdots\sigma(t_{m})}\!M_i[\{\sigma\},\{h\},\{\hh\}]}
\right\}
\vspace*{-3mm}
\label{eq:measure_stage1}
\ee
with
\bd
M_i[\{\sigma\},\{h\},\{\hh\}]=
\pi_0(\sigma(0))~
e^{\sum_t\{i\hh(t)[h(t)-\xi_i^1\theta_i(t)]-i\xi_i^1\psi_i(t)\sigma(t)+ \beta \sigma(t+1) h(t)-\log 2\cosh[\beta
h(t)]\}}
\vspace*{-1mm}
\ed
\bd
~~~~~~~~~~~~~~~~~~~~
\times
e^{-i\sum_t[
\ha(t)\sigma(t)+\hk(t)\hh(t)]-i\sum_{t,t^\prime}[
\hq(t,t^\prime)\sigma(t)\sigma(t^\prime)+\hQ(t,t^\prime)\hh(t)\hh(t^\prime)
+\hK(t,t^\prime)\hh(t)\sigma(t^\prime)]}
\ed
in which the values to be inserted for $\{\hm(t),\hk(t),\hq(t,t^\prime),\hQ(t,t^\prime),\hK(t,t^\prime)\}$
are given by the
saddle-point of (\ref{eq:surface}).
Variation of (\ref{eq:surface}) with respect to all the
original macroscopic objects occurring as arguments (those without the `hats') gives the following
set of saddle-point equations:
\be
\ha(t)=i\partial{\Phi}/\partial a(t)
~~~~~~~~~~~~
\hk(t)=i\partial{\Phi}/\partial k(t)
\label{eq:nohats_onetime}
\ee
\be
\hq(t,t^\prime)=i\partial\Phi/\partial q(t,t^\prime)
~~~~~~~~~~
\hQ(t,t^\prime)=i\partial\Phi/\partial Q(t,t^\prime)
~~~~~~~~~~
\hK(t,t^\prime)=i\partial\Phi/\partial K(t,t^\prime)
\label{eq:nohats_twotimes}
\ee
Variation of (\ref{eq:surface}) with respect to the conjugate macroscopic objects (those with the
`hats'), in turn, and usage of our newly introduced short-hand notation $\bra\ldots\ket_\star$,
gives:
\be
a(t)=\bra \sigma(t)\ket_\star
~~~~~~~~~~~~
k(t)=\bra \hh(t)\ket_\star
\label{eq:hats_onetime}
\ee
\be
q(t,t^\prime)=\bra\sigma(t)\sigma(t^\prime)\ket_\star
~~~~~~~~~~
Q(t,t^\prime)=\bra \hh(t)\hh(t^\prime)\ket_\star
~~~~~~~~~~
K(t,t^\prime)=\bra \hh(t)\sigma(t^\prime)\ket_\star
\label{eq:hats_twotimes}
\ee
The coupled equations
(\ref{eq:nohats_onetime},\ref{eq:nohats_twotimes},\ref{eq:hats_onetime},\ref{eq:hats_twotimes})
are to be solved simultaneously, once we have calculated the term $\Phi[\ldots]$
(\ref{eq:dyn_meanfield}) which depends on the synapses. This
appears to be a formidable task; it can, however, be simplified
considerably upon first deriving the physical meaning of the above
macroscopic quantities. We  apply
(\ref{eq:disav_m},\ref{eq:disav_CG}) to (\ref{eq:Zav_stage3}), using identities such as
\bd
\frac{\partial\Xi[\ldots]}{\partial\psi_j(t)}=
-\frac{i}{N}\xi_j^1\left[\frac{\int\!\{dh
d\hh\}\!\!\sum_{\sigma(0)\cdots\sigma(t_{m})}\!M_j[\{\sigma\},\{h\},\{\hh\}]\sigma(t)}
{\int\!\{dh
d\hh\}\!\!\sum_{\sigma(0)\cdots\sigma(t_{m})}\!M_j[\{\sigma\},\{h\},\{\hh\}]}
\right]
\ed
\bd
\frac{\partial\Xi[\ldots]}{\partial\theta_j(t)}=
-\frac{i}{N}\xi_j^1\left[\frac{\int\!\{dh
d\hh\}\!\!\sum_{\sigma(0)\cdots\sigma(t_{m})}\!M_j[\{\sigma\},\{h\},\{\hh\}]\hh(t)}
{\int\!\{dh
d\hh\}\!\!\sum_{\sigma(0)\cdots\sigma(t_{m})}\!M_j[\{\sigma\},\{h\},\{\hh\}]}
\right]
\ed
\bd
\frac{\partial^2\Xi[\ldots]}{\partial\psi_j(t)\partial \psi_j(t^\prime)}=
-\frac{1}{N}\left[\frac{\int\!\{dh
d\hh\}\!\!\sum_{\sigma(0)\cdots\sigma(t_{m})}\!M_j[\{\sigma\},\{h\},\{\hh\}]\sigma(t)\sigma(t^\prime)}
{\int\!\{dh
d\hh\}\!\!\sum_{\sigma(0)\cdots\sigma(t_{m})}\!M_j[\{\sigma\},\{h\},\{\hh\}]}
\right]
-N\left[\frac{\partial\Xi[\ldots]}{\partial\psi_j(t)}\right]
\left[\frac{\partial\Xi[\ldots]}{\partial\psi_j(t^\prime)}\right]
\ed
\bd
\frac{\partial^2\Xi[\ldots]}{\partial\theta_j(t)\partial \theta_j(t^\prime)}=
-\frac{1}{N}\left[\frac{\int\!\{dh
d\hh\}\!\!\sum_{\sigma(0)\cdots\sigma(t_{m})}\!M_j[\{\sigma\},\{h\},\{\hh\}]\hh(t)\hh(t^\prime)}
{\int\!\{dh
d\hh\}\!\!\sum_{\sigma(0)\cdots\sigma(t_{m})}\!M_j[\{\sigma\},\{h\},\{\hh\}]}
\right]
-N\left[\frac{\partial\Xi[\ldots]}{\partial\theta_j(t)}\right]
\left[\frac{\partial\Xi[\ldots]}{\partial\theta_j(t^\prime)}\right]
\ed
\bd
\frac{\partial^2\Xi[\ldots]}{\partial\psi_j(t)\partial \theta_j(t^\prime)}=
-\frac{i}{N}\left[\frac{\int\!\{dh
d\hh\}\!\!\sum_{\sigma(0)\cdots\sigma(t_{m})}\!M_j[\{\sigma\},\{h\},\{\hh\}]\sigma(t)\hh(t^\prime)}
{\int\!\{dh
d\hh\}\!\!\sum_{\sigma(0)\cdots\sigma(t_{m})}\!M_j[\{\sigma\},\{h\},\{\hh\}]}
\right]
-N\left[\frac{\partial\Xi[\ldots]}{\partial\psi_j(t)}\right]
\left[\frac{\partial\Xi[\ldots]}{\partial\theta_j(t^\prime)}\right]
\ed
and using the short-hand notation (\ref{eq:measure_stage1}) wherever
possible.
Note that the external fields $\{\psi_i(t),\theta_i(t)\}$ occur only in the function $\Xi[\ldots]$, not in $\Psi[\ldots]$
or $\Phi[\ldots]$, and that overall constants
in $\overline{Z[\bpsi]}$ can always be recovered {\em a posteriori},
using $\overline{Z[\bnul]}=1$:
\bd
m(t)
=\lim_{\bpsi\rightarrow\bnul}\frac{i}{N}\sum_{i}\xi_i^1
\frac{\int\!d\ba\ldots d\hbK\left[\frac{N\partial\Xi}{\partial
\psi_i(t)}\right]
e^{N[\Psi+\Phi+\Xi]+\order(\ldots)}}
{\int\!d\ba \ldots d\hbK~
e^{N[\Psi+\Phi+\Xi]+\order(\ldots)}}=\lim_{\bpsi\rightarrow\bnul}\bra\sigma(t)\ket_{\star}
\ed
\bd
C(t,t^\prime)
=-\lim_{\bpsi\rightarrow\bnul}\frac{1}{N}\sum_{i}
\frac{\int\!d\ba\ldots d\hbK\left[\frac{N\partial^2\Xi}{\partial
\psi_i(t)\partial\psi_i(t^\prime)}+ \frac{N\partial\Xi}{\partial
\psi_i(t)}\frac{N\partial\Xi}{\partial
\psi_i(t^\prime)} \right]
e^{N[\Psi+\Phi+\Xi]+\order(\ldots)}}
{\int\!d\ba \ldots d\hbK~
e^{N[\Psi+\Phi+\Xi]+\order(\ldots)}}=\lim_{\bpsi\rightarrow\bnul}\bra\sigma(t)\sigma(t^\prime)\ket_{\star}
\ed
\bd
iG(t,t^\prime)
=-\lim_{\bpsi\rightarrow\bnul}\frac{1}{N}\sum_{i}
\frac{\int\!d\ba\ldots d\hbK\left[\frac{N\partial^2\Xi}{\partial
\psi_i(t)\partial\theta_i(t^\prime)}+ \frac{N\partial\Xi}{\partial
\psi_i(t)}\frac{N\partial\Xi}{\partial
\theta_i(t^\prime)} \right]
e^{N[\Psi+\Phi+\Xi]+\order(\ldots)}}
{\int\!d\ba \ldots d\hbK~
e^{N[\Psi+\Phi+\Xi]+\order(\ldots)}}=\lim_{\bpsi\rightarrow\bnul}\bra\sigma(t)\hh(t^\prime)\ket_{\star}
\ed
Finally we obtain useful identities from the
seemingly trivial statements
$N^{-1}\sum_i\xi_i^1\partial \overline{Z[\bnul]}/\partial \theta_i(t)=0$
and
$N^{-1}\sum_i \partial^2\overline{Z[\bnul]}/\partial
\theta_i(t)\partial\theta_i(t^\prime)=0$:
\bd
0=\lim_{\bpsi\rightarrow\bnul}\frac{i}{N}\sum_{i}\xi_i^1
\frac{\int\!d\ba\ldots d\hbK\left[\frac{N\partial\Xi}{\partial
\theta_i(t)}\right]
e^{N[\Psi+\Phi+\Xi]+\order(\ldots)}}
{\int\!d\ba \ldots d\hbK~
e^{N[\Psi+\Phi+\Xi]+\order(\ldots)}}=\lim_{\bpsi\rightarrow\bnul}\bra\hh(t)\ket_{\star}
\ed
\bd
0
=-\lim_{\bpsi\rightarrow\bnul}\frac{1}{N}\sum_{i}
\frac{\int\!d\ba\ldots d\hbK\left[\frac{N\partial^2\Xi}{\partial
\theta_i(t)\partial\theta_i(t^\prime)}+ \frac{N\partial\Xi}{\partial
\theta_i(t)}\frac{N\partial\Xi}{\partial
\theta_i(t^\prime)} \right]
e^{N[\Psi+\Phi+\Xi]+\order(\ldots)}}
{\int\!d\ba \ldots d\hbK~
e^{N[\Psi+\Phi+\Xi]+\order(\ldots)}}=\lim_{\bpsi\rightarrow\bnul}\bra\hh(t)\hh(t^\prime)\ket_{\star}
\ed
In combination with (\ref{eq:hats_onetime},\ref{eq:hats_twotimes}),
the above five identities simplify our problem
considerably. The dummy fields $\psi_i(t)$ have served their
purpose and will now be put to zero, as a result we can now identify our
macroscopic observables {\em at the relevant saddle-point} as:
\be
a(t)=m(t)~~~~~~ k(t)=0~~~~~~
q(t,t^\prime)=C(t,t^\prime)~~~~~~Q(t,t^\prime)=0~~~~~~
K(t,t^\prime)=iG(t^\prime,t)
\label{eq:identify}
\ee
Finally we make a convenient choice for the external fields,
$\theta_i(t)=\xi_i^1\theta(t)$, with which the effective measure $\bra \ldots\ket_\star$
of (\ref{eq:measure_stage2}) simplifies to
\be
\bra f[\{\sigma\},\{h\},\{\hh\}]\ket_\star=
\frac{\int\!\{dh
d\hh\}\!\!\sum_{\sigma(0)\cdots\sigma(t_{m})}\!M[\{\sigma\},\{h\},\{\hh\}]~f[\{\sigma\},\{h\},\{\hh\}]}
{\int\!\{dh
d\hh\}\!\!\sum_{\sigma(0)\cdots\sigma(t_{m})}\!M[\{\sigma\},\{h\},\{\hh\}]}
\vspace*{-3mm}
\label{eq:measure_stage2}
\ee
with
\bd
M[\{\sigma\},\{h\},\{\hh\}]=
\pi_0(\sigma(0))~
e^{\sum_t\{i\hh(t)[h(t)-\theta(t)]+ \beta \sigma(t+1) h(t)-\log 2\cosh[\beta
h(t)]\}-i\sum_t[
\ha(t)\sigma(t)+\hk(t)\hh(t)]}
\vspace*{-1mm}
\ed
\bd
\times e^{-i\sum_{t,t^\prime}[
\hq(t,t^\prime)\sigma(t)\sigma(t^\prime)+\hQ(t,t^\prime)\hh(t)\hh(t^\prime)
+\hK(t,t^\prime)\hh(t)\sigma(t^\prime)]}
\ed
In summary: our saddle-point equations are given by
(\ref{eq:nohats_onetime},\ref{eq:nohats_twotimes},\ref{eq:hats_onetime},\ref{eq:hats_twotimes}),
and the physical meaning of the macroscopic quantities is given by
(\ref{eq:identify}) (apparently many of them must be
zero). Our final task is finding
(\ref{eq:dyn_meanfield}), i.e. calculating the leading order of
\be
{\cal F}[\{\bsigma\},\{\hbh\}]=\frac{1}{N}\log\overline{ \left[e^{-i\sum_{it}\hh_i(t)\sum_j
J_{ij}\sigma_j(t)}\right]}
\label{eq:modeldependent}
\ee
which is where the properties of the synapses
(\ref{eq:diluted_synapses}) come in.

\subsection{Parallel Dynamics Hopfield Model Near Saturation}

{\em The Disorder Average.}
The fully connected Hopfield \cite{Hopfield} network (here with parallel dynamics)
is obtained upon choosing $c=N$ in the recipe
(\ref{eq:diluted_synapses}), i.e. $c_{ij}=1\minus \delta_{ij}$ and $p=\alpha N$.
The disorder average thus involves only the patterns with $\mu>1$.
In view of our objective to write (\ref{eq:modeldependent}) in the
form (\ref{eq:dyn_meanfield}), we will substitute the observables
defined in (\ref{eq:onetime},\ref{eq:qQ},\ref{eq:K}) whenever
possible. Now (\ref{eq:modeldependent}) gives
\bd
{\cal F}[\ldots]
=\frac{1}{N}\log\overline{ \left[e^{-iN^{-1}\sum_{t}
\sum_{\mu}\sum_{i\neq
j}\xi_i^\mu\xi_j^\mu\hh_i(t)\sigma_j(t)}\right]}
\vspace*{-1mm}
~~~~~~~~~~~~~~~~~~~~~~~~~~~~~~~~~~~~~~~~~~~~~~~~~~~~~~~~~~~~~
\ed
\be
=i\alpha\sum_{t}K(t,t;\{\bsigma,\hbh\})
-i\sum_{t}a(t)k(t)
+\alpha\log\overline{\left[
e^{-i\sum_{t}[\sum_{i}\xi_i\hh_i(t)/\sqrt{N}][\sum_i\xi_i\sigma_i(t)/\sqrt{N}]}
\right]}+\order(N^{-1})
\label{eq:dishop_halfway}
\ee
We concentrate on the last term:
\bd
\overline{\left[
e^{-i\sum_{t}[\sum_{i}\xi_i\hh_i(t)/\sqrt{N}][\sum_i\xi_i\sigma_i(t)/\sqrt{N}]}
\right]}
=\int\!d\bx d\by
~e^{-i\bx\cdot\by}~\overline{\prod_{t}\left\{
\delta[x(t)\minus\frac{\sum_i\xi_i\sigma_i(t)}{\sqrt{N}}]~
\delta[y(t)\minus\frac{\sum_i\xi_i\hh_i(t)}{\sqrt{N}}]\right\}}
\ed
\bd
=\int\!\frac{d\bx d\by d\hbx d\hby}{(2\pi)^{2(t_m+1)}}~
e^{i[\hbx\cdot\bx+\hby\cdot\by-\bx\cdot\by]}~
\overline{\left[e^{-\frac{i}{\sqrt{N}}\sum_i \xi_i\sum_t
[\hx(t)\sigma_i(t)+\hy(t)\hh_i(t)]}\right]}
\ed
\bd
=\int\!\frac{d\bx d\by d\hbx d\hby}{(2\pi)^{2(t_m+1)}}~
e^{i[\hbx\cdot\bx+\hby\cdot\by-\bx\cdot\by]+\sum_i\log\cos\left[\frac{1}{\sqrt{N}}\sum_t
[\hx(t)\sigma_i(t)+\hy(t)\hh_i(t)]\right]}
\ed
\bd
=\int\!\frac{d\bx d\by d\hbx d\hby}{(2\pi)^{2(t_m+1)}}~
e^{i[\hbx\cdot\bx+\hby\cdot\by-\bx\cdot\by]
-\frac{1}{2N}\sum_i\{\sum_t[\hx(t)\sigma_i(t)+\hy(t)\hh_i(t)]\}^2+\order(N^{-1})}
\ed
\bd
=\int\!\frac{d\bx d\by d\hbx d\hby}{(2\pi)^{2(t_m+1)}}~
e^{i[\hbx\cdot\bx+\hby\cdot\by-\bx\cdot\by]
-\frac{1}{2}\sum_{t,t^\prime}\left[\hx(t)\hx(t^\prime)q(t,t^\prime)
+2\hx(t)\hy(t^\prime)K(t^\prime,t)
+\hy(t)\hy(t^\prime)Q(t,t^\prime)\right]
+\order(N^{-1})}
\ed
Together with (\ref{eq:dishop_halfway}) we have now shown that the
disorder average (\ref{eq:modeldependent}) is indeed, in leading
order in $N$, of the form (\ref{eq:dyn_meanfield}) (as claimed), with
\bd
\Phi[\ba,\bk,\bq,\bQ,\bK]=
i\alpha\sum_{t}K(t,t)-i\ba\cdot\bk
+\alpha\log \int\!\frac{d\bx d\by d\hbx d\hby}{(2\pi)^{2(t_m+1)}}~
e^{i[\hbx\cdot\bx+\hby\cdot\by-\bx\cdot\by]
-\frac{1}{2}[\hbx\cdot\bq\hbx+2\hby\cdot\bK\hbx
+\hby\cdot\bQ\hby]}
\ed
\be
=
i\alpha\sum_{t}K(t,t)-i\ba\cdot\bk
+\alpha\log \int\!\frac{d\bu d\bv}{(2\pi)^{t_m+1}}~
e^{-\frac{1}{2}[\bu\cdot\bq\bu+2\bv\cdot\bK\bu-2i\bu\cdot\bv
+\bv\cdot\bQ\bv]}
\label{eq:Phi_fullyconnected}
\ee
(which, of course, can be simplified further).
\vsp

\noindent{\em Simplification of the Saddle-Point
Equations.}
We are now in a position to work out equations
(\ref{eq:nohats_onetime},\ref{eq:nohats_twotimes}). For
the single-time observables this gives
$\ha(t)=k(t)$ and $\hk(t)=a(t)$, and for the two-time ones:
\vspace*{-2mm}
\bd
\hq(t,t^\prime)=-\frac{1}{2}\alpha i~\frac{\int\!d\bu
d\bv~u(t)u(t^\prime)
e^{-\frac{1}{2}[\bu\cdot\bq\bu+2\bv\cdot\bK\bu-2i\bu\cdot\bv
+\bv\cdot\bQ\bv]}}{\int\!d\bu d\bv~
e^{-\frac{1}{2}[\bu\cdot\bq\bu+2\bv\cdot\bK\bu-2i\bu\cdot\bv
+\bv\cdot\bQ\bv]}}
\ed
\bd
\hQ(t,t^\prime)=
-\frac{1}{2}\alpha i~\frac{\int\!d\bu d\bv~v(t)v(t^\prime)
e^{-\frac{1}{2}[\bu\cdot\bq\bu+2\bv\cdot\bK\bu-2i\bu\cdot\bv
+\bv\cdot\bQ\bv]}}{\int\!d\bu d\bv~
e^{-\frac{1}{2}[\bu\cdot\bq\bu+2\bv\cdot\bK\bu-2i\bu\cdot\bv
+\bv\cdot\bQ\bv]}}
\ed
\bd
\hK(t,t^\prime)=-\alpha i~\frac{\int\!d\bu
d\bv~v(t)u(t^\prime)
e^{-\frac{1}{2}[\bu\cdot\bq\bu+2\bv\cdot\bK\bu-2i\bu\cdot\bv
+\bv\cdot\bQ\bv]}}{\int\!d\bu d\bv~
e^{-\frac{1}{2}[\bu\cdot\bq\bu+2\bv\cdot\bK\bu-2i\bu\cdot\bv
+\bv\cdot\bQ\bv]}}-\alpha \delta_{t,t^\prime}
\ed
At the physical saddle-point we can use (\ref{eq:identify}) to express all non-zero objects in terms of
the observables  $m(t)$, $C(t,t^\prime)$ and
$G(t,t^\prime)$, with a clear physical meaning. Thus we find $\ha(t)=0$, $\hk(t)=m(t)$, and

\clearpage
\noindent
\be
\hq(t,t^\prime)=-\frac{1}{2}\alpha i~\frac{\int\!d\bu
d\bv~u(t)u(t^\prime)
e^{-\frac{1}{2}[\bu\cdot\bC\bu-2i\bu\cdot[\one-\bG]\bv]}}{\int\!d\bu d\bv~
e^{-\frac{1}{2}[\bu\cdot\bC\bu-2i\bu\cdot[\one-\bG]\bv]}}=0
\label{eq:find_qhat}
\ee
\be
\hQ(t,t^\prime)=
-\frac{1}{2}\alpha i~
\frac{\int\!d\bu d\bv~v(t)v(t^\prime) e^{-\frac{1}{2}[\bu\cdot\bC\bu-2i\bu\cdot[\one-\bG]\bv]}}{\int\!d\bu d\bv~
e^{-\frac{1}{2}[\bu\cdot\bC\bu-2i\bu\cdot[\one-\bG]\bv]}}
=-\frac{1}{2}\alpha i\left[(\one\minus\bG)^{-1}\bC(\one\minus\bG^\dag)^{-1}\right](t,t^\prime)
\label{eq:find_Qhat}
\ee
\be
\hK(t,t^\prime)+\alpha\delta_{t,t^\prime}=-\alpha i~\frac{\int\!d\bu
d\bv~v(t)u(t^\prime)
e^{-\frac{1}{2}[\bu\cdot\bC\bu-2i\bu\cdot[\one-\bG]\bv]}}{\int\!d\bu d\bv~
e^{-\frac{1}{2}[\bu\cdot\bC\bu-2i\bu\cdot[\one-\bG]\bv]}}
=\alpha
(\one\minus\bG)^{-1}(t,t^\prime)
\label{eq:find_Khatold}
\ee
(with $G^\dag(t,t^\prime)=G(t^\prime,t)$, and using standard manipulations of Gaussian integrals).
Note that we can use the identity $(\one\minus \bG)^{-1}\minus \one=\sum_{\ell\geq
0}\bG^\ell\minus \one=\sum_{\ell>0}\bG^{\ell}=\bG(\one\minus\bG)^{-1}$
to compactify (\ref{eq:find_Khatold}) to
\be
\hK(t,t^\prime)=\alpha
[\bG(\one\minus\bG)^{-1}](t,t^\prime)
\label{eq:find_Khat}
\ee
We have now expressed all our  objects in
terms of the disorder-averaged recall overlap $\bm=\{m(t)\}$ and the disorder-averaged single-site
correlation- and response functions $\bC=\{C(t,t^\prime)\}$ and
$\bG=\{G(t,t^\prime)\}$.
We can next simplify the effective measure
(\ref{eq:measure_stage2}), which plays a crucial role in the
remaining saddle-point equations. Inserting
$\ha(t)=\hq(t,t^\prime)=0$ and $\hk(t)=m(t)$ into
(\ref{eq:measure_stage2}), first of all, gives us
\bd
M[\{\sigma\},\{h\},\{\hh\}]=\pi_0(\sigma(0))~\times
~~~~~~~~~~~~~~~~~~~~~~~~~~~~~~~~~~~~~~~~~~~~~~~~~~~~~~~~~~~~~~~~~~~~~~~~~~~~~~~~~~~~~~~~~~~~~~~~~~~~~~~~
\ed
\be
e^{\sum_t\{i\hh(t)[h(t)-m(t)-\theta(t)-\sum_{t^\prime}\hK(t,t^\prime)\sigma(t^\prime)]
+ \beta \sigma(t+1) h(t)-\log 2\cosh[\beta
h(t)]\}-i\sum_{t,t^\prime}\hQ(t,t^\prime)\hh(t)\hh(t^\prime)}
\label{eq:measureM_stage3}
\ee
Secondly, causality ensures that $G(t,t^\prime)=0$ for $t\leq
t^\prime$, from which, in combination with (\ref{eq:find_Khat}),
it follows that the same must be true for the kernel $\hK(t,t^\prime)$, since
\bd
\hK(t,t^\prime)=\alpha
[\bG(\one\minus\bG)^{-1}](t,t^\prime)
=\alpha\left\{\bG+\bG^2+\bG^3+\ldots\right\}(t,t^\prime)
\ed
This, in turn, guarantees that the function $M[\ldots]$
in (\ref{eq:measureM_stage3})  is already normalised:
\bd
\int\!\{dhd\hh\}\!
\sum_{\sigma(0)\cdots\sigma(t_{m})}\!
M[\{\sigma\},\{h\},\{\hh\}]=1
\ed
One can prove this iteratively. After summation over $\sigma(t_m)$
(which due to causality cannot occur in the term with the kernel
$\hK(t,t^\prime)$) one is left with just a single occurrence of
the field $h(t_m)$ in the exponent, integration over which reduces
to $\delta[\hh(t_m)]$, which then eliminates the conjugate field
$\hh(t_m)$. This cycle of operations is next applied to the
variables at time $t_m\minus 1$, etc. The effective measure
(\ref{eq:measure_stage2}) can now be written simply as
\bd
\bra f[\{\sigma\},\{h\},\{\hh\}]\ket_\star=
\sum_{\sigma(0)\cdots\sigma(t_{m})}\int\!\{dhd\hh\}~M[\{\sigma\},\{h\},\{\hh\}]~
f[\{\sigma\},\{h\},\{\hh\}]
\ed
with $M[\ldots]$ as given in (\ref{eq:measureM_stage3}).
The remaining saddle-point equations to
be solved, which can be slightly simplified by using the identity $\bra\sigma(t)\hh(t^\prime)\ket_\star=
i\partial\bra\sigma(t)\ket_\star/\partial\theta(t^\prime)$, are
\be
m(t)=\bra \sigma(t)\ket_\star~~~~~~~~~~~~
C(t,t^\prime)=\bra\sigma(t)\sigma(t^\prime)\ket_\star
~~~~~~~~~~~~
G(t,t^\prime)=\partial\bra\sigma(t)\ket_\star/\partial\theta(t^\prime)
\label{eq:finalsaddle}
\ee
\vsp

\noindent
{\em Extracting the Physics from the Saddle-Point Equations.}
At this stage we observe in (\ref{eq:finalsaddle}) that we only need to insert functions of spin states into the
effective measure $\bra\ldots\ket_\star$ (rather than fields or conjugate fields),
so the effective measure can again be simplified. Upon inserting (\ref{eq:find_Qhat},\ref{eq:find_Khat})
into the function (\ref{eq:measureM_stage3}) we
obtain
$\bra f[\{\sigma\}]\ket_\star=\sum_{\sigma(0)\cdots\sigma(t_{m})}{\rm Prob}[\{\sigma\}]~f[\{\sigma\}]$,
with
\be
{\rm Prob}[\{\sigma\}]
=\pi_0(\sigma(0))\int\!\{d\phi\}~P[\{\phi\}]~
\prod_{t}\left[\frac{1}{2}[1\plus\sigma(t\plus 1)
\tanh[\beta h(t|\{\sigma\},\{\phi\})]\right]
\label{eq:single_spin_fully}
\ee
\clearpage\noindent
in which $\pi_0(\sigma(0))=\frac{1}{2}[1\plus\sigma(0)m_0]$, and
\be
h(t|\{\sigma\},\{\phi\})= m(t)+ \theta(t)
+\alpha\sum_{t^\prime<t}[\bG(\one\minus \bG)^{-1}](t,t^\prime)\sigma(t^\prime)+ \alpha^{\frac{1}{2}}\phi(t)
\label{eq:effective_field_fully}
\ee
\be
P[\{\phi\}]=
\frac{e^{-\frac{1}{2}\sum_{t,t^\prime}\phi(t)\left[(\one\minus\bG^\dag)\bC^{-1}(\one\minus\bG)\right](t,t^\prime)
\phi(t^\prime)}}
{(2\pi)^{(t_m+1)/2}{\rm
det}^{-\frac{1}{2}}\left[(\one\minus\bG^\dag)\bC^{-1}(\one\minus\bG)\right]}
\label{eq:Gaussianfields_fully}
\ee
(note: to predict neuron states up until time $t_m$ we only need
the fields up until time $t_m\minus 1$). We recognise
(\ref{eq:single_spin_fully}) as describing an effective single
neuron, with the usual dynamics ${\rm Prob}[\sigma(t\plus 1)=\pm 1]=\frac{1}{2}[1\pm \tanh[\beta
h(t)]]$, but with the fields
(\ref{eq:effective_field_fully}). This result is indeed of the
form (\ref{eq:effective_field}), with a retarded self-interaction
kernel $R(t,t^\prime)$ and covariance matrix $\bra\phi(t)\phi(t^\prime)\ket$ of the Gaussian
 $\phi(t)$ given by
\be
R(t,t^\prime)=[\bG(\one\minus \bG)^{-1}](t,t^\prime)
~~~~~~~~~~~~
\bra\phi(t)\phi(t^\prime)\ket=
[(\one\minus\bG)^{-1}\bC(\one\minus\bG^\dag)^{-1}](t,t^\prime)
\label{eq:kernels_fully}
\ee
For $\alpha\!\to\! 0$ we loose all the complicated terms in the local fields, and
recover the type of simple expression we
found earlier for finite $p$: $m(t\plus 1)\!=\!\tanh[\beta (m(t)\plus
\theta(t))]$.

It can be shown \cite{CoolenSherr} (space limitations prevent a demonstration in this
paper) that the equilibrium solutions obtained via replica theory
in replica-symmetric ansatz \cite{FontanariKoberle} can be
recovered as those time-translation invariant solutions\footnote{i.e. $m(t)=m$, $C(t,t^\prime)=C(t\minus t^\prime)$
and $G(t,t^\prime)=G(t\minus t^\prime)$} of the above dynamic equations which
(i) obey the parallel dynamics fluctuation-dissipation theorem,
and (ii) obey $\lim_{\tau\to \infty}G(\tau)=0$. It can also be
shown that the AT \cite{AT} instability, where
replica symmetry ceases to hold, corresponds to a dynamical
instability in the present formalism, where so-called anomalous
response sets in: $\lim_{\tau\to \infty}G(\tau)\neq 0$.

Before we calculate the solution explicitly for the first few time-steps, we first
work out the relevant
averages using (\ref{eq:single_spin_fully}).
Note that always $C(t,t)=\bra\sigma^2(t)\ket_\star=1$
and $G(t,t^\prime)=R(t,t^\prime)=0$ for $t\leq t^\prime$.
As a result  the covariance matrix of the Gaussian fields can be
written as
\bd
\bra\phi(t)\phi(t^\prime)\ket=
[(\one\minus\bG)^{-1}\bC(\one\minus\bG^\dag)^{-1}](t,t^\prime)
=\sum_{s,s^\prime\geq 0}[\delta_{t,s}\plus
R(t,s)]C(s,s^\prime)[\delta_{s^\prime,t^\prime}\plus R(t^\prime,s^\prime)]
\vspace*{-2mm}
\ed
\be
=\sum_{s=0}^{t}\sum_{s^\prime= 0}^{t^\prime}[\delta_{t,s}\plus
R(t,s)]C(s,s^\prime)[\delta_{s^\prime,t^\prime}\plus
R(t^\prime,s^\prime)]
\label{eq:Gaussmoments}
\ee
Considering arbitrary positive integer powers of the response function immediately shows that
\be
(\bG^\ell)(t,t^\prime)=0~~~~~~ {\rm if}~~~~~~ t^\prime>t\minus\ell
\label{eq:Gpowers}
\vspace*{-2mm}
\ee
which, in
turn, gives
\vspace*{-2mm}
\be
R(t,t^\prime)=\sum_{\ell>0}(\bG^\ell)(t,t^\prime)=\sum_{\ell=1}^{t-t^\prime}(\bG^\ell)(t,t^\prime)
\label{eq:Retarded}
\ee
Similarly we obtain from  $(\one\minus\bG)^{-1}=\one\plus \bR$
that for $t^\prime\geq t$:
$(\one\minus\bG)^{-1}(t,t^\prime)=\delta_{t,t^\prime}$.
 To suppress notation we will simply put
$h(t|..)$ instead of $h(t|\{\sigma\},\{\phi\})$; this need not cause any ambiguity.
We notice that
summation over neuron variables $\sigma(s)$ and integration over Gaussian variables $\phi(s)$
with time arguments $s$ higher than than those
occurring in the function to be averaged can always be
carried out immediately, giving (for $t>0$ and $t^\prime<t$):
\be
m(t)=\!\!\sum_{\sigma(0)\ldots\sigma(t-1)}\!\!\pi_0(\sigma(0))\int\!\{d\phi\}P[\{\phi\}]~
\tanh[\beta h(t\minus 1|..)]
\prod_{s=0}^{t-2}
\frac{1}{2}\left[1\plus\sigma(s\plus 1)
\tanh[\beta h(s|..)]\right]
\label{eq:m_worked}
\ee
\bd
G(t,t^\prime)=\beta\left\{C(t,t^\prime\plus 1)-
\!\!\sum_{\sigma(0)\ldots\sigma(t-1)}\!\!\pi_0(\sigma(0))\int\!\{d\phi\}P[\{\phi\}]~
\tanh[\beta h(t\minus 1|..)] \tanh[\beta h(t^\prime|..)]
\vspace*{-2mm}
\right.
\ed
\be
\left.
~~~~~~~~~~~~~~~~~~~~~~~~~~~~~~~~~~~~~~~~~~~~~~~~~~~~~~~~~~~~~
\times\prod_{s=0}^{t-2}
\frac{1}{2}\left[1\plus\sigma(s\plus 1)
\tanh[\beta h(s|..)]\right]
\right\}
\label{eq:G_worked}
\ee
(which we obtain directly for $t^\prime=t\minus 1$, and which follows for times $t^\prime<t\minus 1$ upon using the identity
$\sigma[1\minus\tanh^2(x)]=[1\plus\sigma\tanh(x)][\sigma\minus\tanh(x)]$).
For the correlations we distinguish between $t^\prime=t\minus 1$
and $t^\prime<t\minus 1$:
\be
C(t,t\minus 1)=
\!\!\!\sum_{\sigma(0)\ldots\sigma(t-2)}\!\!\!\pi_0(\sigma(0))\int\!\{d\phi\}P[\{\phi\}]~
\tanh[\beta h(t\minus 1|..)]
\tanh[\beta h(t\minus 2|..)]
\prod_{s=0}^{t-3}
\frac{1}{2}\left[1\plus\sigma(s\plus 1)
\tanh[\beta h(s|..)]\right]
\label{eq:C_worked1}
\ee
whereas for $t^\prime<t-1$ we have
\be
C(t,t^\prime)=
\!\!\!\sum_{\sigma(0)\ldots\sigma(t-1)}\!\!\!\pi_0(\sigma(0))\int\!\{d\phi\}P[\{\phi\}]~
\tanh[\beta h(t\minus 1|..)]\sigma(t^\prime)
\prod_{s=0}^{t-2}
\frac{1}{2}\left[1\plus\sigma(s\plus 1)
\tanh[\beta h(s|..)]\right]
\label{eq:C_worked2}
\ee
Let us finally work out explicitly  the final macroscopic laws
(\ref{eq:m_worked},\ref{eq:G_worked},\ref{eq:C_worked1},\ref{eq:C_worked2}),
with (\ref{eq:effective_field_fully},\ref{eq:Gaussianfields_fully}),
 for the first few time-steps. For arbitrary times our equations will have
to be evaluated numerically; we will see below, however, that this
can be done in an iterative (i.e. easy) manner.
At $t=0$ we just have the two observables $m(0)=m_0$ and $C(0,0)=1$.
\vsp

\noindent{\em The First Few Time-Steps.}
The field at $t=0$ is $h(0|..)=m_0\plus
\theta(0)\plus\alpha^{\frac{1}{2}}\phi(0)$, since
the retarded self-interaction does not yet come into play.
The distribution of  $\phi(0)$ is fully
characterised by its variance, which (\ref{eq:Gaussmoments})
claims to be
\bd
\bra\phi^2(0)\ket
=C(0,0)=1
\ed
Therefore, with
$Dz=(2\pi)^{-\frac{1}{2}}e^{-\frac{1}{2}z^2}dz$, we immediately
find (\ref{eq:m_worked},\ref{eq:G_worked},\ref{eq:C_worked1},\ref{eq:C_worked2})
reducing to
\be
m(1)=\int\!Dz~
\tanh[\beta(m_0\plus \theta(0)\plus z\sqrt{\alpha})]
~~~~~~~~~~~~
C(1,0)=m_0 m(1)
\label{eq:m(1)C(1,0)}
\vspace*{-2mm}
\ee
\be
G(1,0)=\beta\left\{1-\int\!Dz~\tanh^2[\beta(m_0\plus\theta(0)\plus
z\sqrt{\alpha})]\right\}
\label{eq:G(1,0)}
\ee
For the self-interaction kernel this implies, using
(\ref{eq:Retarded}), that $R(1,0)=G(1,0)$.
We now move on to $t=2$. Here equations
(\ref{eq:m_worked},\ref{eq:G_worked},\ref{eq:C_worked1},\ref{eq:C_worked2})
give us
\bd
m(2)=\frac{1}{2}\sum_{\sigma(0)}\int\!d\phi(0)d\phi(1)P[\phi(0),\phi(1)]~
\tanh[\beta h(1|..)][1\plus\sigma(0)m_0]
\ed
\bd
C(2,1)=
\frac{1}{2}\sum_{\sigma(0)}\int\!d\phi(1)d\phi(0)P[\phi(0),\phi(1)]~
\tanh[\beta h(1|..)]\tanh[\beta h(0|..)][1\plus\sigma(0)m_0]
\ed
\bd
C(2,0)=
\frac{1}{2}\sum_{\sigma(0)\sigma(1)}\int\!\{d\phi\}P[\{\phi\}]~
\tanh[\beta h(1|..)]\sigma(0)
\frac{1}{2}\left[1\plus\sigma(1)
\tanh[\beta h(0|..)]\right][1\plus\sigma(0)m_0]
\ed
\bd
G(2,1)=\beta\left\{1-
\frac{1}{2}\sum_{\sigma(0)}\int\!d\phi(0)d\phi(1)P[\phi(0),\phi(1)]~
\tanh^2[\beta h(1|..)]
[1\plus\sigma(0)m_0]
\right\}
\ed
\bd
G(2,0)=\beta\left\{C(2,1)-
\frac{1}{2}\sum_{\sigma(0)}\int\!d\phi(0)d\phi(1)P[\phi(0),\phi(1)]~
\tanh[\beta h(1|..)]\tanh[\beta h(0|..)]
[1\plus\sigma(0)m_0]
\right\}=0
\ed
\clearpage
\noindent
We already know that $\bra\phi^2(0)\ket=1$; the remaining two moments
we need in order to determine $P[\phi(0),\phi(1)]$ follow again from
(\ref{eq:Gaussmoments}):
\bd
\bra\phi(1)\phi(0)\ket=\sum_{s=0}^{1}[\delta_{1,s}\plus \delta_{0,s}
R(1,0)]C(s,0)=C(1,0)+G(1,0)
\vspace*{-2mm}
\ed
\bd
\bra\phi^2(1)\ket
=\sum_{s=0}^{1}\sum_{s^\prime=1}^{1}[\delta_{1,s}\plus\delta_{0,s}
R(1,0)]C(s,s^\prime)[\delta_{s^\prime,1}\plus \delta_{s^\prime,0}R(1,0)]
=G^2(1,0)+2C(0,1) G(1,0)+1
\ed
We now know $P[\phi(0),\phi(1)]$ and can work out all macroscopic objects with $t=2$
explicitly, if we wish. I will not do this here in full, but only point at the
emerging pattern of all calculations at a given time $t$ depending
only on macroscopic quantities that have been calculated at times
$t^\prime<t$, which allows for iterative solution.
Let us just work out $m(2)$ explicitly, in order to compare the first two
recall overlaps $m(1)$ and $m(2)$ with the values found in simulations and
in approximate theories.
We note that calculating $m(2)$ only requires the field $\phi(1)$,
for which we found $\bra\phi^2(1)\ket=G^2(1,0)+2C(0,1) G(1,0)+1$:
\bd
m(2)=\frac{1}{2}\sum_{\sigma(0)}\int\!d\phi(1)P[\phi(1)]~
\tanh[\beta(m(1)\plus\theta(1)\plus\alpha G(1,0)\sigma(0)\plus \alpha^{\frac{1}{2}}\phi(1))]
[1\plus\sigma(0)m_0]
~~~~~~~~~~
\vspace*{-2mm}
\ed
\bd
~~~~~~~~~~
=\frac{1}{2}[1\plus m_0]\int\!Dz~
\tanh[\beta(m(1)\plus \theta(1)\plus \alpha G(1,0)\plus z\sqrt{\alpha[G^2(1,0)\plus 2m_0 m(1) G(1,0)\plus
1]})]
\vspace*{-2mm}
\ed
\bd
~~~~~~~~~~~~~~~~~~~~
+\frac{1}{2}[1\minus m_0]\int\!Dz~
\tanh[\beta(m(1)\plus \theta(1)\minus \alpha G(1,0)\plus z\sqrt{\alpha[G^2(1,0)\plus 2m_0 m(1) G(1,0)\plus
1]})]
\ed
\vsp

\noindent
{\em Exact Results Versus Simulations and Gaussian
Approximations.}
I close this section on the fully connected networks with a
comparison of some of the approximate theories, the (exact) generating
functional formalism, and numerical simulations, for the case $\theta(t)=0$ (no external stimuli at any time).
The evolution of the recall overlap in the
first two
time-steps has been described as follows:
\bd
\begin{array}{llll}
{\sl Naive~Gaussian~Approximation:}
&&  m(1)=& \int\!Dz ~\tanh[\beta(m(0)+z\sqrt{\alpha})]\\[2mm]
&&  m(2)=& \int\!Dz ~\tanh[\beta(m(1)+z\sqrt{\alpha})]
\\[5mm]
{\sl Amari\!-\!Maginu~Theory:}
&&  m(1)=& \int\!Dz ~\tanh[\beta(m(0)+z\sqrt{\alpha})]\\[2mm]
&&  m(2)=& \int\!Dz ~\tanh[\beta(m(1)+z\Sigma\sqrt{\alpha})]\\[2mm]
&&  \Sigma^2=&1 + 2 m(0)m(1)G+ G^2\\
&&  G=&\beta\left[1-\int\!Dz~\tanh^2[\beta(m(0)+z\sqrt{\alpha})]\right]
\\[5mm]
{\sl Exact~Solution:}
&&  m(1)=&\int\!Dz ~\tanh[\beta(m(0)+z\sqrt{\alpha})]\\[2mm]
&&  m(2)=& \frac{1}{2}[1\plus m_0]\int\!Dz~
    \tanh[\beta(m(1)+ \alpha G+ z\Sigma\sqrt{\alpha})]\\[1mm]
&&  & ~+\frac{1}{2}[1\minus m_0]\int\!Dz~
    \tanh[\beta(m(1)- \alpha G+ z\Sigma\sqrt{\alpha})]\\[2mm]
&&  \Sigma^2=&1+ 2m(0) m(1)G+G^2\\
&&  G=&\beta\left[1-\int\!Dz~\tanh^2[\beta(m(0)+z\sqrt{\alpha})]\right]
\end{array}
\ed
\begin{figure}[t]
\begin{center}\vspace*{-10mm}
\epsfxsize=57mm\epsfbox{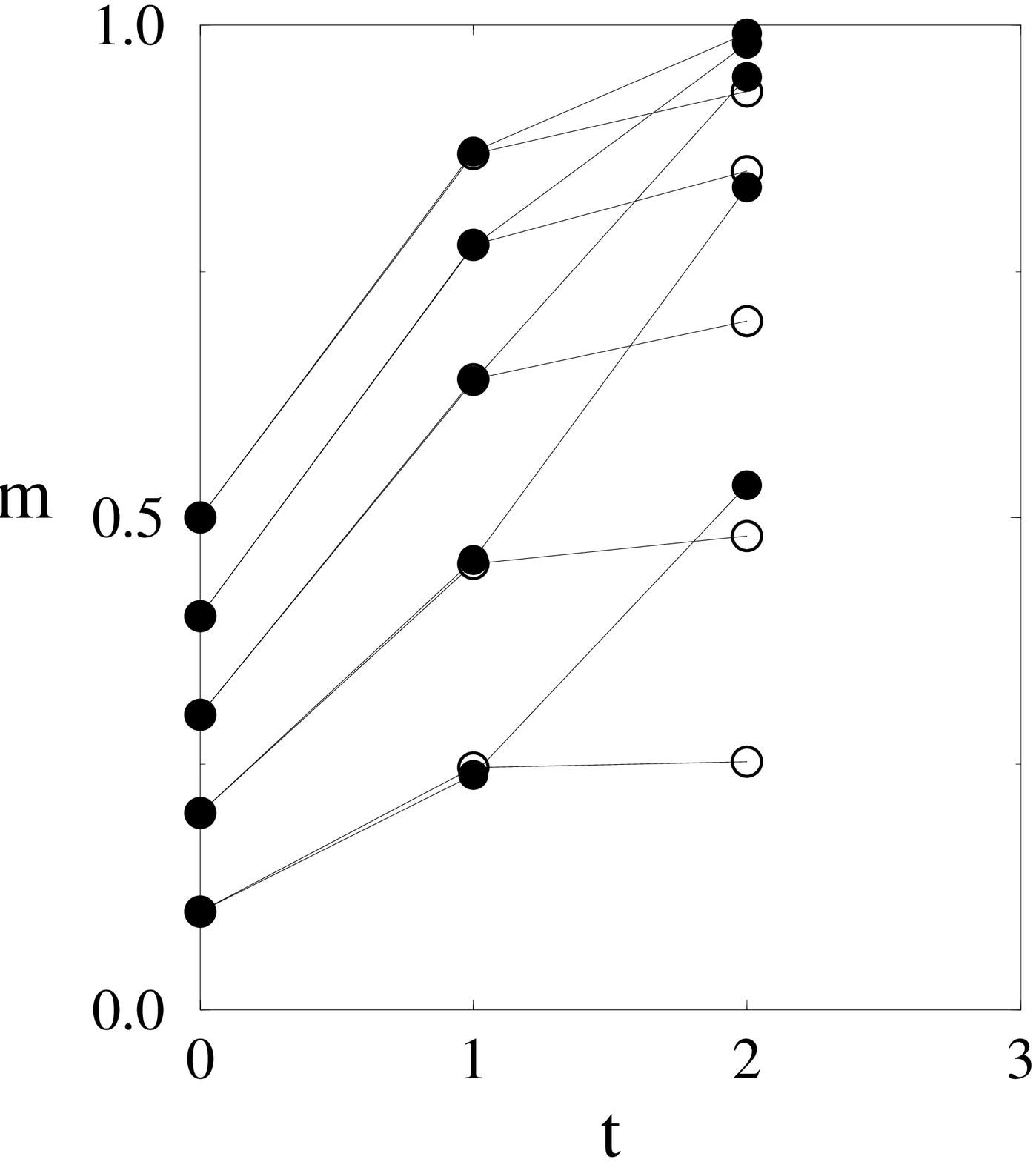}\hspace*{-2mm}
\epsfxsize=57mm\epsfbox{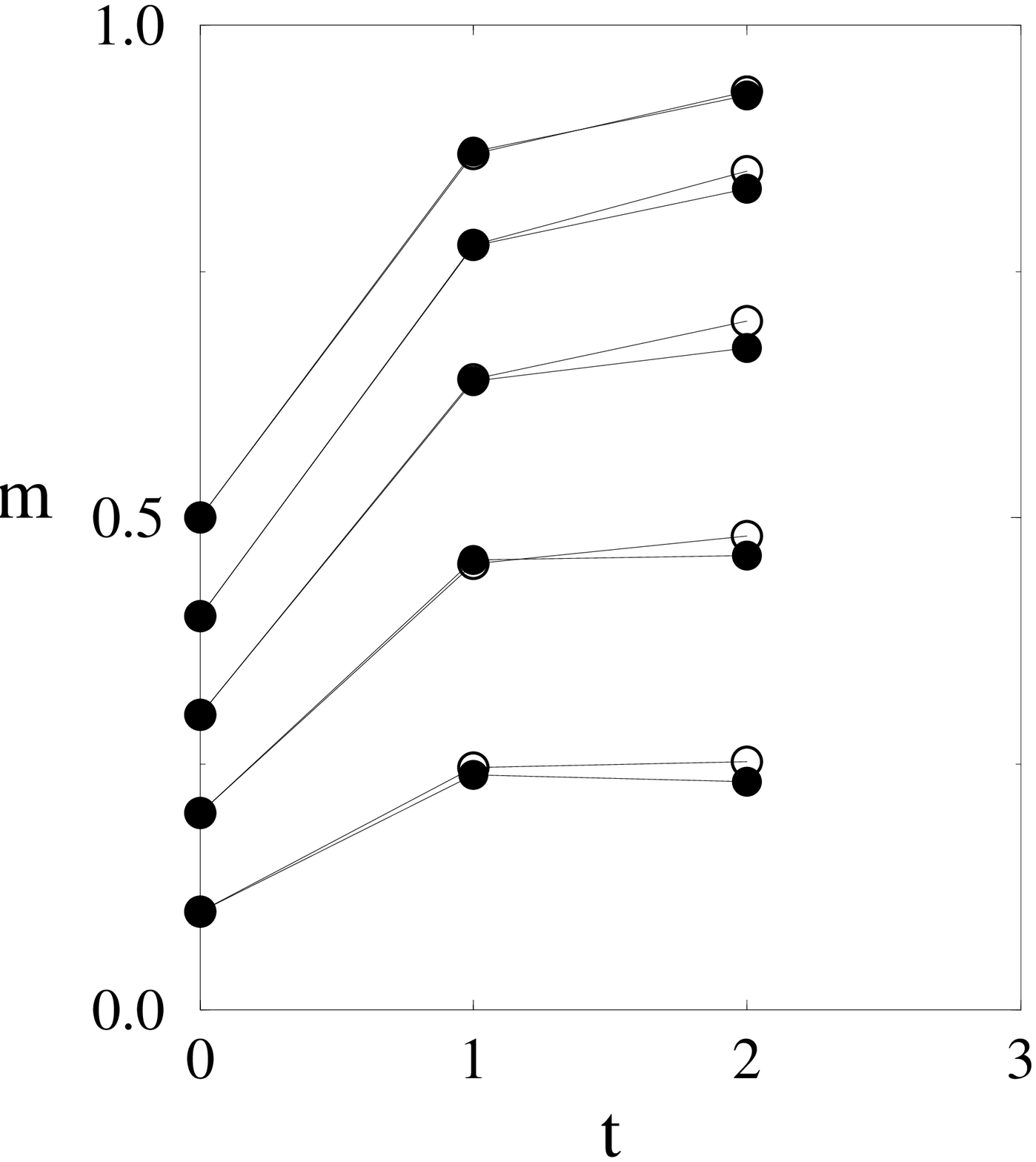}\hspace*{-2mm}
\epsfxsize=57mm\epsfbox{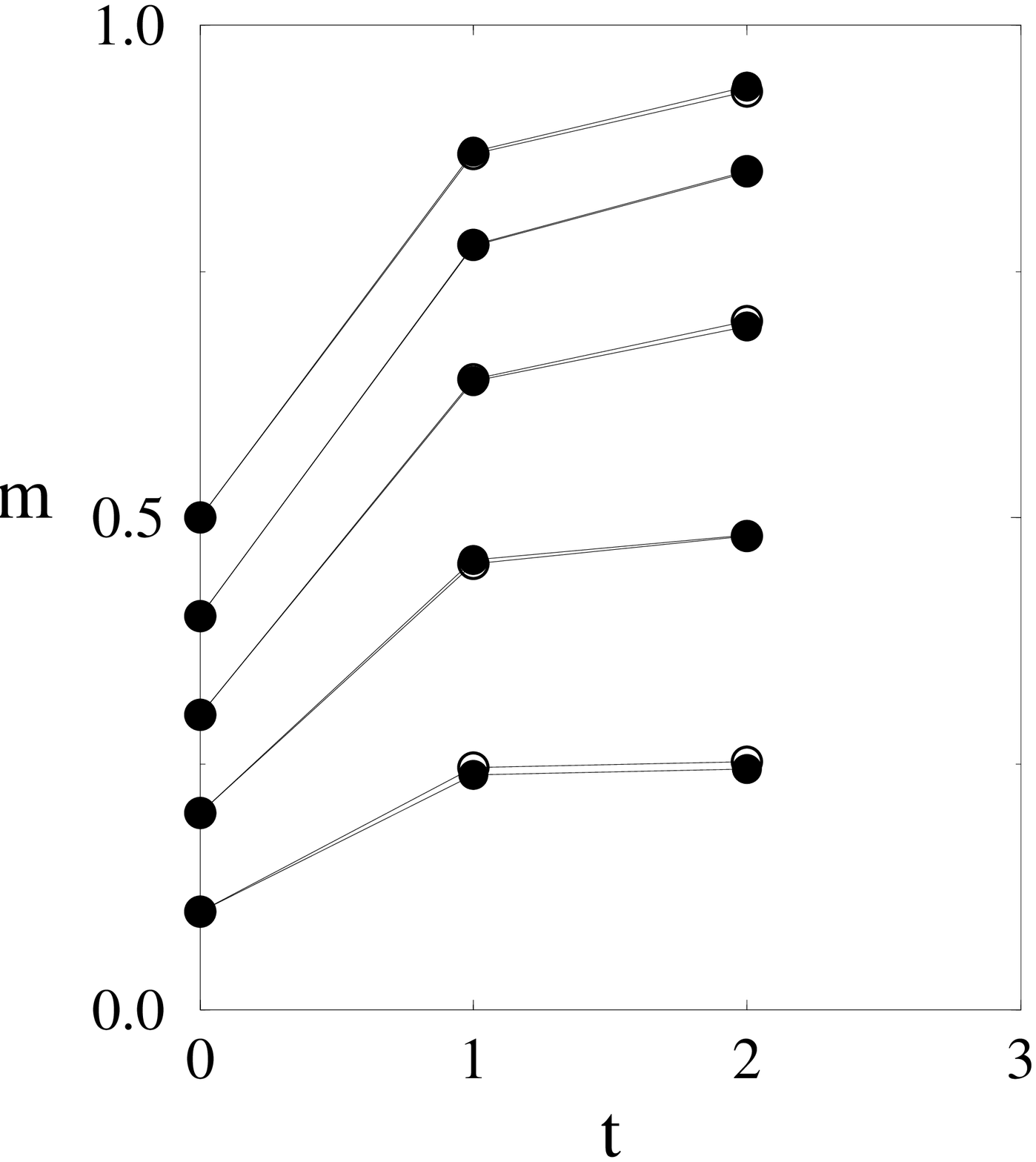}
\end{center}
\vspace*{-9mm}
\caption{The first few time-steps in the evolution of the overlap $m(\bsigma)=N^{-1}\sum_i\sigma_i\xi_i^1$
in a parallel dynamics Hopfield model
with $\alpha\!=\!T\!=\!0.1$ and random patterns, following initial states correlated
with pattern one only.
Left: simulations (o) versus naive Gaussian approximation
($\bullet$). Middle: simulations (o) versus advanced
Gaussian approximation (Amari-Maginu theory, $\bullet$). Right: simulations (o) versus (exact) generating functional
theory ($\bullet$).
All simulations were done with $N\!=\!30,\!000$.}
\label{fig:comparetheories}
\end{figure}
We can now appreciate why the more advanced Gaussian approximation
(Amari-Maginu theory, \cite{AmariMaginu}) works well when the system state
is close to the target attractor.
This theory gets the moments of the Gaussian
part of the interference noise
distribution at $t=1$ exactly right, but not the discrete part, whereas close to the attractor
both the response function $G(1,0)$ and one of the two pre-factors $\frac{1}{2}[1\pm m_0]$ in the exact
expression for $m(2)$ will be very small, and the latter will
therefore indeed approach a Gaussian shape. One can also see why the
non-Gaussian approximation of \cite{HenkelOpper} made sense: in
the calculation of $m(2)$ the interference noise distribution can indeed
be written as the sum of two Gaussian ones (although for $t>2$
this will cease to be true).
Numerical evaluation of these expressions result in explicit
predictions which can be tested against numerical simulations.
This is done in figure \ref{fig:comparetheories}, which
confirms the picture sketched above, and hints that the
performance of the Gaussian approximations is indeed worse for
those initial conditions which fail to trigger pattern recall.

\subsection{Extremely Diluted Attractor Networks Near Saturation}

Extremely diluted attractor networks
are obtained upon choosing $\lim_{N\to\infty}c/N=0$ (while still $c\to\infty$)
in definition
(\ref{eq:diluted_synapses}) of the Hebbian-type synapses.
The disorder average now involves both the patterns with $\mu>1$
and the realisation of the `wiring' variables
$c_{ij}\in\{0,1\}$. Again, in working out the key function (\ref{eq:modeldependent})
we will show that for $N\to\infty$ the outcome can be written in terms of the macroscopic
quantities  (\ref{eq:onetime},\ref{eq:qQ},\ref{eq:K}). We carry
out the average over the spatial structure variables $\{c_{ij}\}$ first:
\bd
{\cal F}[\ldots]
=\frac{1}{N}\log
\overline{\left[e^{-\frac{i}{c}\sum_{i\neq j}c_{ij}\sum_{\mu}\xi_i^\mu\xi_j^\mu
\sum_t \hh_i(t) \sigma_j(t)}\right]}
=\frac{1}{N}\log
\overline{\prod_{i<j} e^{-\frac{i}{c}\sum_{\mu}\xi_i^\mu\xi_j^\mu[
c_{ij}\sum_t \hh_i(t) \sigma_j(t)+c_{ji}\sum_t \hh_j(t)
\sigma_i(t)]}}
\ed
At this stage we have to distinguish between symmetric and asymmetric
dilution.
\vsp

\noindent{\em The Disorder Average.}
First we deal with the case of symmetric dilution: $c_{ij}=c_{ji}$ for all $i\neq j$.
The average over the $c_{ij}$, with the distribution (\ref{eq:symm_dil}), is trivial:
\bd
\overline{\prod_{i<j} e^{-\frac{i}{c}c_{ij}\sum_{\mu}\xi_i^\mu\xi_j^\mu
\sum_t [\hh_i(t) \sigma_j(t)+\hh_j(t)\sigma_i(t)]}}
=
\overline{\prod_{i<j} \left\{1\plus \frac{c}{N}[
e^{-\frac{i}{c}\sum_{\mu}\xi_i^\mu\xi_j^\mu
\sum_t [\hh_i(t)
\sigma_j(t)+\hh_j(t)\sigma_i(t)]}\minus 1]\right\}
}
\ed
\bd
=
\overline{\prod_{i<j} \left\{1\minus \frac{c}{N}\left[
\frac{i}{c}\sum_{\mu}\!\xi_i^\mu\!\xi_j^\mu\!
\sum_t [\hh_i(t)\sigma_j(t)\plus \hh_j(t)\sigma_i(t)]
\plus \frac{1}{2 c^2 }[\sum_{\mu}\!\xi_i^\mu\!\xi_j^\mu\!
\sum_t [\hh_i(t)\sigma_j(t)\plus \hh_j(t)\sigma_i(t)]]^2
\plus \order(c^{-\frac{3}{2}})
\right]\right\}
}
\ed
\bd
=\overline{\prod_{i<j} e^{-\frac{i}{N}\sum_{\mu}\!\xi_i^\mu\!\xi_j^\mu\!
\sum_t [\hh_i(t)\sigma_j(t)\plus \hh_j(t)\sigma_i(t)]
- \frac{1}{2 c N }[\sum_{\mu}\!\xi_i^\mu\!\xi_j^\mu\!
\sum_t [\hh_i(t)\sigma_j(t)\plus \hh_j(t)\sigma_i(t)]]^2
+ \order(\frac{1}{N\sqrt{c}})+ \order(\frac{c}{N^2})}}
\ed
We separate in the exponent the terms where $\mu=\nu$ in the quadratic term
(being of the form $\sum_{\mu\nu}\ldots$), and the
terms with $\mu=1$. Note: $p=\alpha c$. We also use the definitions (\ref{eq:onetime},\ref{eq:qQ},\ref{eq:K})
wherever we can:
\bd
{\cal F}[\ldots]=
-i\sum_t a(t)k(t)
- \frac{1}{2}\alpha\sum_{st}[q(s,t)Q(s,t)\plus K(s,t)K(t,s)]
+ \order(c^{-\frac{1}{2}})
+\order(c/N)+
~~~~~~~~~~~~~~~~~~~~~~~~~~~~~~
\vspace*{-2mm}
\ed
\bd
\frac{1}{N}\log\left\{
\overline{
e^{-\frac{i}{N}\sum_{\mu>1}\sum_t[\sum_{i}\xi_i^\mu
\hh_i(t)][\sum_j\xi_j^\mu\sigma_j(t)]
- \frac{1}{4 c N }\sum_{i\neq j}\sum_{\mu\neq \nu}\sum_{st}
\xi_i^\mu\!\xi_j^\mu \xi_i^\nu\!\xi_j^\nu
[\hh_i(s)\sigma_j(s)\plus \hh_j(s)\sigma_i(s)][\hh_i(t)\sigma_j(t)\plus \hh_j(t)\sigma_i(t)]
}}
\right\}
\ed
Our `condensed ansatz' implies
that for $\mu>1$:
$N^{-\frac{1}{2}}\sum_i\xi_i^\mu\sigma_i(t)=\order(1)$ and
$N^{-\frac{1}{2}}\sum_i\xi_i^\mu\hh_i(t)=\order(1)$. Thus the
first term in the exponent containing the disorder is $\order(c)$,
contributing $\order(c/N)$ to ${\cal F}[\ldots]$. We therefore retain only
the second term in the exponent. However, the same argument
applies to the second term. There all contributions can be seen as
uncorrelated in leading order, so that $\sum_{i\neq j}\sum_{\mu\neq
\nu}\ldots=\order(Np)$, giving a non-leading $\order(N^{-1})$ cumulative contribution
to ${\cal F}[\ldots]$.
Thus, provided $\lim_{N\to\infty}c^{-1}=\lim_{N\to\infty}c/N=0$
(which we assumed), we have shown that the
disorder average (\ref{eq:modeldependent}) is again, in leading
order in $N$, of the form (\ref{eq:dyn_meanfield}) (as claimed), with
\be
{\sl Symmetric:}~~~~~~~~
\Phi[\ba,\bk,\bq,\bQ,\bK]=
-i\ba\cdot\bk
- \frac{1}{2}\alpha\sum_{st}[q(s,t)Q(s,t)\plus K(s,t)K(t,s)]
\label{eq:Phi_symmdiluted}
\ee
Next we deal with the asymmetric case (\ref{eq:asymm_dil}), where $c_{ij}$ and $c_{ji}$ are
independent. Again the average over the $c_{ij}$ is trivial; here
it gives
\bd
\overline{\prod_{i<j}\left\{ e^{-\frac{i}{c}c_{ij}\sum_{\mu}\xi_i^\mu\xi_j^\mu
\sum_t \hh_i(t) \sigma_j(t)}
e^{-\frac{i}{c}c_{ji}\sum_{\mu}\xi_i^\mu\xi_j^\mu
\sum_t \hh_j(t)\sigma_i(t)}\right\}}
~~~~~~~~~~~~~~~~~~~~~~~~~~~~~~~~~~~~~~~~~~~~~~~~~~~~~~~~~~~~~~~~~~~~~~~~~~~~~~~
\vspace*{-1mm}
\ed
\bd
=
\overline{\prod_{i<j} \left\{1\plus \frac{c}{N}[
e^{-\frac{i}{c}\sum_{\mu}\xi_i^\mu\xi_j^\mu\sum_t \hh_i(t)\sigma_j(t)}\minus 1]\right\}
\left\{1\plus\frac{c}{N}[
e^{-\frac{i}{c}\sum_{\mu}\xi_i^\mu\xi_j^\mu\sum_t\hh_j(t)\sigma_i(t)}\minus 1]\right\}
}
\ed
\bd
=
\overline{\prod_{i<j} \left\{1\minus \frac{c}{N}[
\frac{i}{c}\sum_{\mu}\!\xi_i^\mu\!\xi_j^\mu\!\sum_t\hh_i(t)\sigma_j(t)
\plus \frac{1}{2 c^2 }[\sum_{\mu}\!\xi_i^\mu\!\xi_j^\mu\!\sum_t\hh_i(t)\sigma_j(t)]^2
\plus \order(c^{-\frac{3}{2}})
]\right\}
}
~~~~~~~~~~~~~~~~~~~~~~~~~~~~~~~~
\vspace*{-1mm}
\ed
\bd
~~~~~~~~~~~~~~~~~~~~~~~~~~~~~~~~
\overline{\times~
\left\{1\minus \frac{c}{N}[
\frac{i}{c}\sum_{\mu}\!\xi_i^\mu\!\xi_j^\mu\!\sum_t\hh_j(t)\sigma_i(t)
\plus \frac{1}{2 c^2 }[\sum_{\mu}\!\xi_i^\mu\!\xi_j^\mu\!\sum_t\hh_j(t)\sigma_i(t)]^2
\plus \order(c^{-\frac{3}{2}})
]\right\}
}
\ed
(in which the horizontal bars of the two constituent lines are to be read as
connected)
\bd
=\overline{\prod_{i<j} e^{-\frac{i}{N}\sum_{\mu}\!\xi_i^\mu\!\xi_j^\mu\!
\sum_t [\hh_i(t)\sigma_j(t)\plus \hh_j(t)\sigma_i(t)]
- \frac{1}{2 c N }[\sum_{\mu}\!\xi_i^\mu\!\xi_j^\mu\!\sum_t \hh_i(t)\sigma_j(t)]^2
- \frac{1}{2 c N }[\sum_{\mu}\!\xi_i^\mu\!\xi_j^\mu\!\sum_t \hh_j(t)\sigma_i(t)]^2
+ \order(\frac{1}{N\sqrt{c}})+ \order(\frac{c}{N^2})}}
\ed
Again we separate in the exponent the terms where $\mu=\nu$ in the quadratic term
(being of the form $\sum_{\mu\nu}\ldots$), and the
terms with $\mu=1$, and use the definitions (\ref{eq:onetime},\ref{eq:qQ},\ref{eq:K}):
\bd
{\cal F}[\ldots]=-i\sum_t a(t)k(t)- \frac{1}{2}\alpha \sum_{st}q(s,t)Q(s,t)
+ \order(c^{-\frac{1}{2}})+ \order(c/n)
\vspace*{-2mm}
~~~~~~~~~~~~~~~~~~~~~~~~~~~~~~~~~~~~~~~~~~~~~~~~~~~~~~~~~
\ed
\bd
+\frac{1}{N}\log\left\{
\overline{
e^{-\frac{i}{N}\sum_{\mu>1}\sum_t[\sum_i\xi_i^\mu\hh_i(t)][\sum_j\xi_j^\mu\sigma_j(t)]
- \frac{1}{2 c N }\sum_{i\neq j}\sum_{\mu\neq \nu}\!\xi_i^\mu\!\xi_j^\mu\!\!\xi_i^\nu\!\xi_j^\nu\!
\sum_{st} \hh_i(s)\sigma_j(s)\hh_i(t)\sigma_j(t)}}
\right\}
\ed
The scaling arguments given in the symmetric case, based on our
`condensed ansatz', apply again, and tell us that the remaining terms with
the disorder are of vanishing order in $N$.
We have again shown that the
disorder average (\ref{eq:modeldependent}) is, in leading
order in $N$, of the form (\ref{eq:dyn_meanfield}), with
\be
{\sl Asymmetric:}~~~~~~~~
\Phi[\ba,\bk,\bq,\bQ,\bK]=
-i\ba\cdot\bk
- \frac{1}{2}\alpha\sum_{st}q(s,t)Q(s,t)
\label{eq:Phi_asymmdiluted}
\ee
\vsp

\noindent{\em Extracting the Physics from  the Saddle-Point Equations.}
First we combine the above two results (\ref{eq:Phi_symmdiluted},\ref{eq:Phi_asymmdiluted})
in the following way (with $\Delta=1$
for symmetric dilution and $\Delta=0$ for asymmetric dilution):
\be
\Phi[\ba,\bk,\bq,\bQ,\bK]=
-i\ba\cdot\bk
- \frac{1}{2}\alpha\sum_{st}[q(s,t)Q(s,t)\plus \Delta K(s,t)K(t,s)]
\label{eq:Phi_bothdiluted}
\ee
We can now work out equations
(\ref{eq:nohats_onetime},\ref{eq:nohats_twotimes}), and use (\ref{eq:identify}) to
express the result at the physical saddle-point in terms of the trio $\{m(t),C(t,t^\prime),G(t,t^\prime)\}$. For
the single-time observables this gives (as with the fully connected
system)
$\ha(t)=k(t)$ and $\hk(t)=a(t)$; for the two-time ones we find:
\bd
\hat{Q}(t,t^\prime)=-\frac{1}{2}i\alpha C(t,t^\prime)~~~~~~~~~~
\hat{q}(t,t^\prime)=0~~~~~~~~~~
\hat{K}(t,t^\prime)=\alpha \Delta G(t,t^\prime)
\ed
We now observe that the remainder of the derivation followed for
the fully connected network can be followed with only two minor
adjustments to the terms generated by $\hat{K}(t,t^\prime)$ and by $\hat{Q}(t,t^\prime)$:
 $\alpha \bG(\one\minus\bG)^{-1} \to \alpha\Delta\bG$
in the retarded self-interaction, and $(\one\minus\bG)^{-1}\bC(\one\minus\bG^\dag)^{-1}\to\bC$
in the covariance of the Gaussian noise in the effective single
neuron problem. This results in the familiar saddle-point equations
(\ref{eq:finalsaddle}) for an effective single neuron problem, with state
probabilities (\ref{eq:single_spin_fully}) equivalent to
the dynamics ${\rm Prob}[\sigma(t\plus 1)=\pm 1]=\frac{1}{2}[1\pm \tanh[\beta
h(t)]]$, and in which
$\pi_0(\sigma(0))=\frac{1}{2}[1\plus\sigma(0)m_0]$ and
\be
h(t|\{\sigma\}\!,\!\{\phi\})\!= \!m(t)\plus \theta(t)
\plus \alpha\Delta\sum_{t^\prime<t}G(t,t^\prime)\sigma(t^\prime)\plus \alpha^{\frac{1}{2}}\phi(t)
~~~~~~~~
P[\{\phi\}]=
\frac{e^{-\frac{1}{2}\sum_{t,t^\prime}\phi(t)\bC^{-1}\!(t,t^\prime)
\phi(t^\prime)}}
{(2\pi)^{(t_m+1)/2}{\rm
det}^{\frac{1}{2}}\bC}
\label{eq:finalresult_diluted}
\ee
\vsp

\begin{figure}[t]
\begin{center}\vspace*{-6mm}
\hspace*{-4mm}
\epsfxsize=67mm\epsfbox{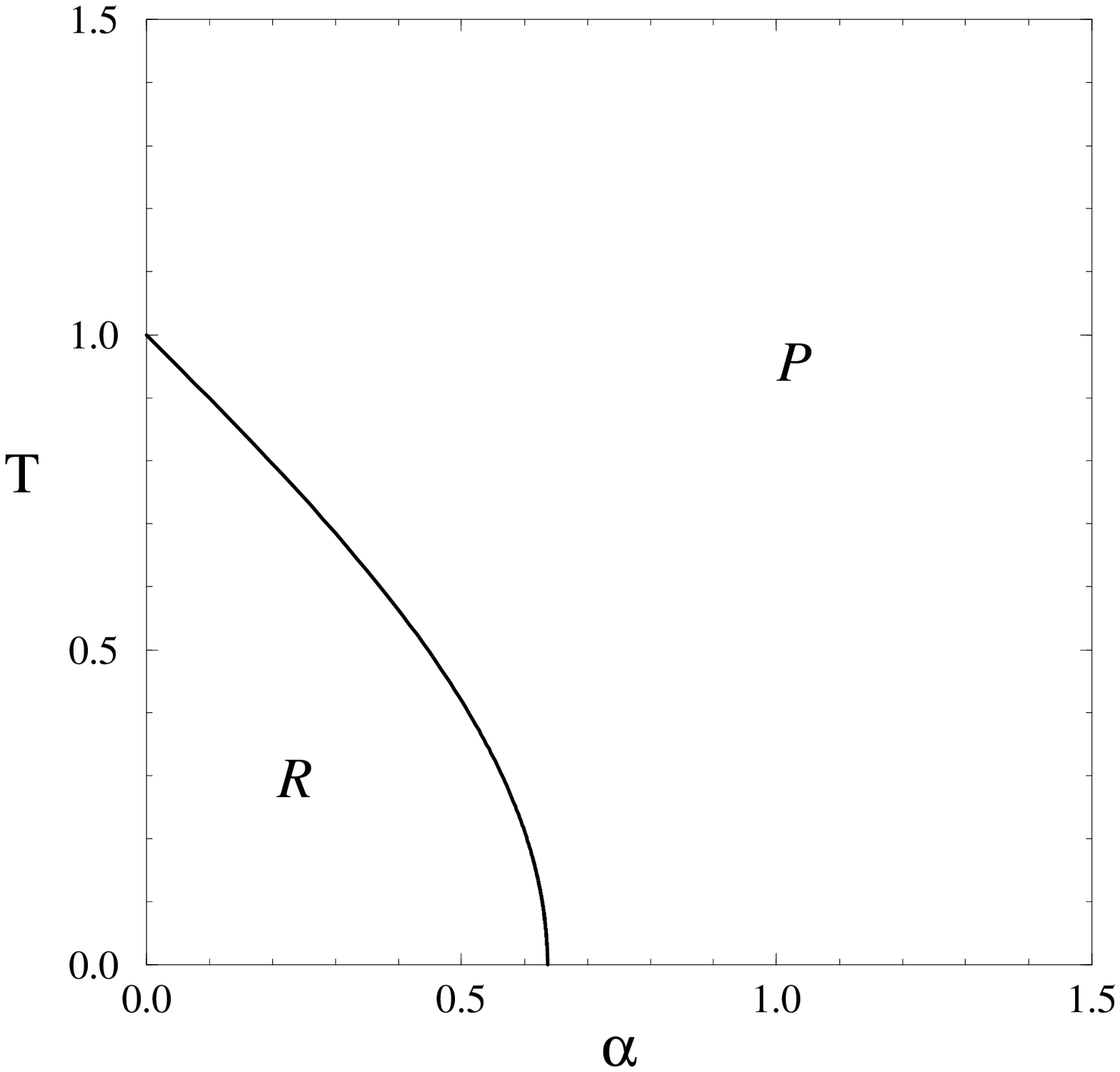}
\epsfxsize=67mm\epsfbox{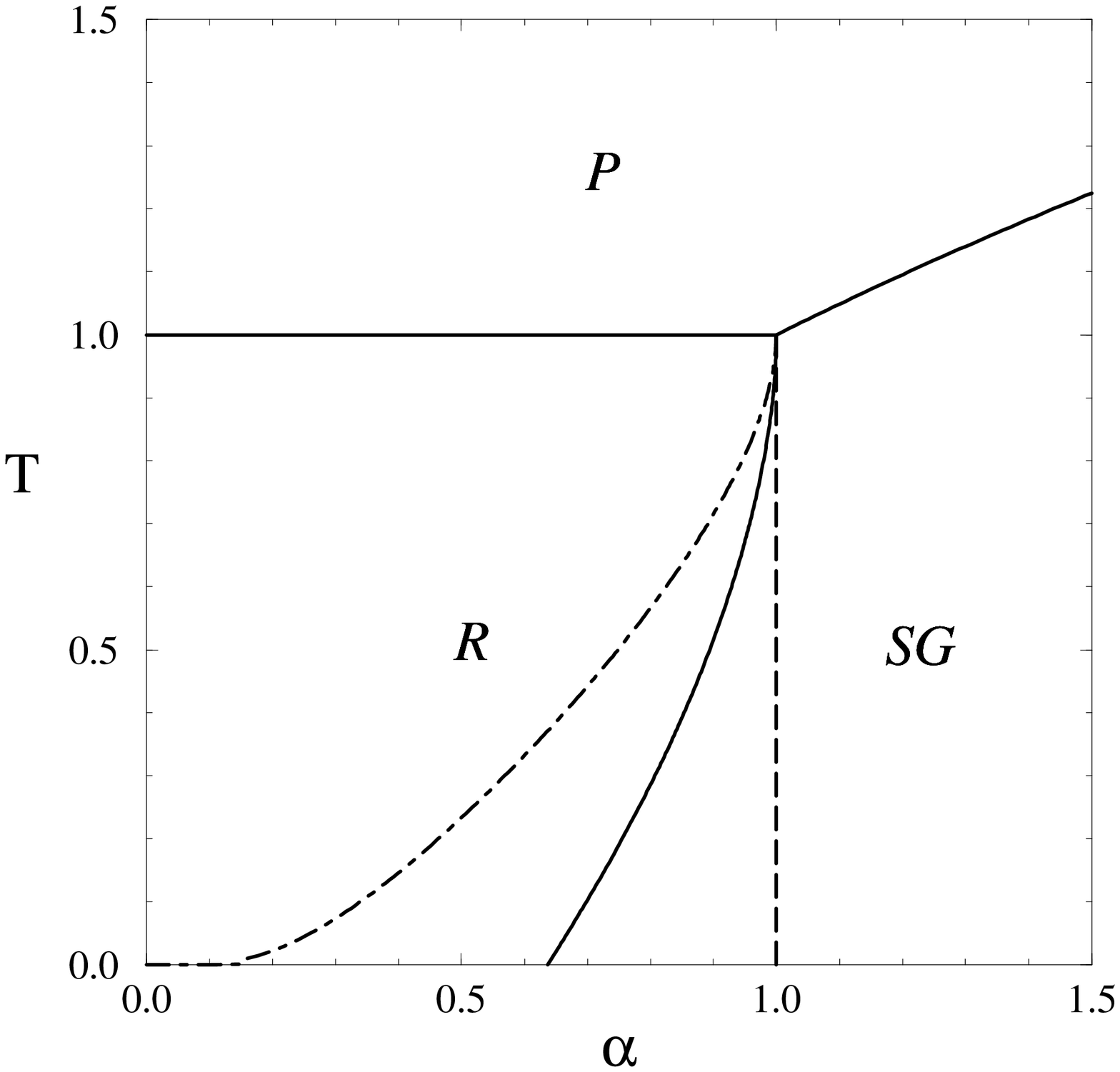}
\end{center}
\vspace*{-10mm}
\caption{
Phase diagrams of extremely diluted attractor networks.
Left: asymmetric dilution, $c_{ij}$ and $c_{ji}$ are statistically independent.
Solid line: continuous transition, separating a non-recall (paramagnetic) region (P) from a
recall region (R). The line reaches $T=0$ at $\alpha_c=2/\pi\approx 0.637$.
Right: symmetric dilution, $c_{ij}=c_{ji}$ for all $i,j$.
Solid lines: continuous transitions, separating a non-recall region (P) from a
recall region (R), for $\alpha<1$, and from a spin-glass region (SG), for $\alpha>1$.
Dashed-dotted line: the AT instability.
The R$\to$SG line (calculated within RS) reaches $T=0$ at $\alpha^{\rm RS}_c=2/\pi\approx 0.637$.
In RSB the latter is replaced by a new (dashed) line, giving a new storage capacity
of $\alpha_c^{\rm RSB}=1$.}
\label{fig:dilutedphases}
\end{figure}

\noindent{\em Physics of Networks with Asymmetric Dilution.}
Asymmetric  dilution corresponds to $\Delta=0$, i.e. there is no retarded self-interaction, and the
response function no longer plays a role. In (\ref{eq:finalresult_diluted}) we now only retain
$h(t|\ldots)= m(t)\plus \theta(t)\plus\alpha^{\frac{1}{2}}\phi(t)$, with
$\bra\phi^2(t)\ket=C(1,1)=1$. We now find (\ref{eq:m_worked}) simply
giving
\bd
m(t\plus 1)\!=\!\!\sum_{\sigma(0)\ldots\sigma(t)}\!\!\pi_0(\sigma(0))\int\!\{d\phi\}P[\{\phi\}]~
\tanh[\beta h(t|\ldots)]
\prod_{s=0}^{t-1}
\frac{1}{2}\left[1\plus\sigma(s\plus 1)
\tanh[\beta h(s|\ldots)]\right]
\vspace*{-3mm}
\ed
\be
=\int\!Dz
~\tanh[\beta(m(t)\plus \theta(t)\plus z\sqrt{\alpha})]
\label{eq:asym_diluted_overlap}
\ee
Apparently this is the one case where the simple Gaussian
dynamical law (\ref{eq:gaussian_simple}) is exact at all times.
Similarly, for $t>t^\prime$ equations (\ref{eq:G_worked},\ref{eq:C_worked1},\ref{eq:C_worked2})
for correlation and response functions reduce to
\bd
C(t,t^\prime)=
~~~~~~~~~~~~~~~~~~~~~~~~~~~~~~~~~~~~~~~~~~~~~~~~~~~~~~~~~~~~~~~~~~~~~~~~~~~~~~~~~~~~~~~~~~~~~~~~
~~~~~~~~~~~~~~~~~~~~~~~~~~~~~~~~~~~~~~~~~~~~~~~
\vspace*{-3mm}
\ed
\be
\int\!\frac{d\phi_a d\phi_b~e^{-\frac{1}{2}\frac{\phi_a^2+\phi_b^2-2C(t-1,t^\prime-1)\phi_a
\phi_b}{1-C^2(t-1,t^\prime-1)}}}
{2\pi\sqrt{1\minus C^2(t\minus 1,t^\prime\minus 1)}}
~\tanh[\beta( m(t\minus 1)\plus \theta(t\minus 1)\plus \phi_a\sqrt{\alpha})]
\tanh[\beta ( m(t^\prime\minus 1)\plus \theta(t^\prime\minus 1)\plus\phi_b\sqrt{\alpha})]
\label{eq:asym_diluted_correlation}
\ee
\be
G(t,t^\prime)=\beta\delta_{t,t^\prime+1}
\left\{1-
\int\!Dz
~\tanh^2[\beta(m(t\minus 1)\plus \theta(t\minus 1)\plus z\sqrt{\alpha})] \right\}
\ee
Let us also inspect the stationary state $m(t)=m$, for $\theta(t)=0$. One easily proves
that $m=0$ as soon as $T>1$, using
$m^2=\beta m\int_0^m dk[1\minus \int\!Dz\tanh^2[\beta(k\plus z\sqrt{\alpha})]]\leq \beta
m^2$.
A continuous bifurcation
occurs from the $m=0$ state to an $m>0$ state when $T=1\minus \int\!Dz~\tanh^2[\beta
z\sqrt{\alpha}]$. A parametrisation of this transition line in the
$(\alpha,T)$-plane is given by
\bd
T(x)=1\minus \int\!Dz~\tanh^2(zx),~~~~~~~~\alpha(x)=x^2
T^2(x),~~~~~~~~x\geq 0
\ed
For $\alpha=0$ we just jet $m=\tanh(\beta m)$
so $T_c=1$. For $T=0$ we obtain the equation $m={\rm erf}[m/\sqrt{2\alpha}]$,
giving a continuous transition to $m>0$ solutions at $\alpha_c=2/\pi\approx 0.637$.
The remaining question concerns the nature of the $m=0$ state.
Inserting $m(t)=\theta(t)=0$  (for all $t$) into (\ref{eq:asym_diluted_correlation})
tells us that $C(t,t^\prime)=f[C(t\minus 1,t^\prime\minus 1)]$ for
$t>t^\prime>0$, with `initial conditions' $C(t,0)=m(t)m_0$, where
\bd
f[C]=
\int\!\frac{d\phi_a d\phi_b}{2\pi\sqrt{1\minus C^2}}
~e^{-\frac{1}{2}\frac{\phi_a^2+\phi_b^2-2C\phi_a
\phi_b}{1-C^2}}
\tanh[\beta\sqrt{\alpha}\phi_a]
\tanh[\beta \sqrt{\alpha}\phi_b]
\ed
In the $m=0$ regime we have $C(t,0)=0$ for any $t>0$, inducing $C(t,t^\prime)=0$
for any $t>t^\prime$, due to $f[0]=0$.
Thus we conclude that $C(t,t^\prime)=\delta_{t, t^\prime}$ in the $m=0$
phase, i.e. this phase is para-magnetic rather than of a spin-glass
type. The resulting phase diagram is given in figure
\ref{fig:dilutedphases}, together with that of symmetric dilution
(for comparison).
\vsp

\noindent{\em Physics of Networks with Symmetric Dilution.}
This is the more complicated situation. In spite of the extreme dilution, the interaction
symmetry makes sure that the spins still have a sufficient number of common ancestors
for complicated correlations to build up in finite time. We have
\be
h(t|\{\sigma\}\!,\!\{\phi\})\!= \!m(t)\plus \theta(t)
\plus \alpha\sum_{t^\prime<t}G(t,t^\prime)\sigma(t^\prime)\plus \alpha^{\frac{1}{2}}\phi(t)
~~~~~~~~
P[\{\phi\}]=
\frac{e^{-\frac{1}{2}\sum_{t,t^\prime}\phi(t)\bC^{-1}\!(t,t^\prime)
\phi(t^\prime)}}
{(2\pi)^{(t_m+1)/2}{\rm
det}^{\frac{1}{2}}\bC}
\label{eq:symmdil1}
\ee
The effective single neuron problem (\ref{eq:single_spin_fully},\ref{eq:symmdil1}) is
found to be exactly of the form found also for the Gaussian model in \cite{part1} (which, in turn, maps onto
the parallel dynamics SK model \cite{SK}) with the synapses $J_{ij}=J_0 \xi_i\xi_j/N+J z_{ij}/\sqrt{N}$
(in which the $z_{ij}$ are symmetric zero-average and unit-variance Gaussian
variables, and $J_{ii}=0$ for all $i$), with the identification:
\bd
J\to\sqrt{\alpha}~~~~~~~~~~
J_0\to 1
\ed
(this becomes clear upon applying the generating functional analysis to the Gaussian model,
page limitations prevent me from explicit demonstration here).
Since one can show that for $J_0>0$ the parallel dynamics SK model gives the same
equilibrium state as the sequential one,
we can now immediately write down the stationary solution of our
dynamic equations which corresponds to the FDT regime, with $q=\lim_{\tau\to\infty}\lim_{t\to\infty} C(t,t\plus
\tau)$:
\be
q=\int\!Dz ~\tanh^2[\beta ( m\plus z\sqrt{\alpha q})]~~~~~~~~~~
m=\int\!Dz ~\tanh[\beta ( m\plus z\sqrt{\alpha q})]
\ee
These are neither identical to the equations for the fully
connected Hopfield model, nor to those of the asymmetrically diluted model.
Using the equivalence with the  (sequential and parallel) SK model \cite{SK} we can immediately
translate the phase transition lines as well, giving:
\bd
\begin{array}{lccc}
& {\sl SK~model} && {\sl Symmetrically~Diluted~Model} \\[2mm]
P\to F:                & T=J_0 ~~~~~~~~~~ {\rm for}~J_0>J  && T=1 ~~~~~~~ {\rm for}~\alpha<1\\[1mm]
P\to SG:               & T=J   ~~~~~~~~~~~ {\rm for}~J_0<J  && T=\sqrt{\alpha} ~~~~~ {\rm for}~\alpha>1 \\[1mm]
F\to SG~({\rm in~RS}): & T=J_0(1\minus q)~~{\rm for}~T<J_0 && T=1\minus q ~~ {\rm for}~T<1 \\[1mm]
F\to SG~({\rm in~RSB}):& J_0=J ~~~~~~~~~~~~ {\rm for}~T<J && \alpha=1 ~~~~~~ {\rm for}~T<\sqrt{\alpha} \\[1mm]
AT-{\rm line}:         & T^2\!=\!J^2\int\!Dz\cosh^{-4}\!\beta[J_0m\plus Jz\sqrt{q}]  &&
T^2\!=\!\alpha\int\!Dz\cosh^{-4}\!\beta[m\plus z\sqrt{\alpha q}]
\end{array}
\ed
where $q=\int\!Dz~\tanh^2\!\beta[m\plus z\sqrt{\alpha q}]$.
Note that for $T=0$ we have $q=1$, so that the equation for $m$
reduces to the one found for asymmetric dilution: $m={\rm
erf}[m/\sqrt{2\alpha}]$. However, the phase diagram shows that the
line $F\to SG$ is entirely in the RSB region and describes physically unrealistic re-entrance (as in the SK model),
so that the true transition must be calculated using Parisi's replica-symmetry breaking (RSB) formalism (see e.g.
\cite{Mezardetal}),
giving here $\alpha_c=1$.

The extremely diluted models analysed here were first studied in
\cite{DGZ}  (asymmetric dilution) and \cite{WatkinSherr}
(symmetric dilution). We note that it is not extreme dilution
which is responsible for a drastic simplification in the macroscopic
dynamics in the complex regime (i.e. close to saturation), but
rather the absence of synaptic symmetry. Any finite degree of
synaptic symmetry, whether in a fully connected or in an extremely
diluted attractor network, immediately generates an effective
retarded self-interaction in the dynamics, which is ultimately
responsible for highly non-trivial `glassy' dynamics.

\section{Epilogue}

In this paper I have tried to explain how the techniques from
non-equilibrium statistical mechanics can be used to solve the
dynamics of recurrent neural networks. As in the companion paper
on statics in this volume, I have restricted myself to relatively simple models,
where one can most clearly see the potential and restrictions of
these techniques, without being distracted by details.
I have dealt with binary neurons and graded response neurons, and
with fully connected and extremely diluted networks, with
symmetric but also with non-symmetric synapses.
Similar calculations could have been done for neuron models which
are not based on firing rates, such as coupled oscillators or
integrate-and-fire type ones, see e.g.
\cite{Gerstner}.
My hope is that bringing together methods and results that have so far been
mostly scattered over research papers, and by presenting these
in a uniform language to simplify comparison, I will have made the
area somewhat more accessible to the interested outsider.

At another level I hope to have compensated somewhat for the
incorrect view that has sometimes surfaced in the past that
statistical mechanics applies only to recurrent networks with
symmetric synapses, and is therefore not likely to have a lasting
impact on neuro-biological modeling. This was indeed true for
equilibrium statistical mechanics, but it is not true for
non-equilibrium statistical mechanics.
This does not mean that there are no practical restrictions in the
latter; the golden rule of there not being any free lunches is
obviously also valid here. Whenever we wish to incorporate more
biological details in our models, we will have to reduce our
ambition to obtain exact solutions, work much harder, and turn
to our computer at an earlier stage. However,
the practical restrictions in dynamics are of a quantitative
nature (equations tend to become more lengthy and messy), rather
than of a qualitative one (in statics the issue of detailed
balance decides whether or not we can at all start a calculation).
The main stumbling block that remains is the issue of spatial
structure. Short-range models are extremely difficult to handle, and
this is likely to remain so for a long time.
In statistical mechanics the state of the art in short-range
models is to be able to identify phase transitions, and calculate critical exponents, but this is
generally not the type of
information one is interested in when studying the operation of
recurrent neural networks.

Yet, since dynamical techniques are still far less hampered by the need to impose
biologically dubious (or even unacceptable) model constraints than equilibrium techniques, and since
there are now well-established and efficient methods and techniques
to obtain model solutions in the form of macroscopic laws for large
systems
(some are exact, some are useful
approximations), the future in the statistical mechanical
analysis of biologically more realistic recurrent neural networks
is clearly in the non-equilibrium half of the statistical mechanics playing field.

\subsection*{Acknowledgements}

I is my pleasure to thank Heinz Horner, David Sherrington and  Nikos Skantzos for
their direct and indirect contributions to this review.


\end{document}